\documentclass[%
preprint,
 amsmath,amssymb,
 aps,
]{revtex4-1}

\usepackage{graphicx}
\usepackage{dcolumn}
\usepackage{bm}

\usepackage{color} 

\usepackage{adjustbox}

\begin{document}

\newcommand\MI{{MI} }

\preprint{PRE}

\title{Impact of Embedding View on Cross Mapping Convergence}
\thanks{Distribution Statement A: Approved for Public Release; Distribution is Unlimited. PA Number \#18905}%

\author{Robert Martin}\email{robert.martin.101@us.af.mil}
\author{Justin Koo}%
\affiliation{%
  In-Space Propulsion Branch, Air Force Research Laboratory
}%


\author{Daniel Eckhardt}
\affiliation{
  Postdoctoral Fellow\\
  National Research Council\\
  Air Force Research Laboratory\\
}%


\date{\today}

\begin{abstract}
  Convergent cross mapping (CCM) provides a powerful new technique for exploring causal relationships in nonlinear coupled systems.  The method relies on Takens' theorem by exploiting the fact that any general observation of a smooth nonlinear coupled system contains the full system dynamics which may be recovered through time delay embeddings of sufficiently long observations of the system.  Though true in general, when dealing with finite length data that is potentially corrupted by quantization and noise, not every view of the embedding is equally useful at identifying the system dynamics. Classically, the heuristic of the first minimum of mutual information has been proposed as a means to select appropriate lags for time delay embedding methods. This criteria tends to be sensitive to additive noise and known to fail for some systems where mutual information monotonically decreases.  In this work, alternative heuristics that use mutual information as a metric for identifying useful embedding views are explored further. The impact of coordinate system and noise level on the identification of a useful time lag representative of the structure in chaotic data is studied with respect to several common low dimensional dynamical systems. 
The impact of selecting the first minimum of mutual information relative to several alternate heuristics for the appropriate time lag  in the context of of CCM method is then presented for both the simple dynamical systems as well as experimental data derived from a variety of observations of a Hall Effect Thruster (HET) plasma propulsion device.  
It is found that the shorter of the two global maxima of mutual information from the pair of signals in a stretched discrete Legendre orthonormal coordinate system is a more robust option relative to the alternatives for selecting embedding lag for use with the CCM method. This is demonstrated by the choice's ability to identify time lags commensurate with the maximum bidirectional CCM correlation results generated by sweeping over a wide range of potential time lags. This enhanced performance results from a decreased sensitivity to noise and the resulting fluctuations in the estimates of mutual information when compared to lags derived from local criteria on mutual information.
  
\end{abstract}

\maketitle


\section{Introduction}
\label{sec:intro}

The use of time delay shadow manifold reconstruction as a mechanism to elucidate causality in complex dynamic systems has had a reinvigoration in the recent works of Sugihara et. al. \cite{Sugihara1227079}.  These works build from Takens' theorem \cite{Takens} which states that the entire state-space dynamics of a coupled non-linear system is embedded within the resulting shadow manifold structure constructed from sufficiently high dimensional lags of a single time series trace given appropriate constraints on the smoothness of state-space dynamics.  The resulting analysis tool, convergent cross mapping (CCM), relies on the observation that if parameter X causes Y in such a system, then the near neighbors in that shadow manifold resulting from lags of Y can be used to reconstruct the original X given sufficiently large amounts of data.  This requirement for sufficient data stems from the needed densification of the state-space shadows ensuring that neighboring points on the manifold represent nearby states of the original system and therefore identify times when the system dynamics were similar\footnote{This point is covered well in section 2.2 of Reference \cite{Krakovska} where a review of the concept of non-linear observability as an extension to the common concept of linear observability from control theory is described.}.

In practice, both an appropriate dimensionality and choice of time lag values, collectively referred to as the embedding parameters, strongly affect the performance of the method for finite, noisy real-world signals.  With regards to the appropriate dimensionality of an embedding, Takens' theorem guarantees the existence of an embedding in $\mathbb{R}^{(2d_a+1)}$ for a $d_a$-dimensional attractor.  For smooth low-dimensional problems, it is also possible to apply the False Nearest Neighbor (FNN)\cite{KennelFNN} algorithm to identify the dimensionality of a system.  However for arbitrary high dimensional systems, the dimensionality of the attractor generally remains unknown since the amount of data required to densely populate the attractor grows with the power of its dimension.  Moreover, the existence of finite noise resulting from either truly stochastic behavior or dynamics decoupled from the system can obfuscate how reliably near neighbors converge in higher dimensions. These challenges to applying the FNN method were addressed by Cao in Reference \cite{CaoMethod}, but the results obtained remain practically dependant on first identifying an appropriate time lag. For the use of the CCM technique, finite data constraints typically restrict the embedding to relatively few dimensions despite the ubiquity of noise, although techniques such as multiview embedding\cite{Ye922} demonstrate potential avenues to extend the analysis methods to higher dimensional systems.

More relevant to practical application of the CCM technique is the identification of appropriate time lags for each of these potential finite dimensional embeddings.  From theory, as the amount of data $n$ approaches infinity for a finite dimensional attractor, the lag, $\tau$, used for the embedding becomes irrelevant as the density on the points on the finite dimensional attractor becomes high.  However, in real-world data, the shortest time lags ($\tau \rightarrow 0$) available from measurement are generally the most influenced by experimental noise, while arbitrarily long time lags ($\tau \rightarrow \infty$) eventually destroys the predictive power of CCM. This is due to chaotic effects obfuscating the causal relationships in the data. At long lags, the subsequent points appear only stochastically related rather than the result of causal dynamics.  For widespread application of this technique, it is highly desirable to provide a mechanism for automatic selection of lag for time delay embedding that removes the arbitrary selection of the time lag parameter from the CCM process.  Therefore, the primary goal of this paper is to demonstrate an algorithm to reliably identify a time lag that performs well relative to an optimized value which falls between these two asymptotic limits for finite, noisy data. The goal is to establish a method of identifying a lag likely to produce a ``good embedding'' in that it works effectively with the CCM technique to identify causality in real-world systems.

This work demonstrates a new criteria for identifying an optimal time lag based on the global maximum mutual information (\MI) of an orthogonalized system view.  Background on the construction of time delay embedding, evaluation of embedding quality through the numerical evaluation of the mutual information measure, and motivation for improvements are provided in Section~\ref{sec:background}.  The development of alternative embedding views (i.e. an orthogonalized embedding) to better numerically assess \MI, along with application to canonical chaotic systems, is provided in Section~\ref{sec:alt}.  The impact of these choices are shown to be particularly important in the presence of noise.  Based on these efforts, the impact of a new algorithm for automatic time lag selection is presented and demonstrated on CCM investigations of both canonical and experimental datasets in Section~\ref{Sec:Impact}.

\section{Background}
\label{sec:background}

The CCM methodology is enabled by a nonlinear representation of signal dynamics using a time delay embedding technique.  This technique takes a single dimensional temporal signal and recasts it as a set of symbols in a higher dimensional state space, providing a clear connection to a core goal of information theory, the accurate and efficient encoding of information in sequences of symbols.  Shifting to an information theory perspective opens up numerous possibilities, such as using mutual information (\MI) to evaluate the information content of a temporal signal by studying probability distributions in a higher dimensional embedding.  This idea of using information theory concepts as a route towards understanding high dimensional embeddings of system dynamics has an interesting parallel to the original view of lossless communication as a sequence of corrupted but distinguishable symbols in a high dimensional state space from Reference \cite{ShannonNoise}.  This has direct implications to the concept of what a `near-neighbor' truly means for corrupted noisy observations.  This section explores the background tenets of time delay embedding, \MI, and the connection between the two.

\subsection{Time Delay Embedding Approach}
\label{subsec:time_delay}

Given a discrete time series $\left\{ X\right\}=\left\{x_1,x_2,\dots,x_{n}\right\}$ which represents samples of a continuous time series $s(t)$, a $d$-dimensional phase portrait which represents a shadow of the attractor manifold of some higher dimensional system dynamics can be constructed from vectors of sequential points lagged by $k$-timesteps as shown in Equation \ref{Eq:LagVector}.

\begin{equation}
  \left\{\vec{X}^d\right\}=\begin{Bmatrix} %
  \left\{x_1,x_{1+k},\dots,x_{1+(d-1)k}\right\},\\
  \left\{x_{2},x_{2+k},\dots,x_{2+(d-1)k)}\right\},\\
  \vdots\\
  \left\{x_{n-kd},x_{n-k(d-1)},\dots,x_n\right\} \end{Bmatrix}=\begin{Bmatrix}%
  \vec{x}_1^d,\\
  \vec{x}_2^d,\\
  \vdots\\
  \vec{x}_{n-kd}^d \end{Bmatrix}
  \label{Eq:LagVector}
\end{equation}

Clearly, this is an embedding of the discrete 1D time signal into a high dimensional space populated with symbols (i.e. each entry of $\vec{X}^d$) representing a location in state-space.  For a given dimensionality and defined lag values, this time delay embedding is used directly\footnote{For instance, the CCM method requires is identification of near-neighbors in state space.  This is accomplished efficiently by tracking the Euclidean distance between symbols.} as part of the CCM analysis.  However, to address the question of an optimal lag, it is helpful to consider the combined behavior of the system as a whole.  This is accomplished through a stochastic representation whereby state-space is binned discretely and the probability of visiting each region of state-space is evaluated.  While the resulting so-called phase portraits lose some of the uniqueness of the exact time delay embedding, the probabilistic nature of the phase portraits permits ready application of the \MI metric.

\subsection{Information Theory}

The idea of mutual information descends from Shannon's information entropy as defined in Equation \ref{Eq:Ent}  where $p(\vec{x}^d)$ is the probability density of the distribution at point $\vec{x}^d$.

\begin{equation}
  H(\vec{X}^d)=-\frac{1}{d} \int p(\vec{x}^d) log(p(\vec{x}^d) d\vec{x}^d
  \label{Eq:Ent}
\end{equation}

Though in the case of continuous probability densities where an infinitesimal volume, $dx^d$, has meaning, estimating a probability density from a finite number of discrete observations requires that the observations be grouped into bins of finite volume as in any density estimation.   The $k^{th}$-lagged observation vector, $\vec{x}_k^d$, can then be said to belong to the $i^{th}$ discrete class, $\mathtt{x}_i$, if the point lays within a compact finite bin such that $\vec{x}_i^j \le \vec{x}_k^j < (\vec{x}_i^j + \Delta \vec{x}^j) $ for every dimension $j\in d$.  The probability density for the $i^{th}$ class can then be estimated as the number of observations within the class divided by the total number of observations.  This results in a useful definition of the discrete entropy is shown in Equation \ref{Eq:discEnt} in units of bits where the summation is over the discrete classes of observation\footnote{Note that a $d$-dimensional entropy more akin to Shannon's original notation would be identical to Equation \ref{Eq:discEnt} with the caveat that the observation would lay in the $i^{th}$ class if the $d$-dimensional observation vector lays within a finite $d$-dimensional volume as $\vec{x}^d_i \le \vec{x}^d_k < (\vec{x}^d_i + \Delta \vec{x}^d) $.}.  This discrete entropy is really just an integration quadrature approximating the continuous entropy where the observation delta functions are assumed to span finite phase-space bins.

\begin{equation}
  H(X)=-\sum_{\mathtt{x}_i \in\left\{\mathtt{x}_i\right\}} p(\mathtt{x}_i)\; log_2 (p(\mathtt{x}_i))
  \label{Eq:discEnt}
\end{equation}

In two dimensions, this is equivalent to the joint entropy, $H(X,Y)$, resulting from simply summing over permutations of $\mathtt{x}^d_i$ and $\mathtt{y}^d_j$ with a joint probability density, $p(x,y)$.  The mutual information $I(X;Y)$ can then be defined through a variety of ways where the most accessible in this context is simply $I(X;Y)=H(X)+H(Y)-H(X,Y)$ which can similarly be extended via higher dimensional analogs.  The reader is referred to Reference \cite{CoverBook} for a more exhaustive review of these concepts.

For purely stochastic signals where the $i^{th}$ $x$-class and $j^{th}$ $y$-class are visited independently at random, the joint probability density is flat across all permutations of classes and the mutual information is identically zero.  In the case where the $Y$ observations are in fact identical to the $X$ observation, the joint probability density is simply a diagonal line where $i=j$ with $H(X,X)=H(X)$. This is because the probability density along the one dimensional line is the same as the one dimensional probability distribution, $p(x)$, which makes the mutual information in this case is exactly the one dimensional entropy $H(X)$. The mutual information of any signal real signal must fall between these extremes. 

\subsection{Relationship between Time Delay Embedding and \MI}

In the context of time delay embeddings, $d$-dimensional vectors of observations created from lagging of the time series can be binned into phase portraits and the \MI of these can be calculated directly.  For lag of zero, H(X)=H(Y)=H(X,Y), and the resulting \MI is maximized at a value of the entropy of the original binned one-dimensional signal.  This maximum \MI means that each point from the original signal is uniquely matched to the same point on the lagged signal since the lag is zero.  This embedding in higher dimensions adds no information value beyond the original system and is referred to as \textit{redundancy}\cite{Krakovska} or \textit{redundance}\cite{CASDAGLI199152}.  Moving to finite but very short lags, the \MI decreases as redundancy diminishes. However, in the case of many experimental signals, the \MI remains fairly high as short lags bias the embedding towards strong sensitivity to high frequency correlations.  A challenge in addressing these short time lags is that the relatively high \MI values reflect common information content about the quantized high frequency noise in the signal rather than the underlying signal dynamics themselves.  At a time delay approaching infinity for any finite amplitude real system that is not the perfectly periodic ideal, the signal and its lag eventually become statistically independent of each other.  This leads to a minimum mutual information that asymptotes to the entropy of a perfectly random stochastic signal without correlations in phase space.  This phenomena is referred to as \textit{irrelevance}\cite{Krakovska,CASDAGLI199152}.

The two asymptotic limits of irrelevance and redundancy can theoretically be avoided as the amount of data $n$ approaches infinity for a finite dimensional attractor.  In this case of very abundant data, the lag $\tau$ between subsequent point of the series is no longer significant as long as the finite dimensional attractor becomes dense faster than the lag approaches zero.  However, for finite, chaotic data, the lag chosen has had significant impact on the quality of the embedding and has been the subject of numerous investigations as outlined in Reference \cite{CellucciComparative} where the performance of a few of the most popular methods are compared.  The best performing, and of particular note with respect to the investigations of this work, was the use of Fraser's ``first minimum of mutual information'' \cite{Fraser} criteria to determine the optimal lag in the time delay in conjuction with the false nearest neighbors method for detecting the dimensionality of the attractor.  Mutual information is a powerful tool to quickly evaluate the relationship between a signal and it's lag.  For timescales shorter than the Lyapunov exponent, the \MI is expected to be relatively high since there is a strong deterministic dependence of the signal on its lag. However, at very long lags, chaotic systems become unpredictable and thus the \MI drops towards zero.  The heuristic chosen by Fraser, the first minimum of mutual information, provides a generally good choice of lag for time delay embeddings because it distinguishes a point where correlations that are purely the result of short sampling time redundance are overcome by the \MI contribution from relevant characteristic dynamic timescales of the system.  However, this tradeoff is a delicate balance and the existence of such a minima is not guaranteed.

\section{Alternative Embeddings}
\label{sec:alt}

The crux of the Fraser criteria is the existence of a local minimum of mutual information to identify an optimal lag.  However, this local minimum is hard to interpret.  Mutual information measures shared information relative to random independence and therefore implies some compactness of the data in phase space relative to independently distributed data.  The first local minimum in lag then represents the lag at which the data stops spreading out in phase space, if such a point exists. A more intuitive choice might be maximizing mutual information to emphasize the underlying signal structure relative to random noise, but this is not practical in practice because of strong short-time autocorrelations in the data.  Regardless of the criteria (maximum or minimum) for selecting an optimal lag, the discrete evaluation of \MI is subject to significant variability depending on the binning strategy employed on the joint probability distribution.  In this section, the use of a difference embedding and related orthogonalized embedding are shown to alleviate the issue of redundance at short time lags in the \MI measure, opening up the possibility of lag identification based on maximum \MI.  This is followed by an investigation of the impact of a non-uniform binning strategy on evaluation of the \MI measure.  Application is demonstrated on both the canonical R\"{o}ssler and Lorenz systems.

\subsection{Difference Embeddings}

Given any sequence of observations, $\{X\}$, a two-dimensional alternative to constructing the time delay phase portrait is to construct an $x$-$v$ phase portrait using the finite difference formula. This approach is part of a more general class of difference embeddings.  In practice, the simplest difference embedding can be accomplished by finite difference of subsequent observations $v= (x(t)-x(t-\Delta t))/\Delta t +\mathcal{O}(\Delta t)$.  A more accurate discrete evaluation of $v$ can be accomplished by using the subsequent observation of $x$ to calculate $v=(x(t+\Delta t)-x(t-\Delta t))/2 \Delta t +\mathcal{O}(\Delta t^2)$, i.e. via a second order central difference of the velocity at position $\bar{x}$ for the same time, $\bar{t}$.

For smooth data, this derivative is most accurate as $\Delta t$ approaches zero, so it seems intuitive that for such a system, an $x$-$v$ phase diagram made via finite difference from lags of the time series should provide the same amount of information about the series as the original lag portrait.  In fact, as shown in the appendix of Reference \cite{Kraskov}, mutual information is independent of reparameterization for homeomorphisms.   
Since the linear transformation between time delay and $x$-$v$ coordinates, shown in Eq.~\ref{Eq:XXtoXV}, is a homeomorphism, the mutual information is indeed preserved for difference embedding.

\begin{equation}
  \begin{bmatrix} x(t-\tau/2) \\ v(t-\tau/2) \end{bmatrix} = \begin{bmatrix} 1/2 & 1/2 \\ 1/\tau & -1/\tau \end{bmatrix} \begin{bmatrix} x(t) \\ x(t-\tau) \end{bmatrix}
  \label{Eq:XXtoXV}
\end{equation}

The fact that the mutual information should be preserved under linear transformation opens up other possibilities for other linearly transformed embeddings.  It is quite easy to perform a simple rotation of the time delay data into the coordinate system defined by the eigenvectors of the symmetric positive definite $d \times d$ covariance matrix, $C=(X^d)(X^d)^\intercal$. Implemented using the eigendecomposition $C=V \Lambda V^\intercal$, this embedding results in an orthogonal coordinate system for $X'^d$ using the eigenvectors as $X'^d=VX^d$. This is equivalent to the principal component analysis (PCA) algorithm for decomposing multidimensional data into linearly independent coordinates.

In exploring the results of rotating the time-delay data into these PCA coordinates, it was noted that the eigenvectors for 2D lags corresponded closely to the $x$-$v$ coordinates up to a normalization apparently independent of the system being studied.  It turns out that this is a direct consequence of the so-called ``Small-window solution'' relating PCA and the discrete Legendre coordinates as given in Reference \cite{GIBSON19921} for bounded analytic functions with non-zero amplitude variance and energy in the limit $T\rightarrow \infty$ as shown in Equation \ref{Eq:Bounded}.   These requirements restricts application of these coordinates to continuous systems with finite, non-stationary solutions, but this is generally compatible with a broad class of physically motivated coupled nonlinear systems with non-trivial behavior.   These discrete Legendre coordinates will be referred to as orthogonal coordinates and denoted with the $\perp$ symbol for the remainder of this work.  In the case of the 2D, this corresponds to a simple rescaling of the $x$-$v$ difference embedding, but, as discussed in Section \ref{SubSubSec:Higher}, these coordinates diverge from a simple difference embedding in higher dimensions. 

\begin{equation}
  \begin{aligned}
    &x(t) \text{ is analytic on } t \in \mathbb{R} \\
    &x(t) \text{ is bounded and its derivates are bounded for } t \in \mathbb{R}\\
    &\lim_{T\rightarrow \infty}T^{-1}\int_{-T}^{T} \left[x(t)\right]^2 dt \neq 0\\
    &\lim_{T\rightarrow \infty}T^{-1}\int_{-T}^{T} \left[\frac{dx}{dt}(t)\right]^2 dt \neq 0\\
  \end{aligned}
  \label{Eq:Bounded}
\end{equation}

Prior to binning, the \MI of the time delay and orthogonal embedding are mathematically identical. However, the binning strategy (i.e. the discretization of the joint PDF) used in the practical evaluation of \MI has a strong effect on the value estimated\footnote{Note that in Reference \cite{Kraskov}, a bin independent method for estimating the mutual information based on k-nearest neighbors was described. Though this suggests a potential alternate route to the methods studied in this work for improved mutual information estimates used for selecting time lags, the investigation of the applicability of this method is left to future work.}.    
For the time delay embedding, as $\tau \rightarrow 0$, the tight clustering of data along the diagonal $i=j$ suggests that the square bins along the diagonal are self-similarly inadequate for estimating the true density and therefore mutual information.  No matter how fine of square bins are used as $\tau \rightarrow 0$, the data remains tightly clustered along a small volume along the diagonal of the square making the average density of the square a poor estimate of the probability density.  When the view is linearly transformed using the orthogonal coordinates, not only can the dimensions be scaled independent of $\tau$ to address the challenge of representing diagonal lines with square bins, as explored further in Section \ref{SubSec:Nonuniform}, but also the impact of short-time redundance on the estimate of \MI can be directly eliminated.  Critically, this means that the mutual information is no longer maximal for the smallest possible time lag, enabling new search strategies for optimal time lags. 

\subsection{Performance of orthogonal embedding of smooth canonical systems}

To see the impact of this transformation for smooth data, we first revisit the often used simple 3-parameter Lorenz and R\"{o}ssler systems  given in Equations \ref{Eq:Lorenz} and \ref{Eq:Rossler} respectively where $\sigma$, $r$, $a$, $b$, and $c$ are constants affecting the dynamics.

\begin{align}%
  \dot{x}&=\sigma(y-x),\nonumber \\
  \dot{y}&=x(r-z)-y, \label{Eq:Lorenz}\\
  \dot{z}&=xy-bz\nonumber
\end{align}
\begin{align}
  \dot{x}&=-z-y,\nonumber \\
  \dot{y}&=x+ay,\label{Eq:Rossler}\\
  \dot{z}&=b+z(x-c)\nonumber 
  \end{align}

To compare the influence of lag on the mutual information, a simple $k^d$ uniform binning of phase space is used first to see the trends in mutual information and shadows of the dynamics. This is done for the lagged and orthogonalized coordinates with the understanding that these certainly represent a significantly sub-optimal choice of binning as shown in Section II of Reference \cite{CellucciMICalc}. The same parameters for the Lorenz ($\sigma=10$, $b=8/3$, $r=28$) and R\"{o}ssler ($a=0.2$,$b=0.4$,$c=5.7$) systems were used from Reference \cite{CellucciMICalc}.  The systems were run to 1\textsc{e}7 iterations using $dt=0.01$ for Lorenz and $dt=0.1$ for R\"{o}ssler using the default parameters of the lsode stiff ODE solver with absolute and relative tolerance of 1.49012e-8 as implemented in the GNU Octave code \cite{Octave}.

\begin{figure}
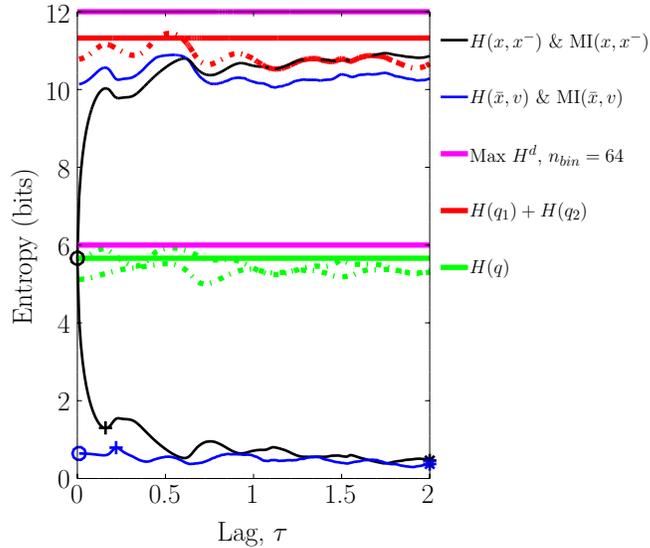

  \scalebox{0.5}{\include{ShadowStraightVsEig}}
  \caption{Entropy and mutual information of Lorenz-$x$ variable shadows on $64\times 64$ uniform bins for standard and orthogonalized coordinate systems relative limits from one-dimensional probability density entropies and d-dimensional maximums, Max $H^d$. The entropies are calculated for normalized transformed state observations as denoted in the legend where $q$ can be any of $x=(x(t))$, $x^-=x(t-\tau)$, $\bar{x}=(x+x^-)$, and $v=(x-x^-)$. Note that constant scaling is irrelevant due to normalization of coordinates in $[0,1]$. Markers denote first, first minima (or global maxima) of mutual information, and final state points to be discussed further.  Dotted lines result from $H(\bar{x})$, $H(v)$ and their sum, $H(\bar{x})+H(v)$, which are non-constant relative to their nearly constant counterpart quantities derived from the original $H(x)$, $H(x^-)$, and $H(x)+H(x^-)$ shown in solid lines of the same color. }
  \label{Fig:ShadowStraight}
\end{figure}

Figure \ref{Fig:ShadowStraight} depicts the variation of entropy and mutual information with lag for the $x$ values of the Lorenz system. In the figure, the rotated frame variables $\bar{x}$ and $v$ are denoted as such because the eigendecomposition of the covariance matrix naturally results in initial eigenvectors of $v_1=\{\sqrt{2},\sqrt{2}\}/2$ and $v_2=\{\sqrt{2},-\sqrt{2}\}/2$ which is equivalent to Equation \ref{Eq:XXtoXV} up to scaling constants. Note that the order of the eigenvalues is arbitrary which results in arbitrary permutation of the eigenvector order.  This does not impact the calculation of entropies, but potentially adds additional mirroring across both axes in addition to the rotation in phase portraits when calculated from numerically estimated PCA eigenvectors. As the scaling constants are subsequently removed in the binning process when the data is normalized, the phase portraits are equivalent to those of $\bar{x}$ and $v$.  Note also that this is simply a $45^\circ$-rotation of the data.  This results from the alignment of the principle component of a two dimensional covariance ellipse with the 1:1-line.  It is reasonable to expect this transformation as the natural result of the orthogonalization as the derivative of a parameterized unit vector is orthogonal to the vector and as $\hat{x}\cdot \hat{v}=0$.

\begin{figure}
  \includegraphics[width=0.3\textwidth]{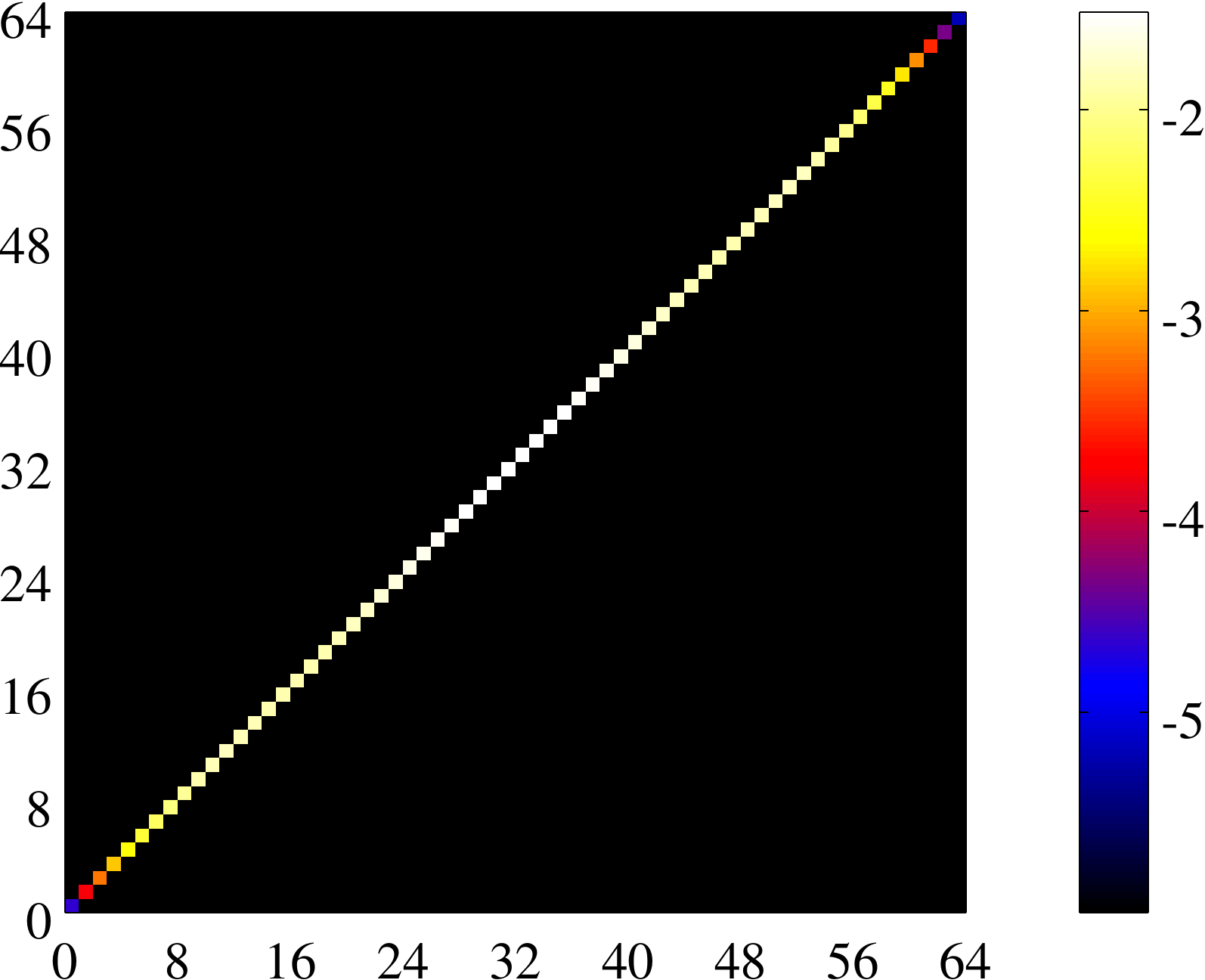}\hspace{16pt}%
  \includegraphics[width=0.3\textwidth]{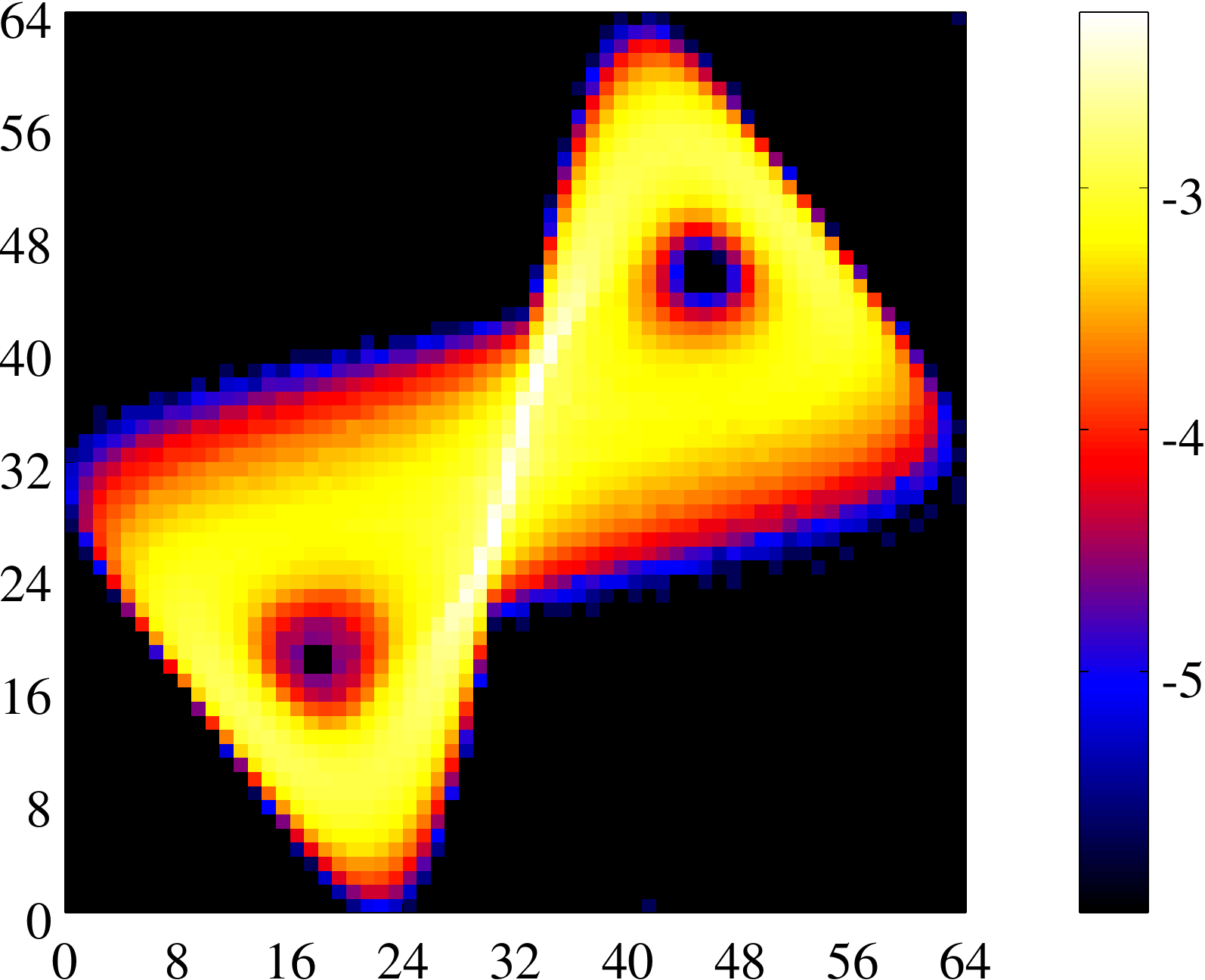}\hspace{16pt}%
  \includegraphics[width=0.3\textwidth]{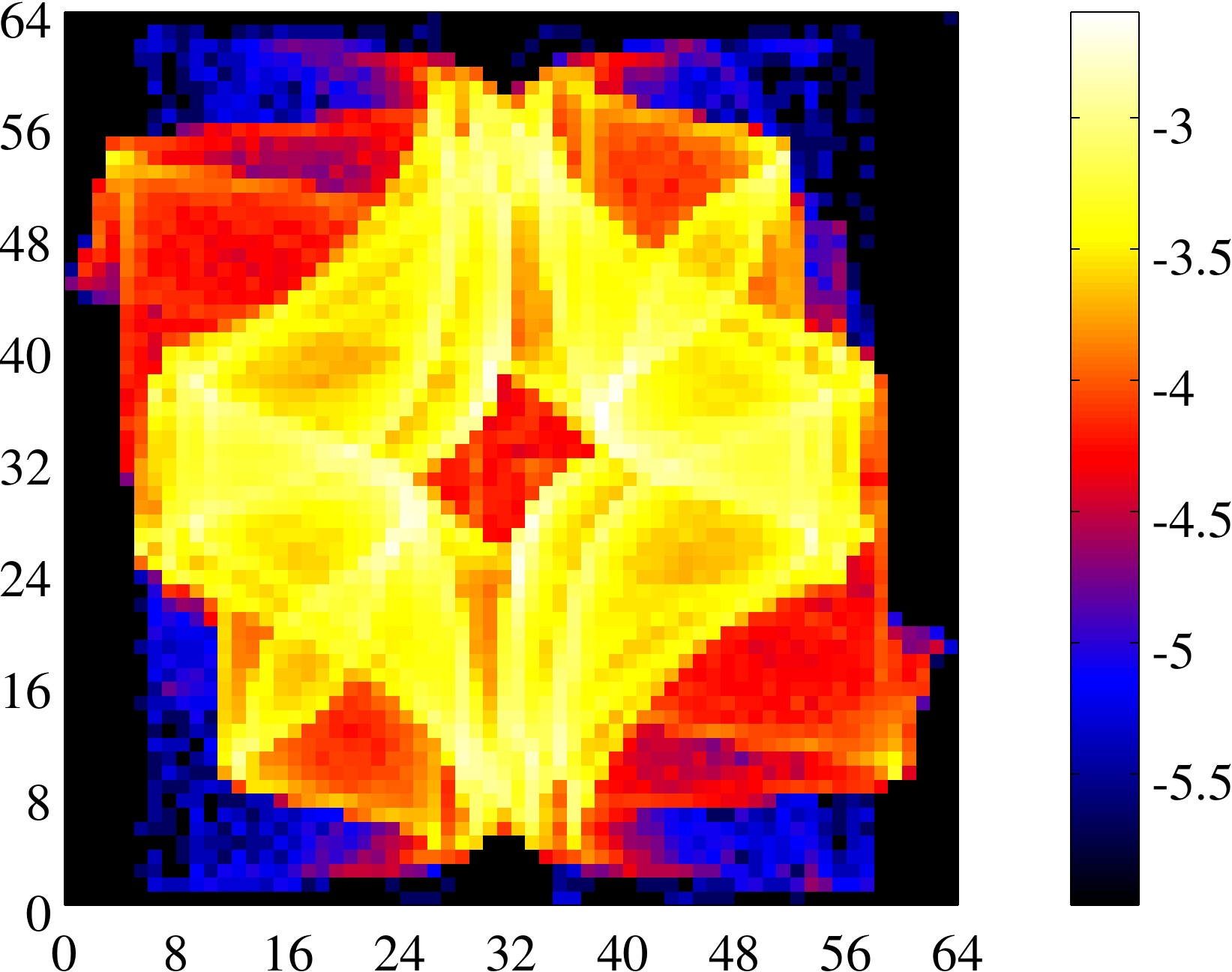}\\
  \includegraphics[width=0.3\textwidth]{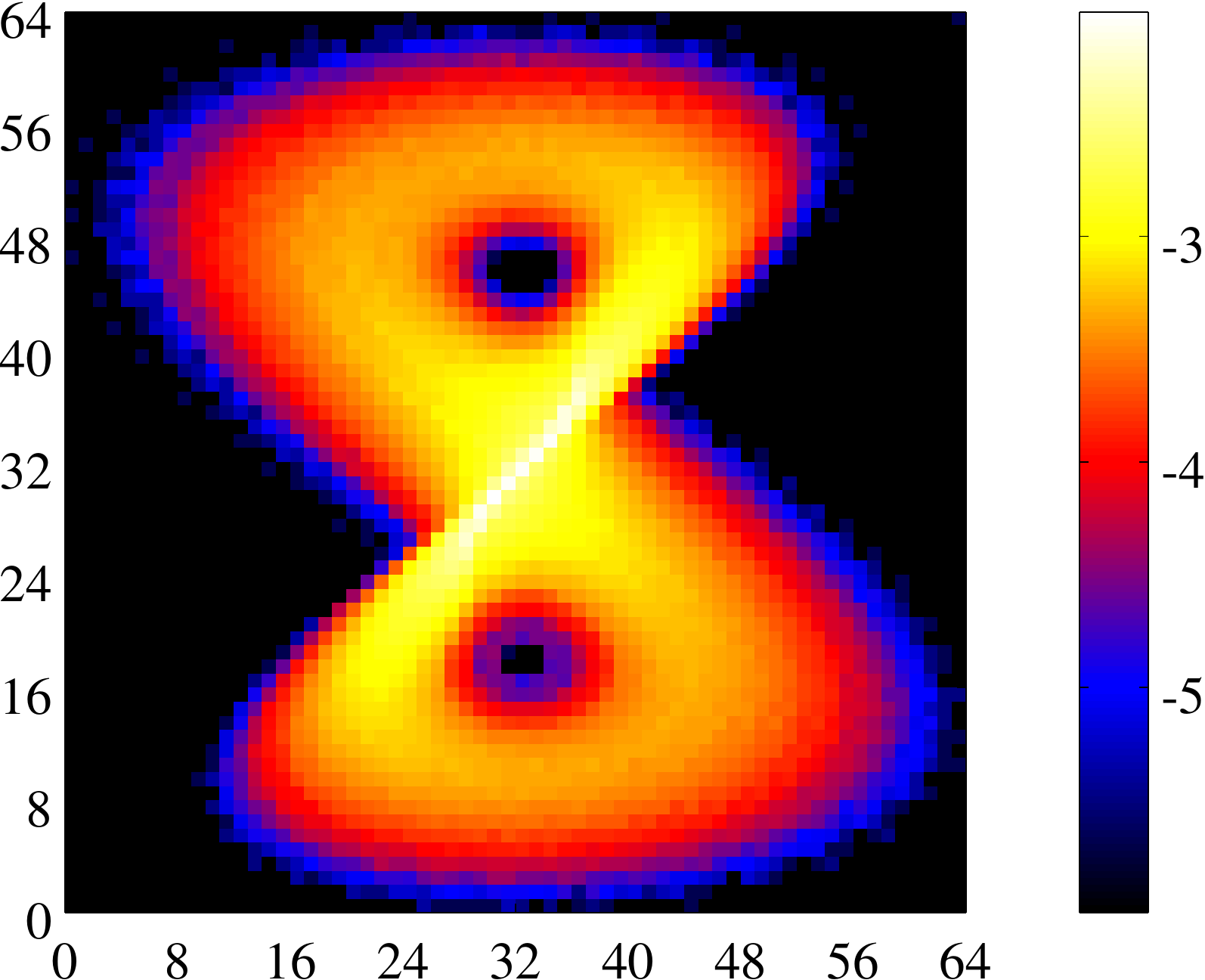}\hspace{16pt}%
  \includegraphics[width=0.3\textwidth]{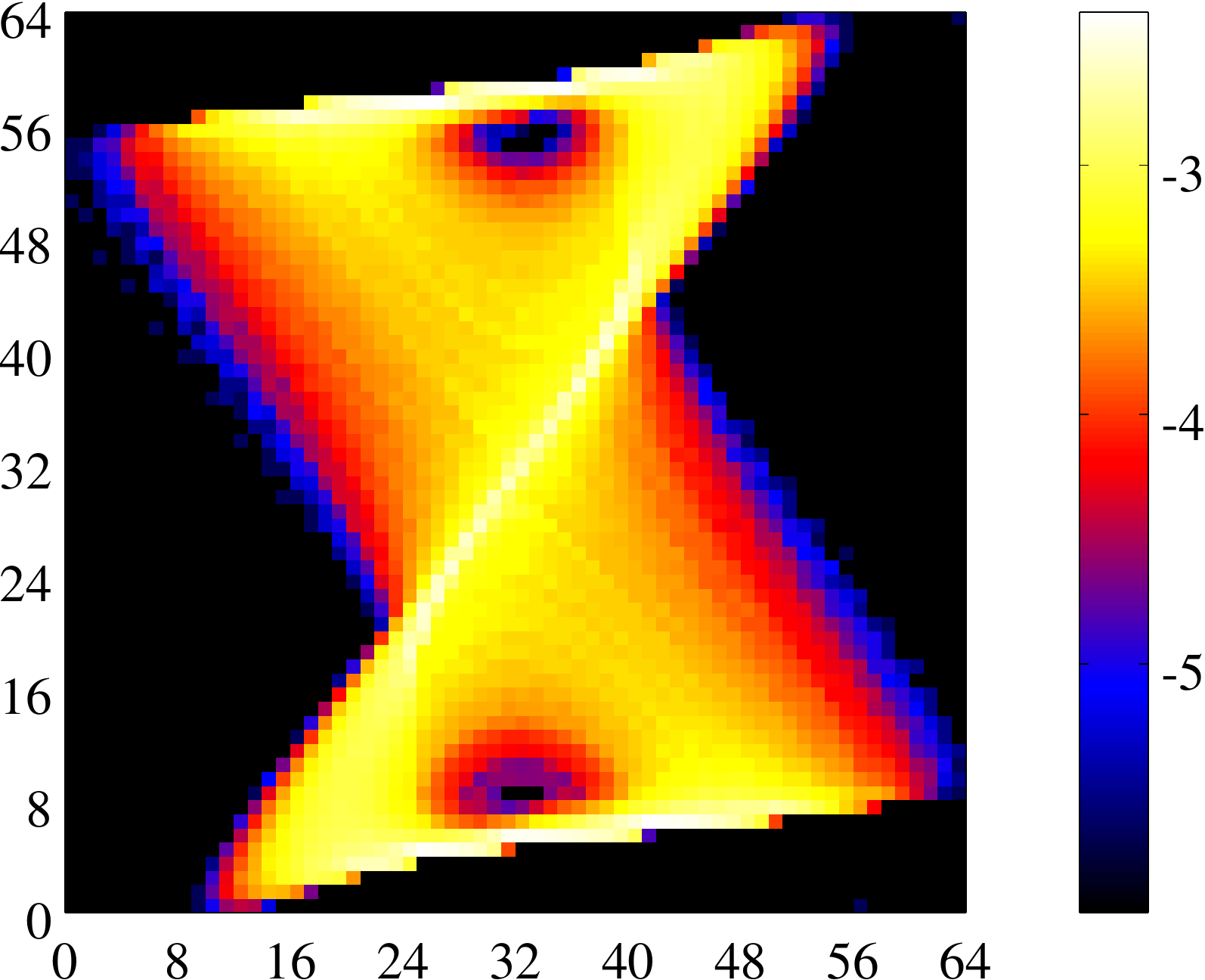}\hspace{16pt}%
  \includegraphics[width=0.3\textwidth]{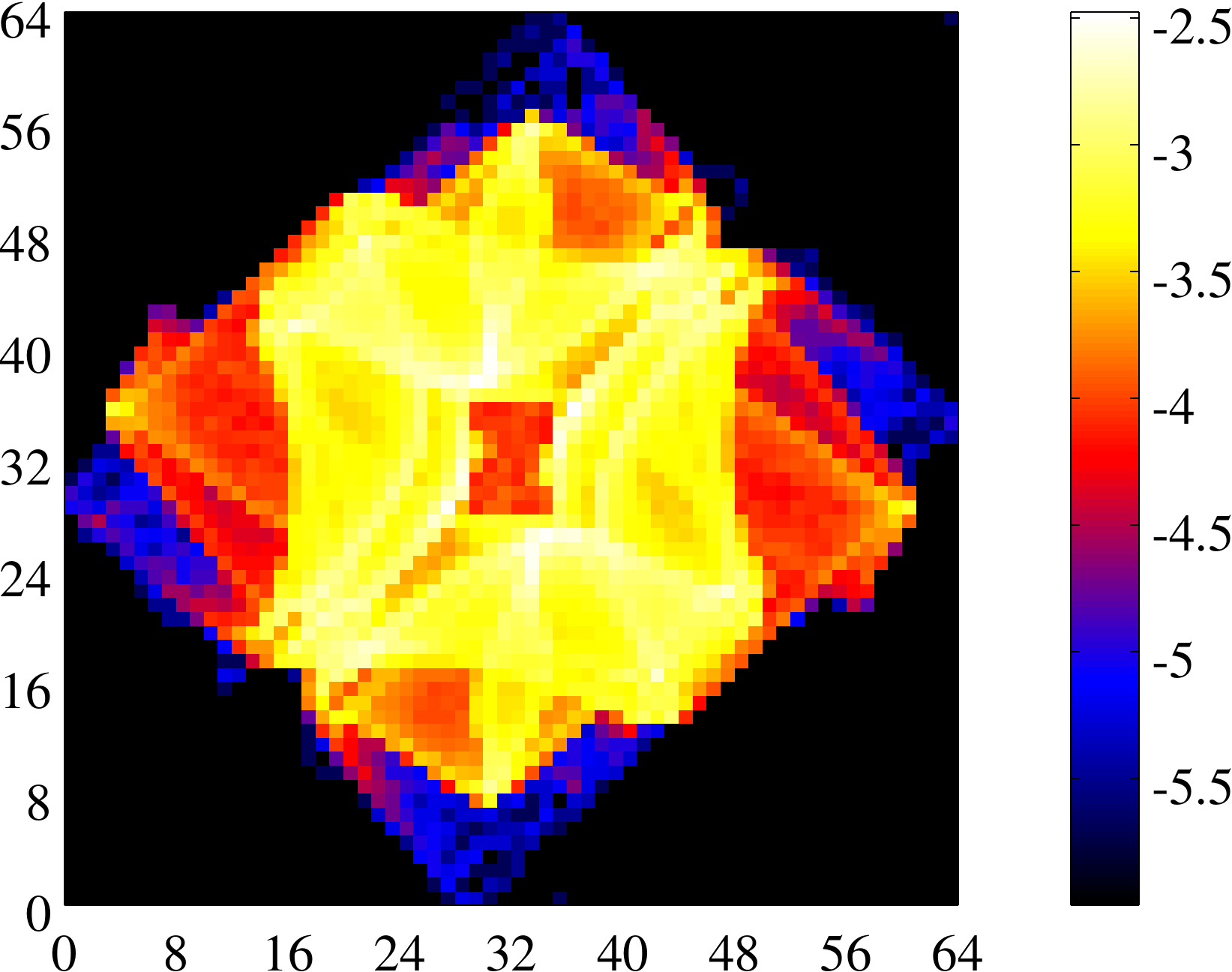}\\  
  \caption{Phase portraits from $x$ variable of lagged Lorenz data. Top row is original lagged coordinates and bottom row is in rotated orthogonal coordinates.  The left and right figures show first and last portraits denoted by ``$\circ$'' and ``$*$'' symbols on entropy and mutual information plot.  Middle figure shows views from ``$+$'' marked points due to the associated criteria on mutual information.  Colors are $log_{10}$ of the bin frequency with a floor of approximately 1\textsc{e}-7 added representing a count of 1 observation per bin to avoid $-\infty$ in the color axis.}
  \label{Fig:ShadowsOfStraight}
\end{figure}

In the Figure \ref{Fig:ShadowStraight}, markers denote a few lagged views of particular interest.  The first, ``$\circ$'', markers are the first points represented by the methods.  Note that the $\tau=0$ point is skipped because one of the eigenvalues is zero when all the covariances are equal resulting in a degenerate distribution of all $x-x^-=0$ points that cannot be scaled to the size of the box.  The last $\tau=2$ point equivalent to the 200 steps from Reference \cite{Kraskov} is marked by the ``$*$''-symbol. Finally, the ``$+$''-symbol is used to denote the first local minima of mutual information in the case of the original lagged coordinates and the global maxima of mutual information in the case of the rotated coordinates with the idea that the maximum mutual information may be a more useful view than a local minima if artificial correlation are first removed from the data via the orthogonalization process.  Figure \ref{Fig:ShadowsOfStraight} depicts the normalized probability densities on the $64\times 64$ mesh for these points.

\begin{figure}
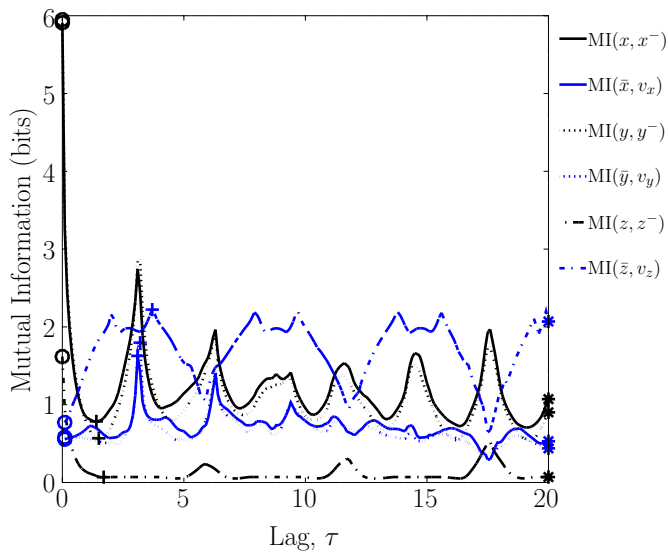

  \scalebox{0.5}{\include{ShadowStraightVsEigRossler}}
  \caption{Mutual information with respect to lag using standard lagged (black) and orthogonalized (blue) coordinates for the R\"{o}ssler system.  Line styles indicate results from the system's $x$, $y$, and $z$ state variables.}
  \label{Fig:RosslerMI}
\end{figure}

Looking at the mutual information of the R\"{o}ssler system identifies significantly different views for the lagged and orthogonalized views. Figure \ref{Fig:RosslerMI} shows the quantity of mutual information based on the R\"{o}ssler system's three state variables as studied in Reference \cite{CellucciMICalc}. As in the standard lagged view, the $z$-variable results in the most significantly different mutual information curves of the three in either view.

\begin{figure}
  \includegraphics[width=0.3\textwidth]{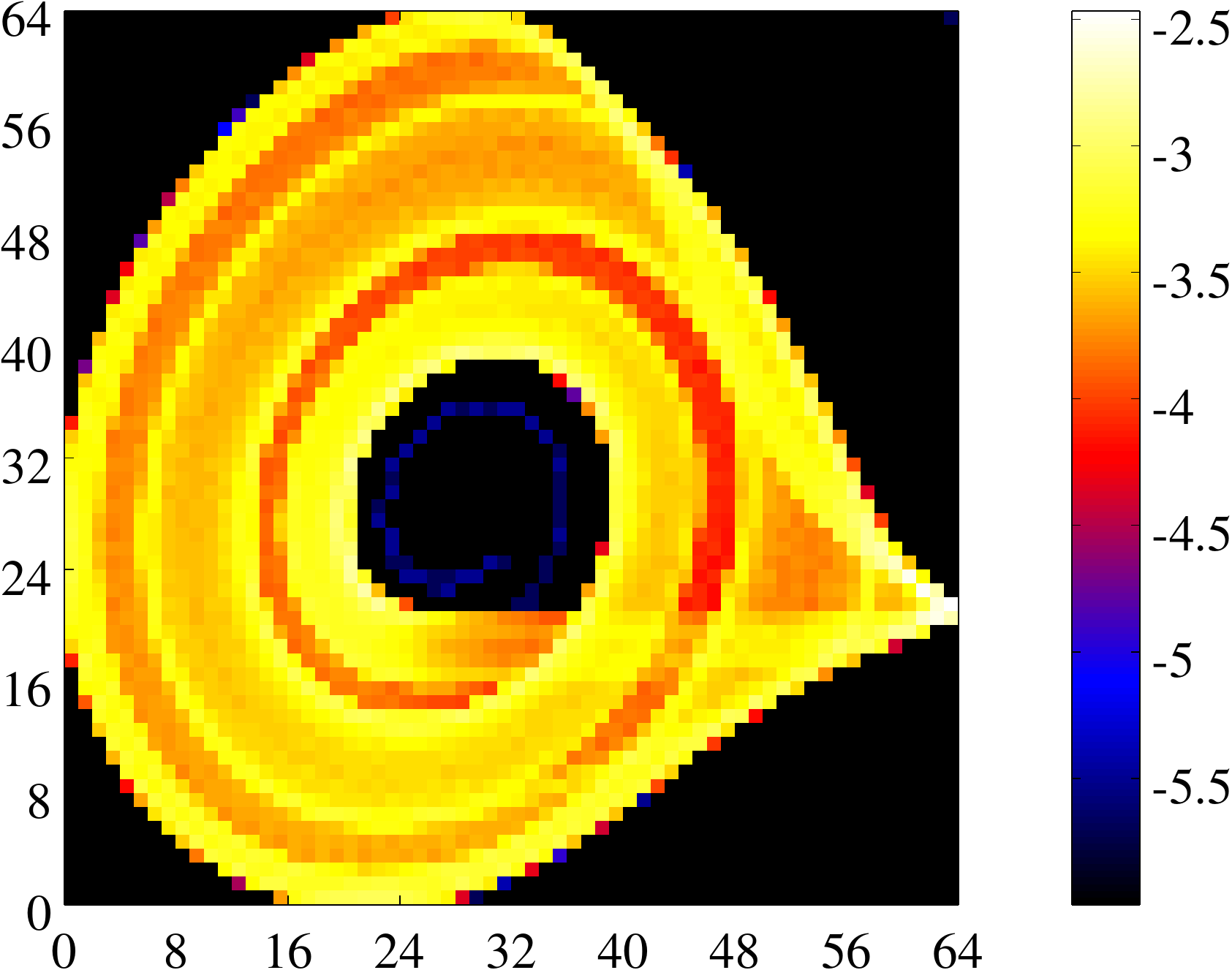}\hspace{16pt}%
  \includegraphics[width=0.3\textwidth]{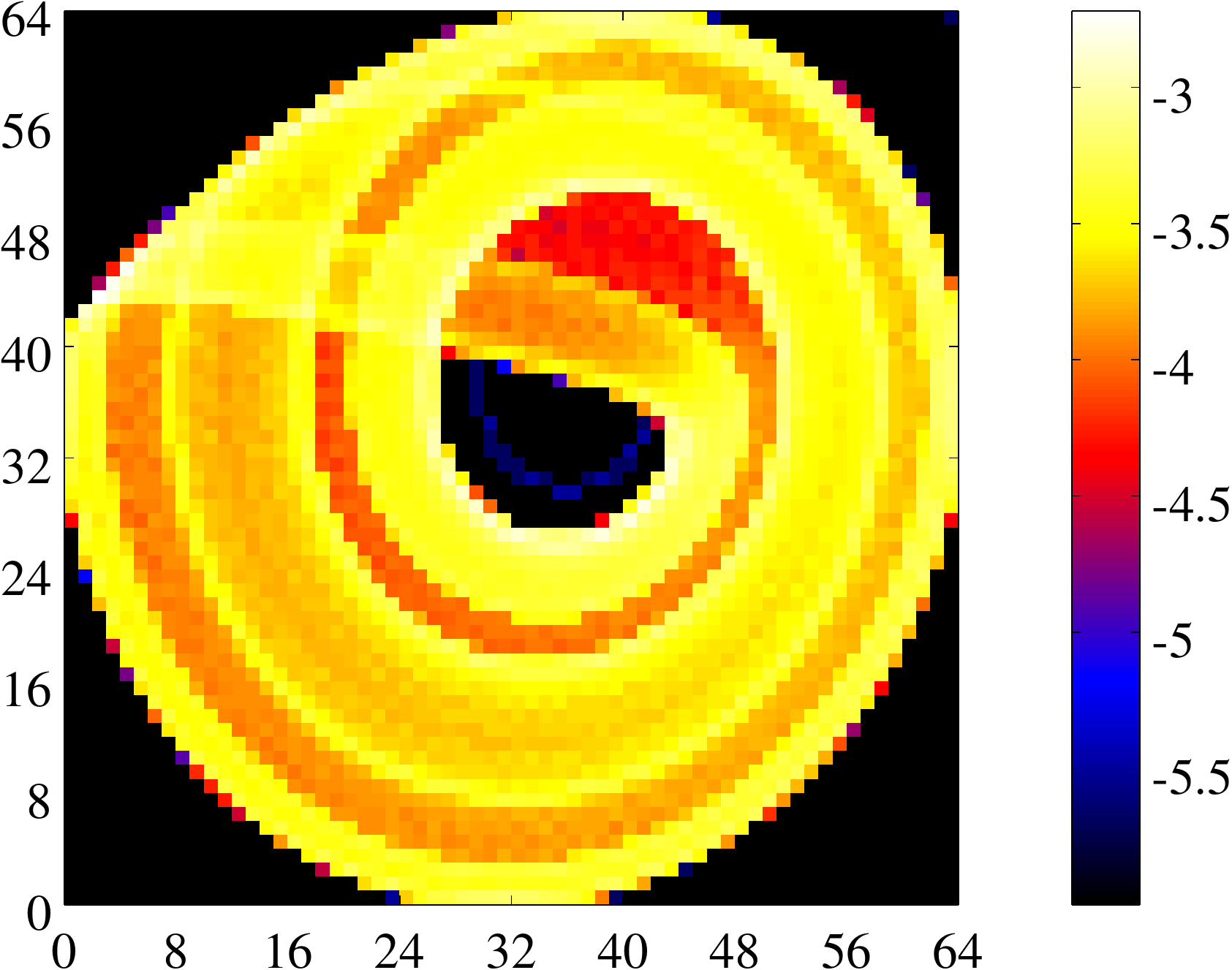}\hspace{16pt}%
  \includegraphics[width=0.3\textwidth]{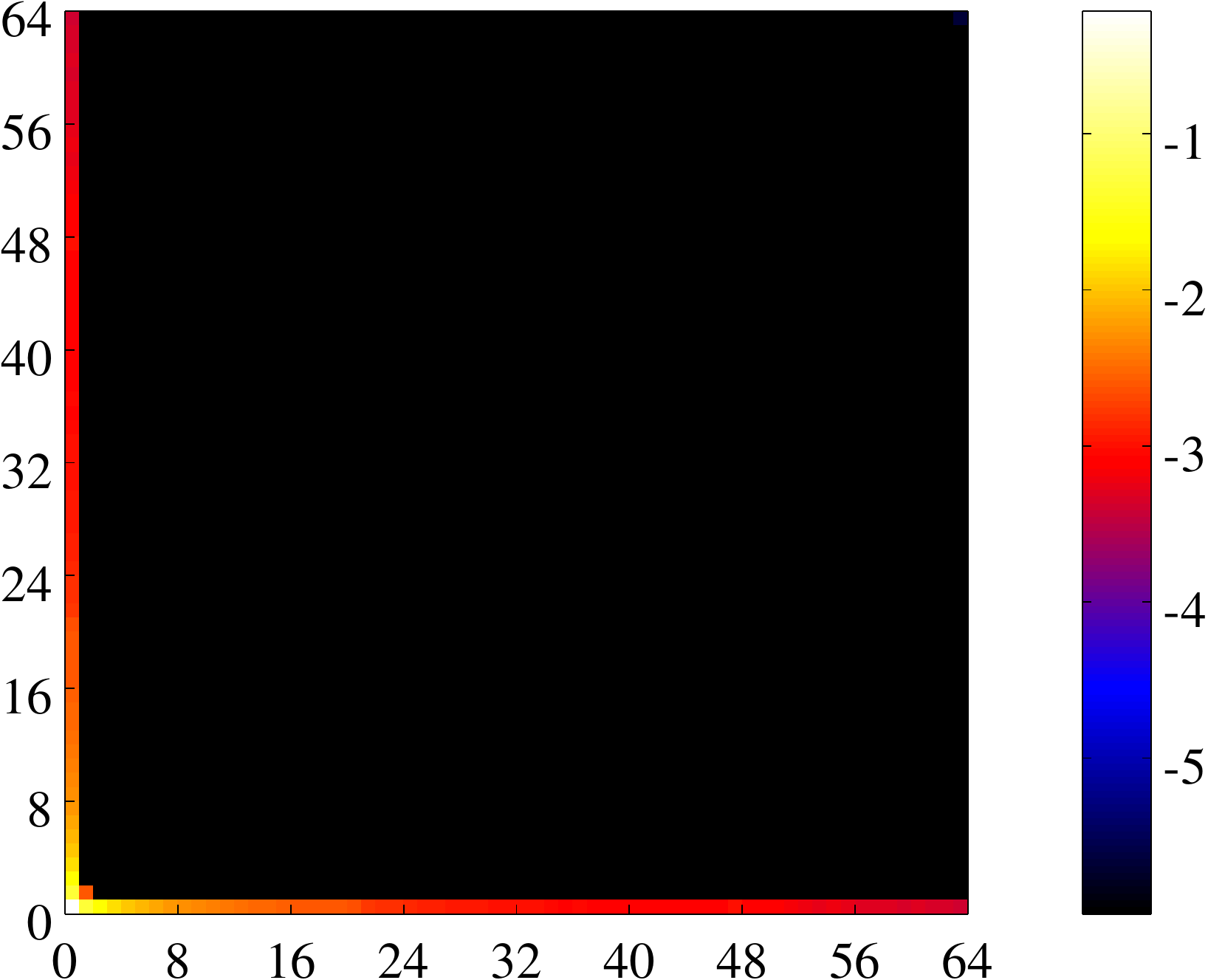}\\
  \includegraphics[width=0.3\textwidth]{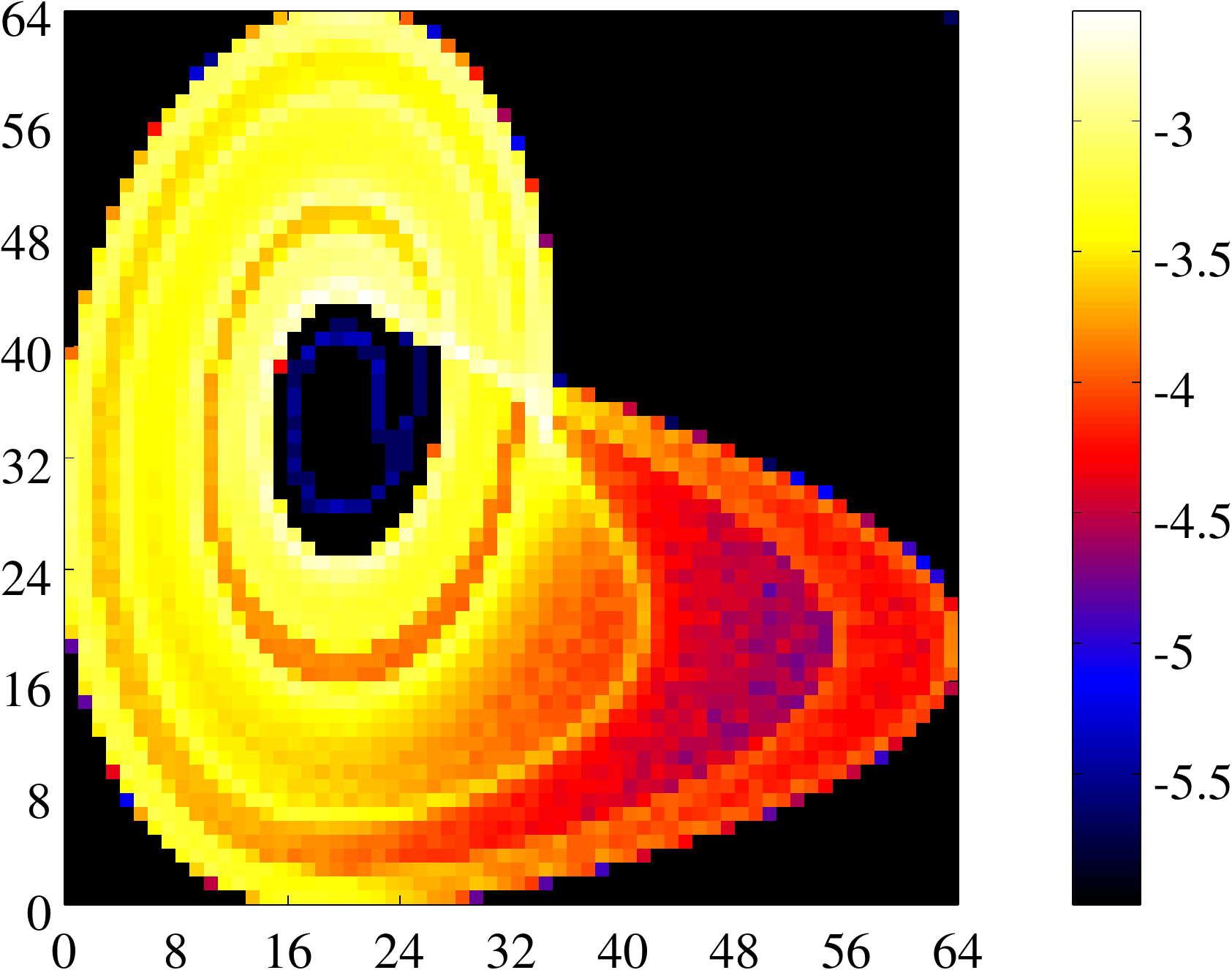}\hspace{16pt}%
  \includegraphics[width=0.3\textwidth]{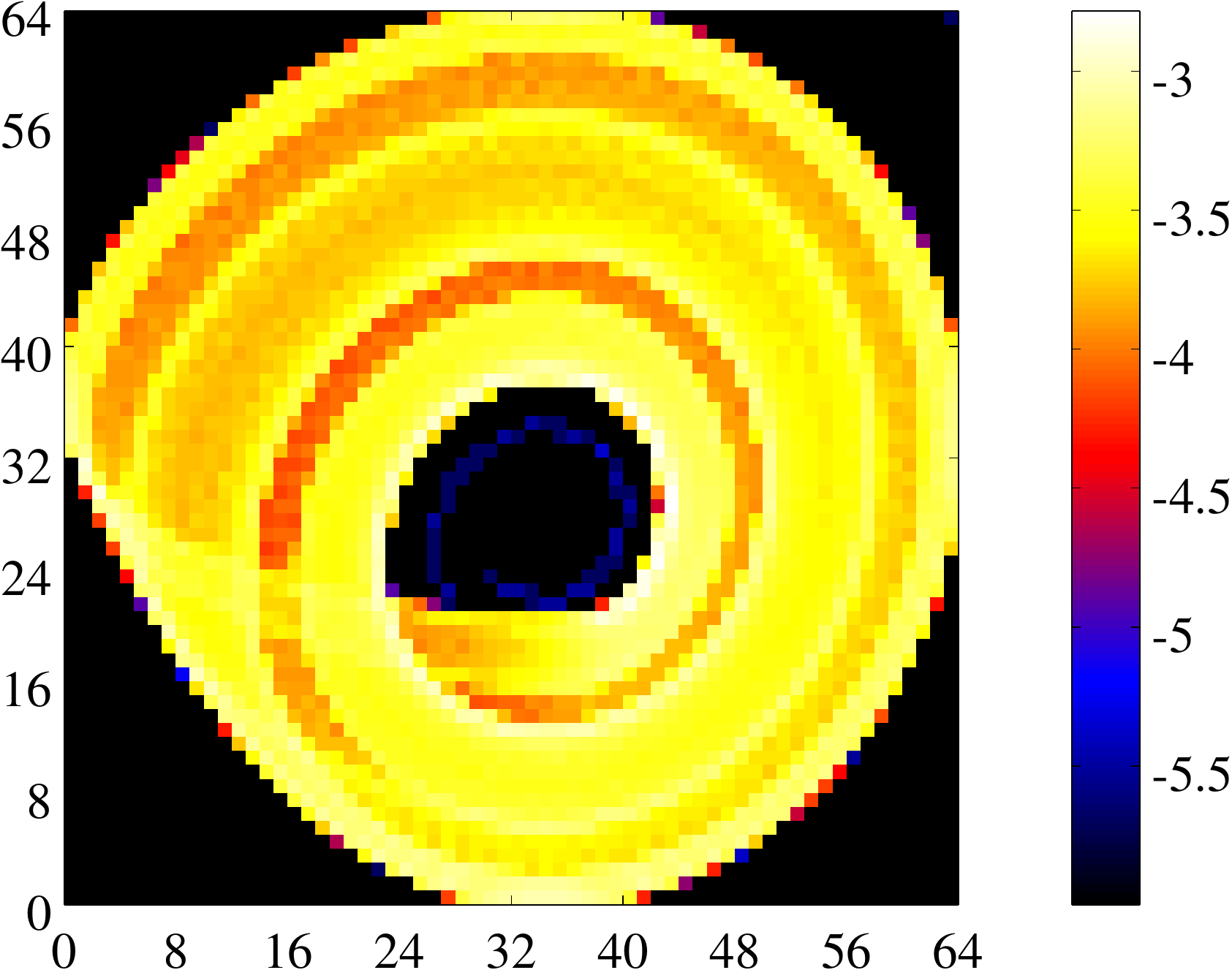}\hspace{16pt}%
  \includegraphics[width=0.3\textwidth]{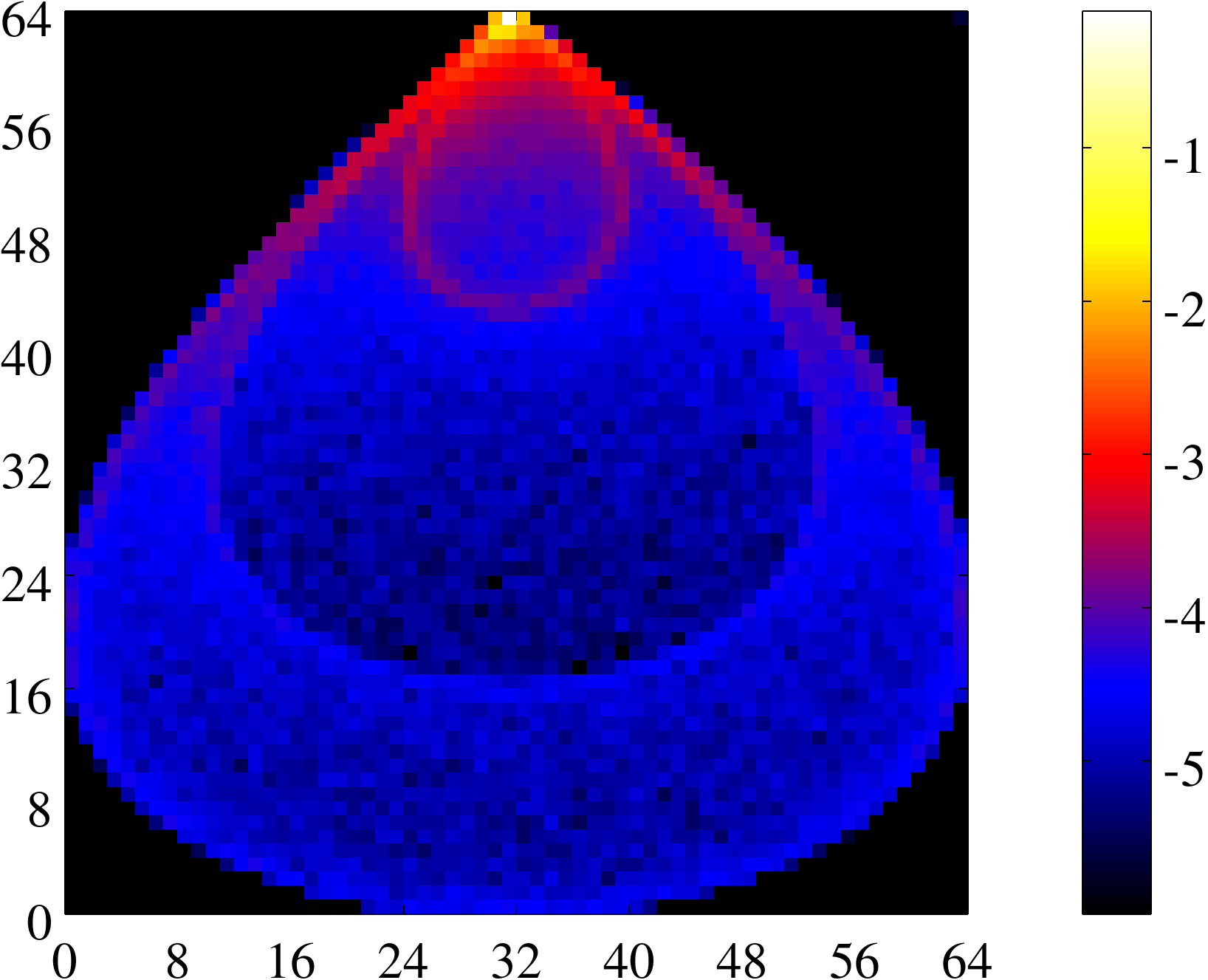}\\
  \includegraphics[width=0.3\textwidth]{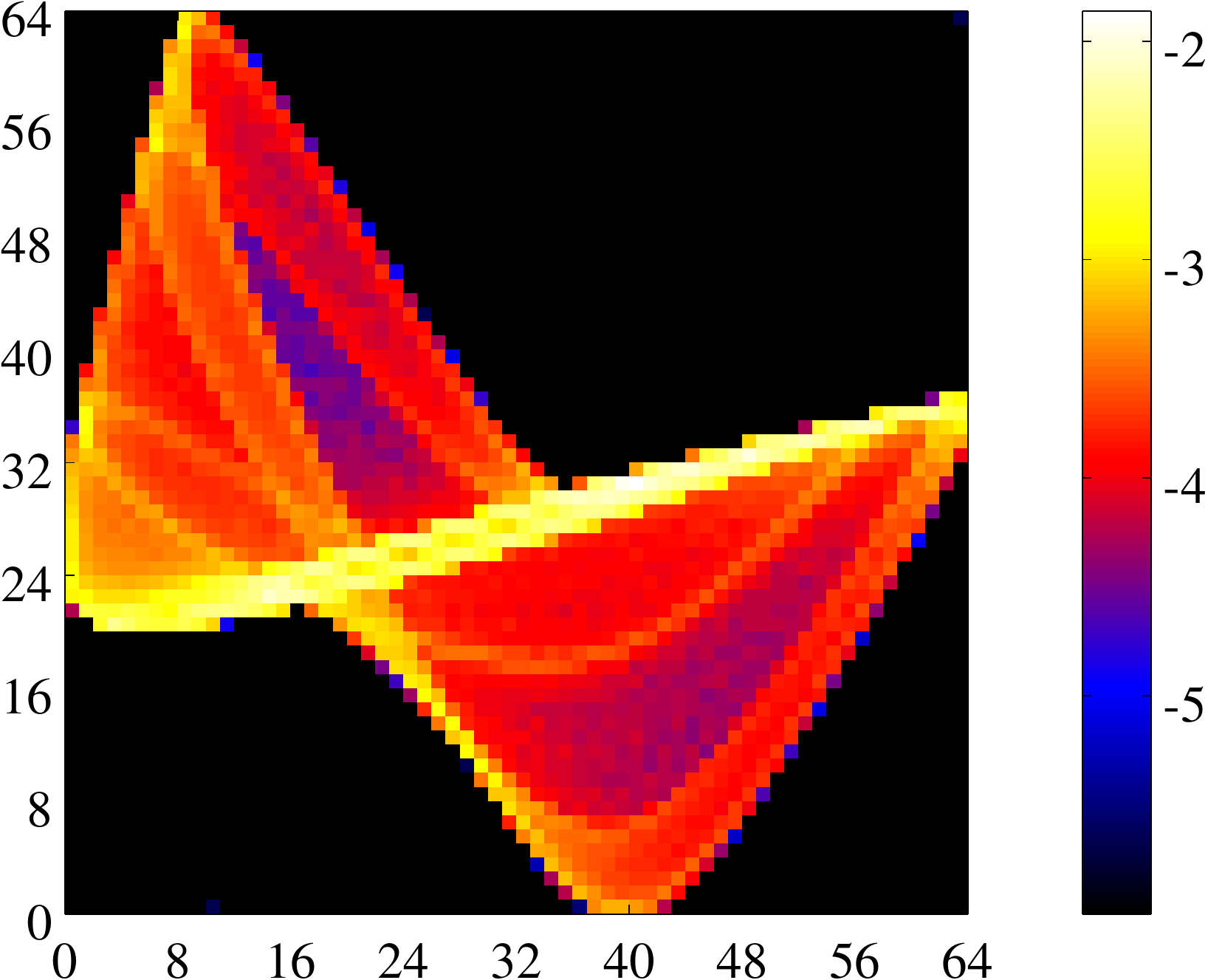}\hspace{16pt}%
  \includegraphics[width=0.3\textwidth]{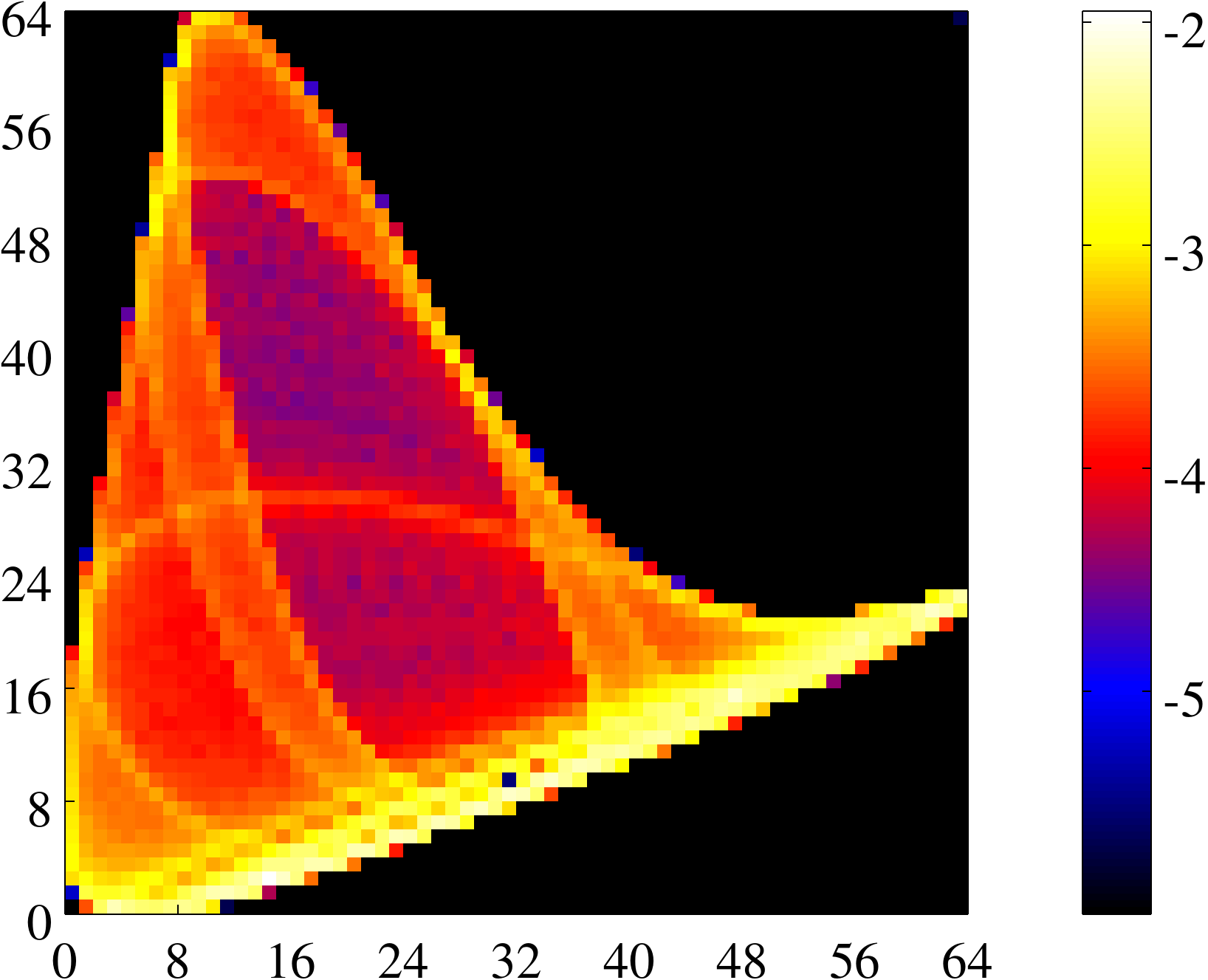}\hspace{16pt}%
  \includegraphics[width=0.3\textwidth]{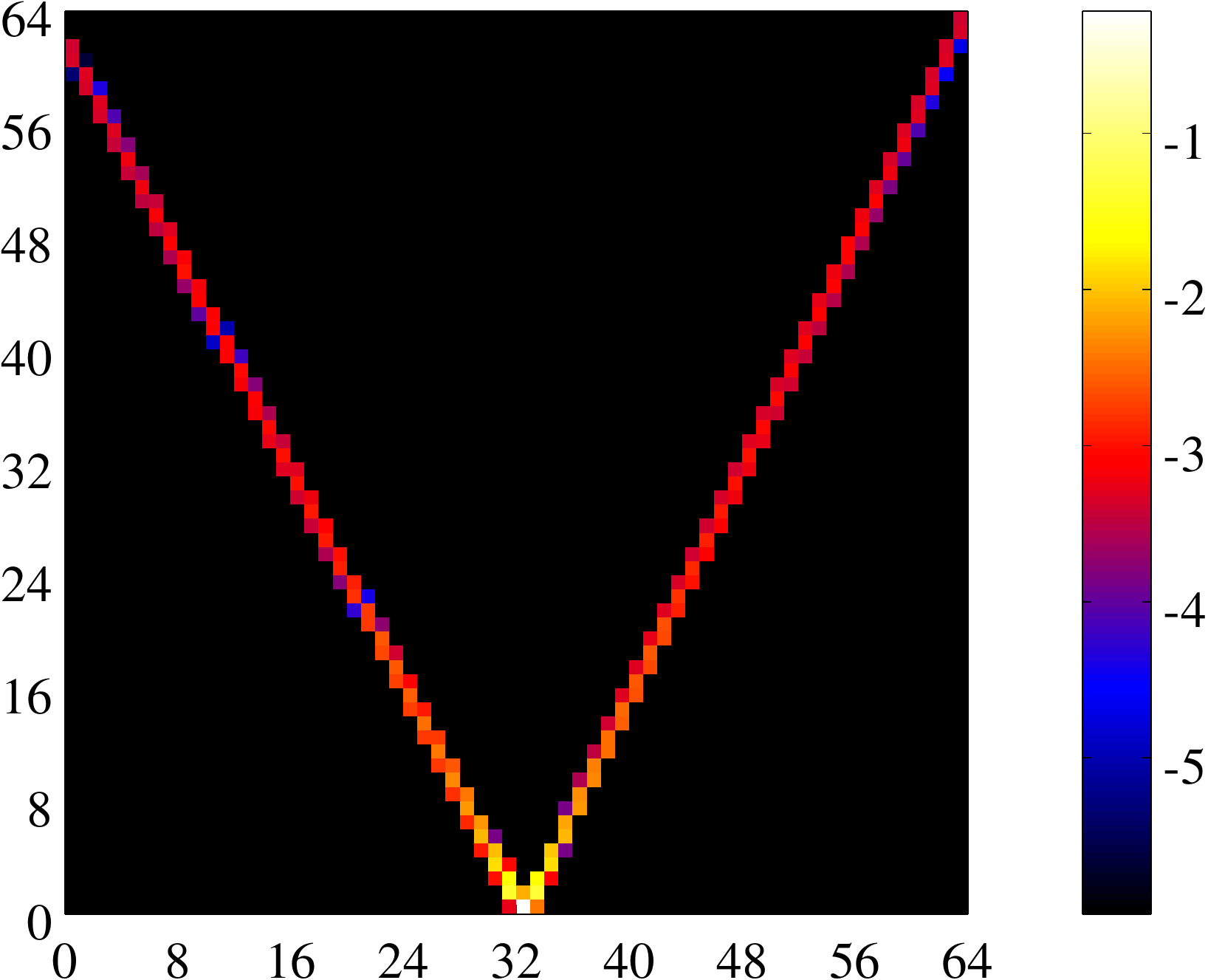}\\ 
  \caption{Several significant phase portraits for R\"{o}ssler system. Portraits are arranged by state variable $x$ (left), $y$ (center), and $z$ (right).  The portraits of standard lag view selected from first local minima of \MI by variable are shown in the top row.  The middle row depicts the minimum lag of a single iteration in the orthogonalized views. The bottom row shows the maximum \MI orthogonalized views.}
  \label{Fig:ShadowsOfStraightRossler}
\end{figure}

Figure \ref{Fig:ShadowsOfStraightRossler} shows the nine most significant views resulting from the first minima using standard lags, first orthogonal view, and maximum mutual information orthogonal view as denoted by the respective markers on Figure \ref{Fig:RosslerMI}.

It is readily observed that for the variables $x$ and $y$, the minimum lag orthogonalized view conveys similar dynamics as the first local minimum criteria on the standard time-delay embedding (up to the rotation).  In contrast, of these two lag selection techniques, the minimum lag orthogonalized view appears to provide significantly more insight into the dynamics of the $z$ variable for this finite dataset than the standard first local minimum.  The maximum mutual information view, collapses the data into the minimum number of bins (maximizing the difference from uniform and random), emphasizing the topology of signal excursions rather than the mean signal itself.  This is most clearly visible in the $z$ representation which, like in the case of the time-delay view, has no clear relationship between near neighbors as it traces out straight lines rotated from the first minimum lagged view.

The behavior of all three views on $z$ are significantly different than for the $x$ and $y$ variables despite the fact that the entire dynamics of the system is guaranteed to be embedded in lags of a single variable as $t\rightarrow \infty$ as indicated by Takens \cite{Takens}.  This is not entirely surprising, given that the underlying system is three dimensional so embedding a variable in only two-dimensions is in no way guaranteed to remove ambiguity.  In Reference \cite{CellucciComparative}, this example was shown to demonstrate that a different binning was necessary for $z$ than for $x$ and $y$ based on minimizing stochastic complexity.  Further investigation of alternative binning will be provided later in this section.

A more critical criterion will be determining the quality of embedding in the range of dimensions $d=[3$-$6]$ where embeddings are known to be possible between the original system dimension up to the
dimension where it is guaranteed for time lags by Takens' theorem.  This issue will be discussed further in Section \ref{SubSec:Opt}. 
Regardless of the impact of the embedding dimension, it appears that the orthogonalized view of the data is practical, at least, at removing some ambiguity due to the redundancy of the short
time lags. 

\subsection{Nonuniform binning}
\label{SubSec:Nonuniform}
Though in theory any monotonic transformation should preserve the amount of mutual information, it is clear from the simple affine transformation into orthogonalized coordinates from the prior sections that the coordinate system has a dramatic impact on how the mutual information is actually numerically estimated using regular discretized bins. This is an artifact of the discretized estimate for finite data.  With infinite data, the bins for the standard lagged coordinates could resolve the structure collapsed tightly along the diagonal of the joint pdf with infinitesimal bin sizes. Attempting to resolve that structure with finite data results in tiny bins containing either zero or one observation that are useless in estimating the actual probability density. 

As early as the original Fraser and Swinney approach of Reference \cite{Fraser} from which the first minimum criteria was introduced, it was recognized that uniform width bins were a poor choice for finite data.  In that work, equiprobable bins in one dimension and equiprobable bins via balanced partitioning in two dimensions were explored.  These issues were investigated in greater detail in Reference \cite{CellucciMICalc}. There, an alternate method is proposed for adaptive partitioning as well as true statistical tests to ensure that the identified information is statistically distinct from noise. In particular, the authors provide a theorem that mutual information is independent of arbitrary coordinate transformation functions $h_X$ and $h_Y$ so long as the functions are monotone and increasing. This means that $I(X,Y)=I(h_X(X),h_Y(Y))$ even if the transformations are nonlinear which is in agreement with the appendix of Reference \cite{Kraskov}.

The orthogonalized view is effectively a clever bounding box on the lagged view that recognizes the short time delay data falls on the diagonal.  The view rotates the data accordingly so that the bounding box can be aligned with the data. This results in some stability issues for the bounding box as the outliers from numerical derivatives have a larger impact on the extrema of the observations used for the bounding box as a function of lag length than the few points truncated at the ends of the observations when lag is changed in standard coordinates.  The instability of the bounding box can be corrected through the use of transformation functions designed to flatten the marginal probability density along each axis of the joint probability density.  This can be done very easily by simply inverting the cumulative probability distribution function along each axis. A uniform mesh in cumulative probability translates into a stretched physical mesh in axis values such that each of the bin symbols has an equal probability of $1/n_{bin}$.  In practice, this is done by simply sorting the observations along each axis and marking every $(n/n_{bin})^{th}$ observation as the edge of a piecewise linear transformation, $h_x(x)$.  Once the observation vectors have been transformed into these equi-informative symbols, the joint pdf is stabilized and saturates theoretical bounds reliably.

In the case of time delay phase portraits, the one-dimensional statistics of the $X$ and $Y$ coordinate data coming from observations of $x(t)$ and $x(t+\tau)$ respectively which results in nearly identical cumulative distributions differing only by the observations that are not repeated at the beginning and end of the series.  This suggests the use of the single adaptive partition as described in Reference \cite{CellucciMICalc} should essentially result in an equivalent to constructing the transformation functions, $h_x$ and $h_y$, independently in both directions.  In the case that the data are indeed independent, the single transformation results in a uniform flat two dimensional probability density estimate and mutual information is statistically indistinguishable from zero.  However, the short time lag data remains highly correlated in the two dimensions. The short time data still falls on a line along the diagonal of the two axes after the transformation. The density of the line just becomes uniform instead of varying with the one-dimensional relative rates of observations observed in the uniform bins. As the lags increase, the probability distribution opens up resulting in mutual information curves that rapidly decline from an initial peak value equal to the one dimensional maximum entropy.  For chaotic data, the mutual information then oscillates while decaying as the lag becomes large.  For purely periodic signals the mutual information structure is repeating cusped peaks with a period equal to the oscillation frequency.  The structure of the mutual information curves remain similar as what was obtained with uniform binning though in a more consistent manner which saturates the theoretical bounds based on the number of bins used.

For the orthogonalized data, the redundancy effects have been removed.  Unlike the traditional lagged coordinates where the short lag data continues to fall along the diagonal of the axes resulting in fictitiously high mutual information estimates, for the orthogonalized view with the minimum time lag, each axis has a different and independent cumulative distribution functions. The data can then be stretched independently in both directions to fill most of the joint probability density unit square. Therefore, any deviations from a flat probability density in this coordinate system is directly a measure of the structure of the system with the exception of fluctuation errors due to finite data as discussed later in Section \ref{SubSubSec:Floor}.  The construction of the independent monotone one-dimensional transformations $h_{\bar{x}}$ and $h_v$ is quite straightforward following the same general description as in the lagged view case.  After orthogonalization, the data $\bar{X}$ and $V$ are sorted.  The equiprobable bins are then constructed by dividing the $n$ observations (i.e. $\{v_0,\dots,v_{n-1}\}$) into $n_{bin}$ groups.  The coordinates of the $n_{bin}$ partition are obtained from the sorted observations by selecting the $k=[0-n_{bin}]$-edges, $e_k$, by rounding $k(n-1)/(n_{bin})$ such that the $k^{th}$ bin falls between the $(k-1)$ and $k^{th}$ edge. The transformation functions $h_{\bar{x}}$ and $h_v$ can then be constructed by piecewise linear interpolation between the edges though this does not impact the calculation of the dimensionless joint pdf where the probability $P(i,j)$ of the $(i,j)^{th}$ bin is simply the count of the observations in the range $(e_{\bar{x},i-1} \le \bar{x} < e_{\bar{x},i})$ and $(e_{v,j-1}\le v < e_{\bar{v},j})$ divided by the number of observations.  In terms of the joint probability distribution, the calculation of entropy can now be thought of in terms of a $n_{bin}$ symbol alphabet where the distribution of two letter symbols is only uniform if the mutual information of the two signals is zero.

\begin{figure}
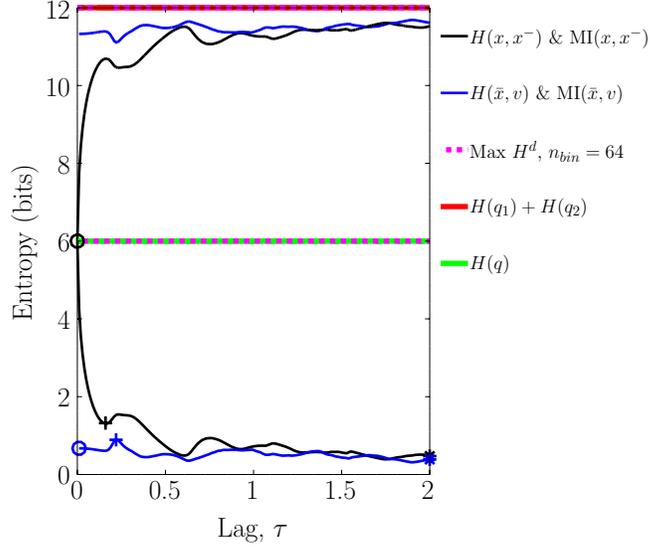

  \scalebox{0.5}{\include{ShadowStraightVsEigEquibin}}
  \caption{Entropy and mutual information of Lorenz-$x$ variable shadows on $64\times 64$ equiprobable bins in standard and orthogonalized coordinate systems. Maximum entropy by dimension $d$ from $n_{bin}$ is marked with dashes to show saturation of limits by the equiprobable distributions.}
  \label{Fig:ShadowStraightEquibin}
\end{figure}

To investigate the impact of this transformation of the joint probability density function on the mutual information for various lags of the system dynamics, the $x$ variable of the Lorenz system is
revisited first in Figure \ref{Fig:ShadowStraightEquibin}.  Note that, as opposed to Figure \ref{Fig:ShadowStraight}, the one-dimensional entropies $H(x)$, $H(x^-)$, $H(\bar{x})$, and $H(v)$ are all saturated at the theoretical limit of 6-bits for the 64 bins. This makes the sum of the respective pairs saturate the 12-bit limit for two dimensions.

\begin{figure}
  \includegraphics[width=0.3\textwidth]{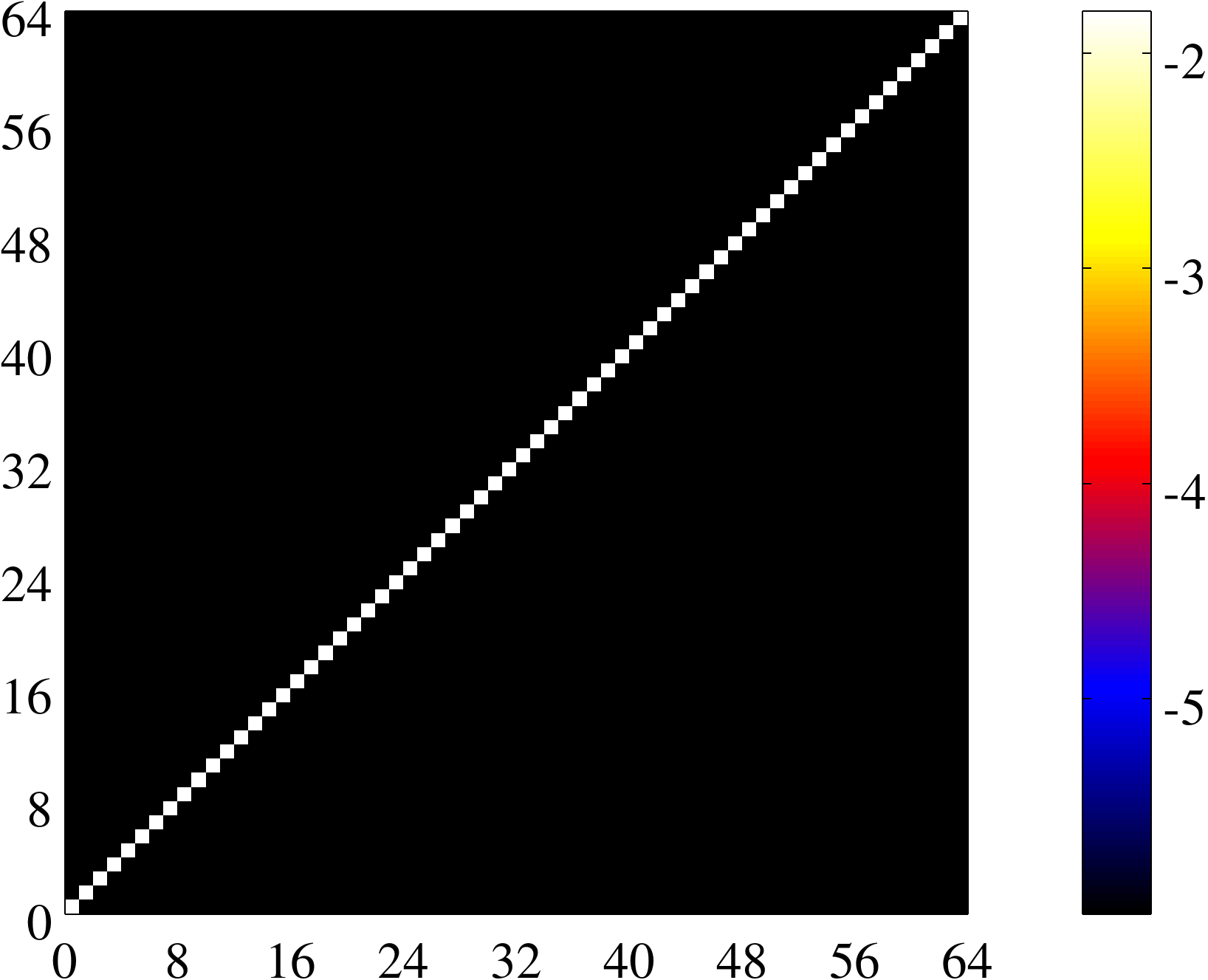}\hspace{16pt}%
  \includegraphics[width=0.3\textwidth]{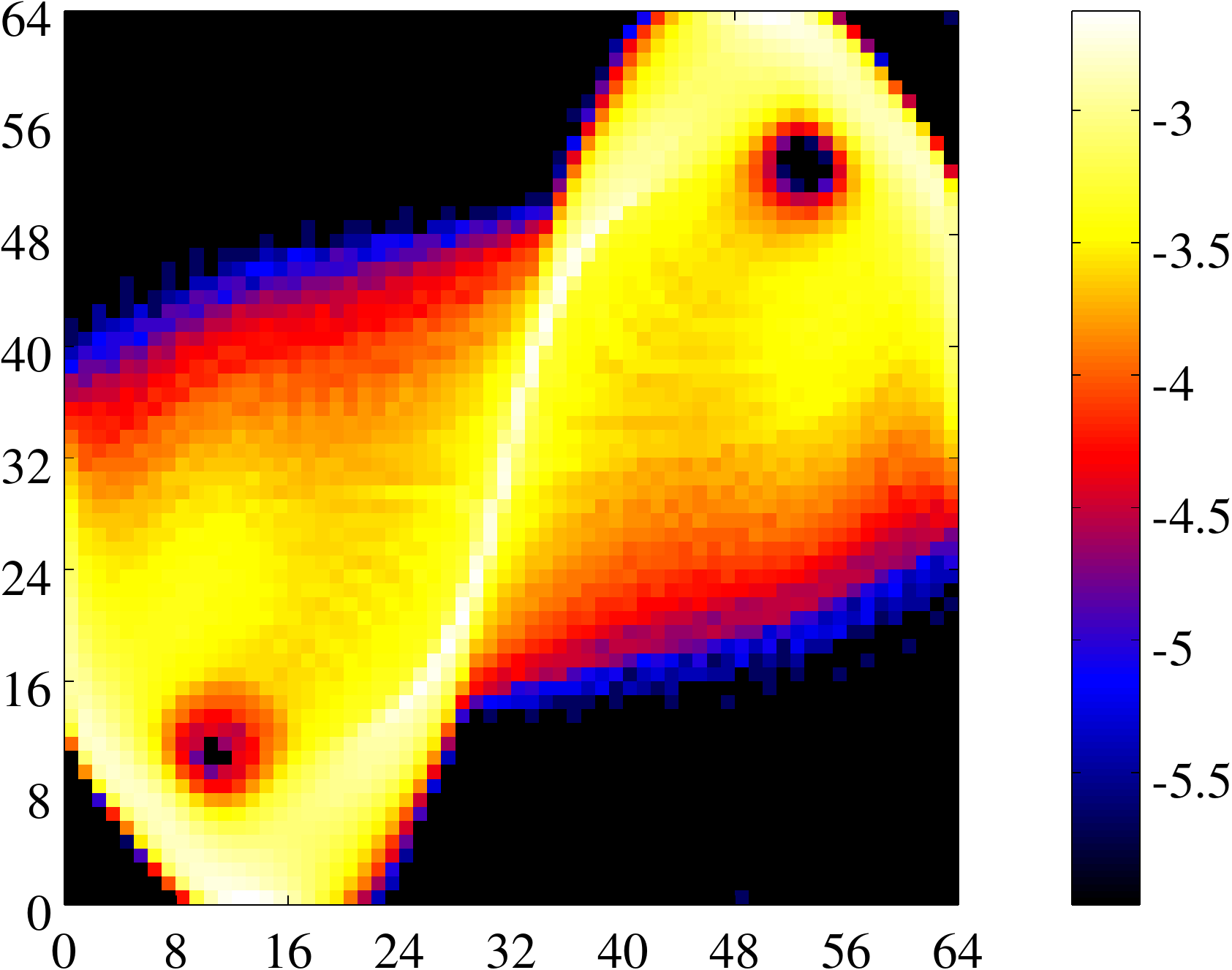}\hspace{16pt}%
  \includegraphics[width=0.3\textwidth]{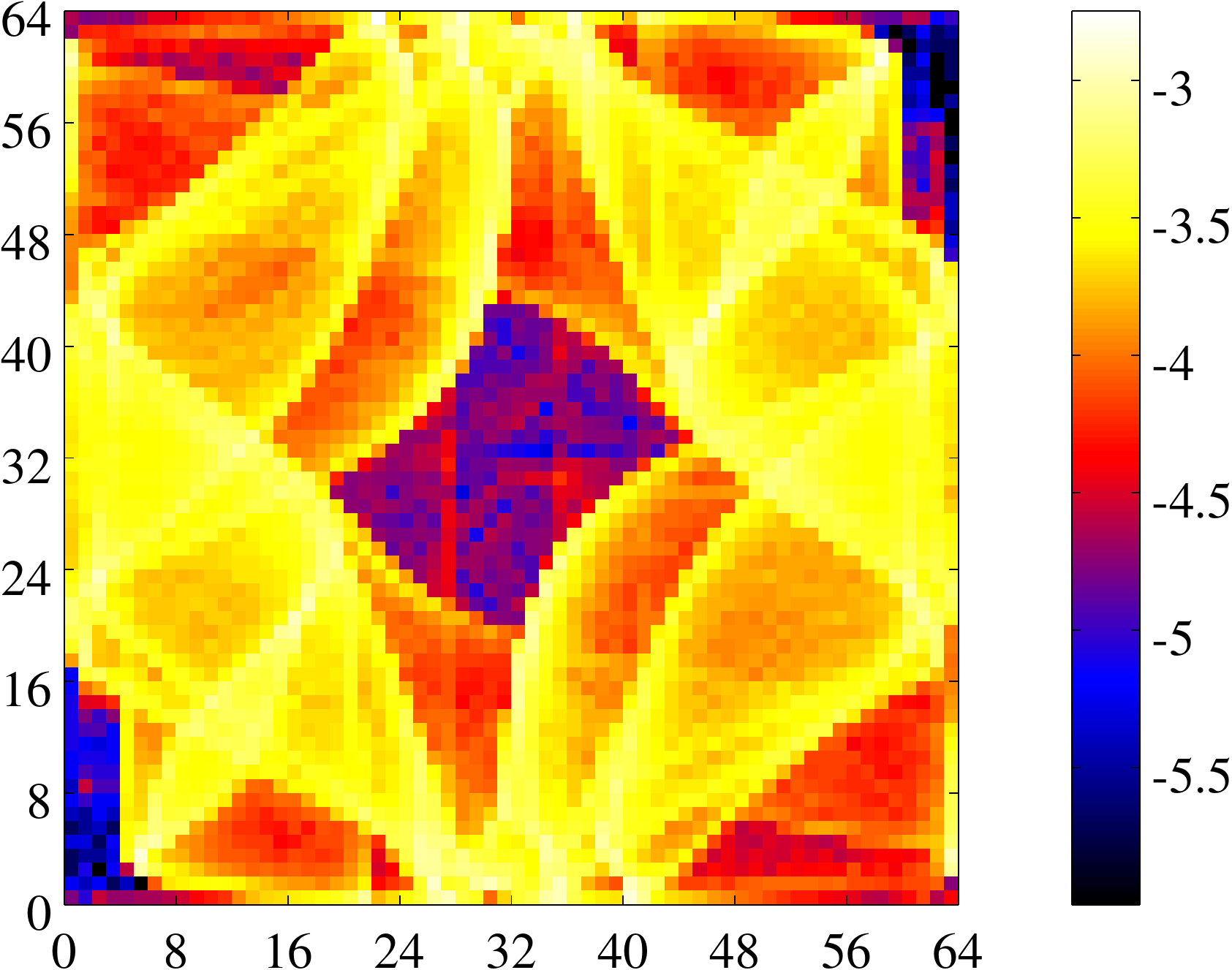}\\
  \includegraphics[width=0.3\textwidth]{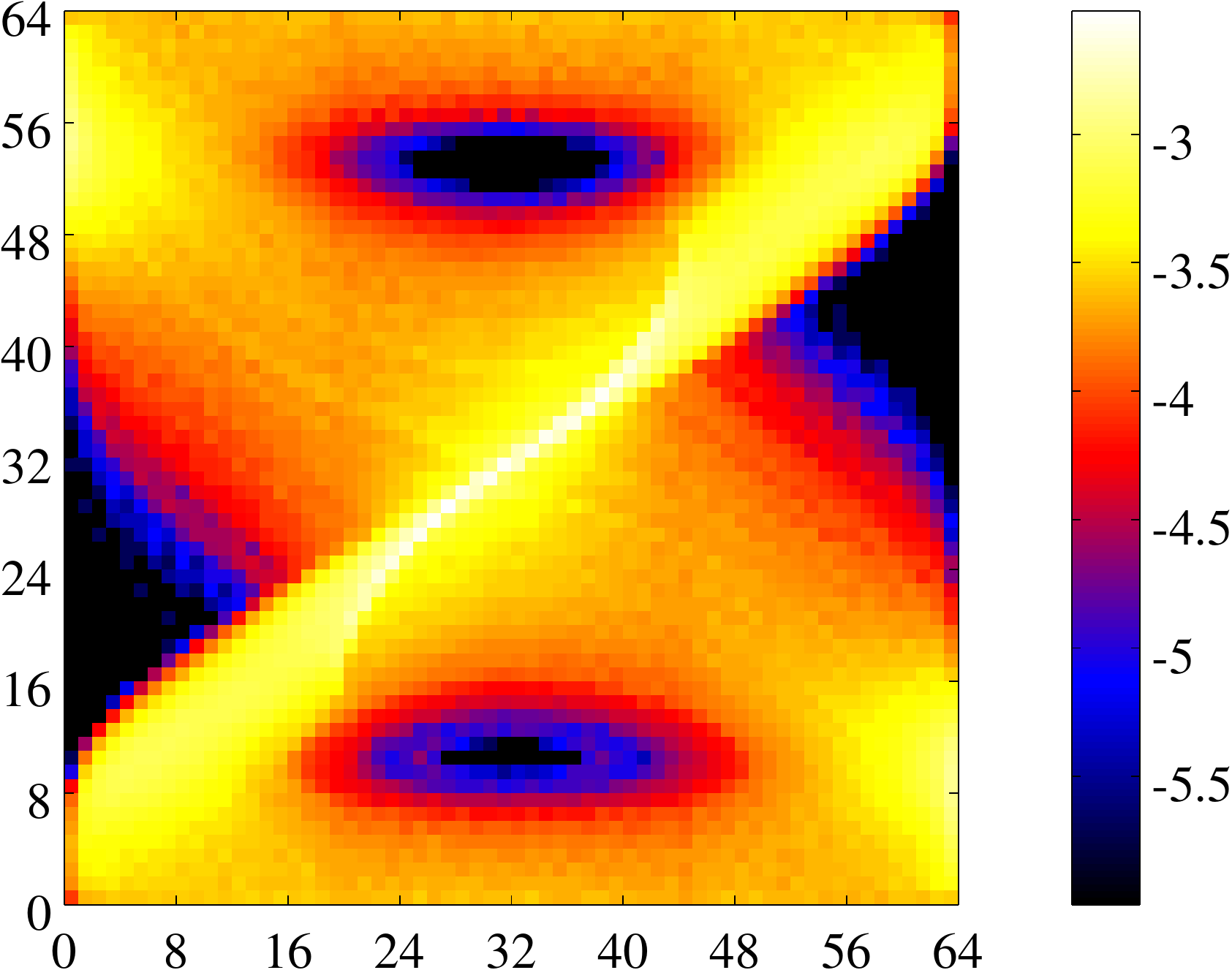}\hspace{16pt}%
  \includegraphics[width=0.3\textwidth]{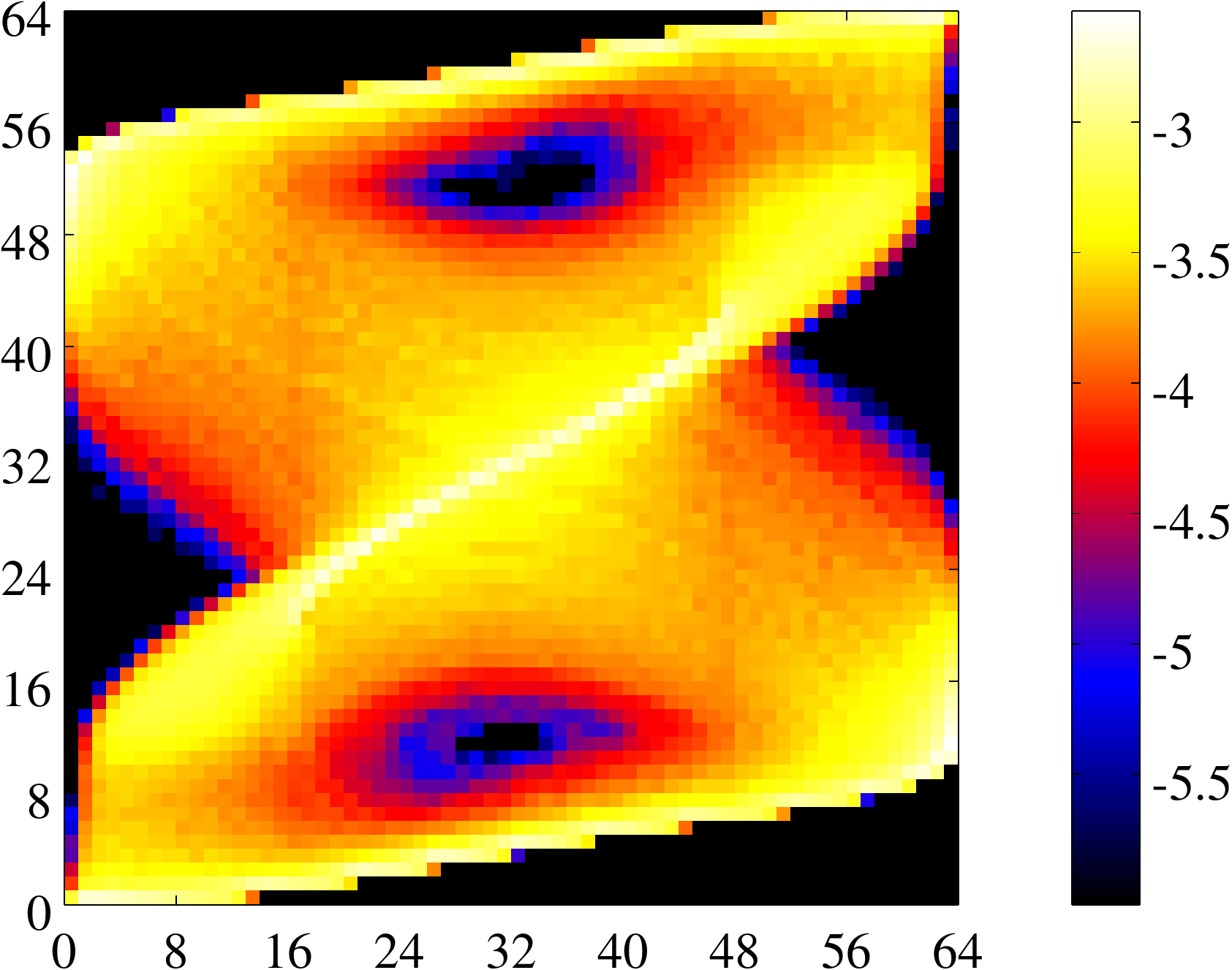}\hspace{16pt}%
  \includegraphics[width=0.3\textwidth]{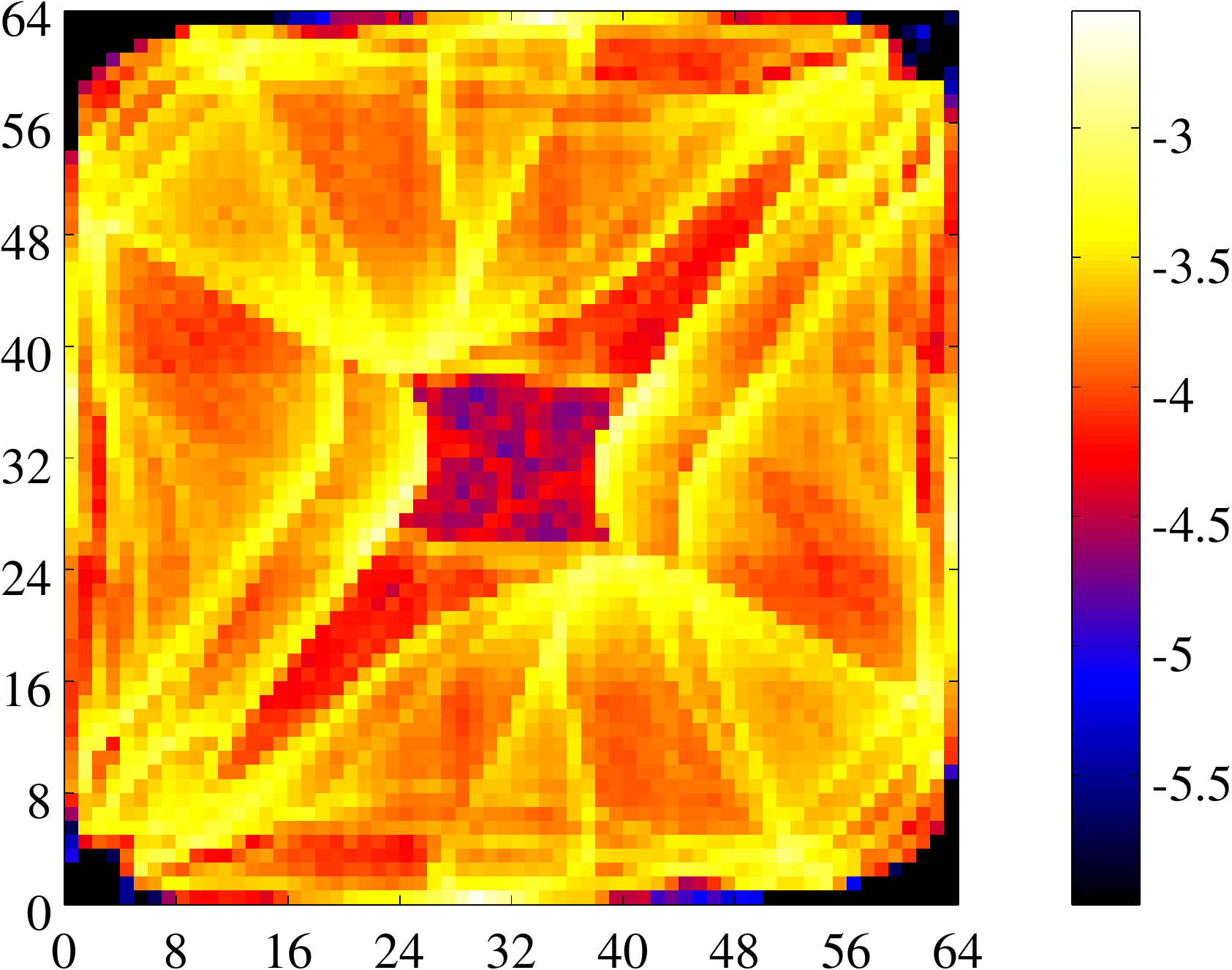}\\  
  \caption{Phase portraits from $x$ variable of lagged Lorenz data in equiprobable binning. Top row is original lagged coordinates and bottom row is in rotated orthogonal coordinates.  The left and right figures show first and last portraits denoted by ``$\circ$'' and ``$*$'' symbols on entropy and mutual information plot.  Middle figure shows views from ``$+$'' marked points due to the associated criteria on mutual information.}
  \label{Fig:ShadowsOfStraightEquibin}
\end{figure}

Figure \ref{Fig:ShadowsOfStraightEquibin} shows the phase portraits from the stretched equiprobable bins comparable to Figure \ref{Fig:ShadowsOfStraight}.  It is clear the transformation serves to more uniformly fill all of the bins, at least as $\tau \rightarrow \infty$.  Note that the standard view of the zero lag data has uniform density along the diagonal instead of a density peaked towards the center in the uniform binning case.  In each of the figures, the rows and columns of the figure all sum to an equivalent $1/64^{th}$ of the total data.  Interestingly, for this case, the same lags are identified using both uniform and equiprobable binning.  

The mesh stretching transforms applied in Figure \ref{Fig:ShadowsOfStraightEquibin} are depicted below in Figure \ref{Fig:Transforms}.  Note that the top row are identical because the original and lagged components use an identical transformation based on the original distribution of data.  The bottom row shows how the transformation evolves with respect to lag in the orthogonalized coordinates.  In the minimum lag view, the transform of the first component is nearly identical to the transform from the standard view.  This is because the component represents the $\bar{x}$ component which has a nearly identical distribution to the original smooth data for $\tau \rightarrow 0$.  The transform in the other direction is significantly different.  This is because there is no reason the $v$ distribution should resemble the $x$ distribution. By stretching the coordinates independently, the phase portrait is stretched to fill the entire box even as $\tau \rightarrow 0$.  This coordinate transformation emphasizes the deviation of the data from the null hypothesis of a uniform flat probability density, as would be expected from statistically independent data.

\begin{figure}
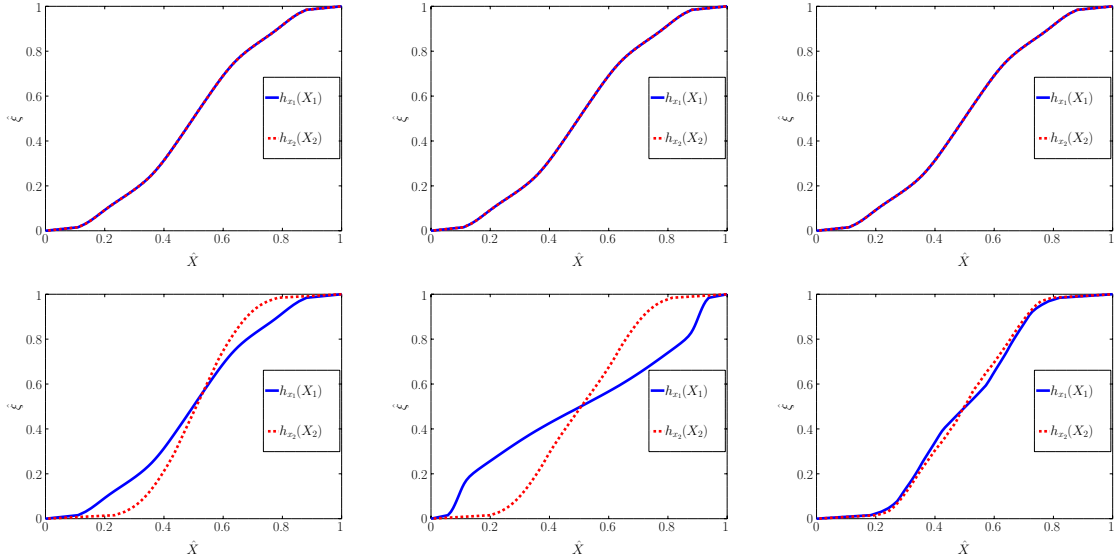

  \scalebox{0.25}{\include{ShadowEquibinTransform_0Equibin}}\hspace{0pt}%
  \scalebox{0.25}{\include{ShadowEquibinTransform_16Equibin}}\hspace{0pt}%
  \scalebox{0.25}{\include{ShadowEquibinTransform_200Equibin}}\\
  \scalebox{0.25}{\include{ShadowEigEquibinTransform_1Equibin}}\hspace{0pt}%
  \scalebox{0.25}{\include{ShadowEigEquibinTransform_22Equibin}}\hspace{0pt}%
  \scalebox{0.25}{\include{ShadowEigEquibinTransform_200Equibin}}\\  
  \caption{Normalized mesh transformation functions for equiprobable bins. Top row is original lagged coordinates and bottom row is in rotated orthogonal coordinates.  The left and right figures show first and last portraits denoted by ``$\circ$'' and ``$*$'' symbols on entropy and mutual information plot.  Middle figure shows views from ``$+$'' marked points due to the associated criteria on mutual information.}
  \label{Fig:Transforms}
\end{figure}

\begin{figure}
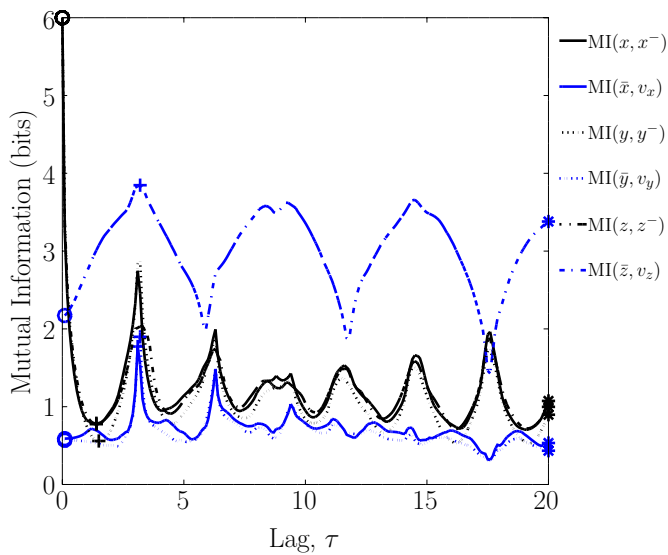

  \scalebox{0.5}{\include{ShadowStraightVsEigEquibinRossler}}
  \caption{Entropy and mutual information of Lorenz-$x$ variable shadows on $64\times 64$ equiprobable bins in standard and orthogonalized coordinate systems. Maximum entropy by dimension $d$ from $n_{bin}$ is marked with dashes to show saturation of limits by the equiprobable distributions.}
  \label{Fig:EquibinRosslerMI}
\end{figure}

Re-running the R\"{o}ssler system with the equiprobable binning improves the ability to identify distinct criteria on the mutual information in all views.  Figure \ref{Fig:EquibinRosslerMI} shows the impact of equiprobable binning on the mutual information as a function of lag for the $x$, $y$, and $z$ state variables.  The peaks and local minima are much more distinct with these transformed coordinates than in the original views from Figure \ref{Fig:RosslerMI}.  This is particularly true of the $z$-coordinate.  The spread in optimal lags between $x$ and $z$ coordinates was reduced from $0.6$ in orthogonalized view and $0.3$ in the standard view to a maximum of $0.1$ difference representing a single timestep using the equiprobable bins. Note that, as seen in Figure \ref{Fig:RosslerMI}, the $z$ \MI curves in the uniform bin cases remain near the lag criteria values in a range from $1.7$ to nearly $5$ which contrasts to the enhanced sharpness of these criteria in the $z$ equiprobable bin curves.  Similar improvements were attained in the adaptive binning used in Reference \cite{CellucciMICalc}, and so the improvement in lag consistency for the standard view is not surprising. It is interesting to note that the peak of mutual information in the orthogonalized coordinates clearly identifies a different lag than the first minimum criteria in the original lag coordinate system.  The efficacy of this global maximum criteria in the orthogonalized coordinates for CCM will be investigated further in Section \ref{Sec:Impact}.

\begin{figure}
  \includegraphics[width=0.3\textwidth]{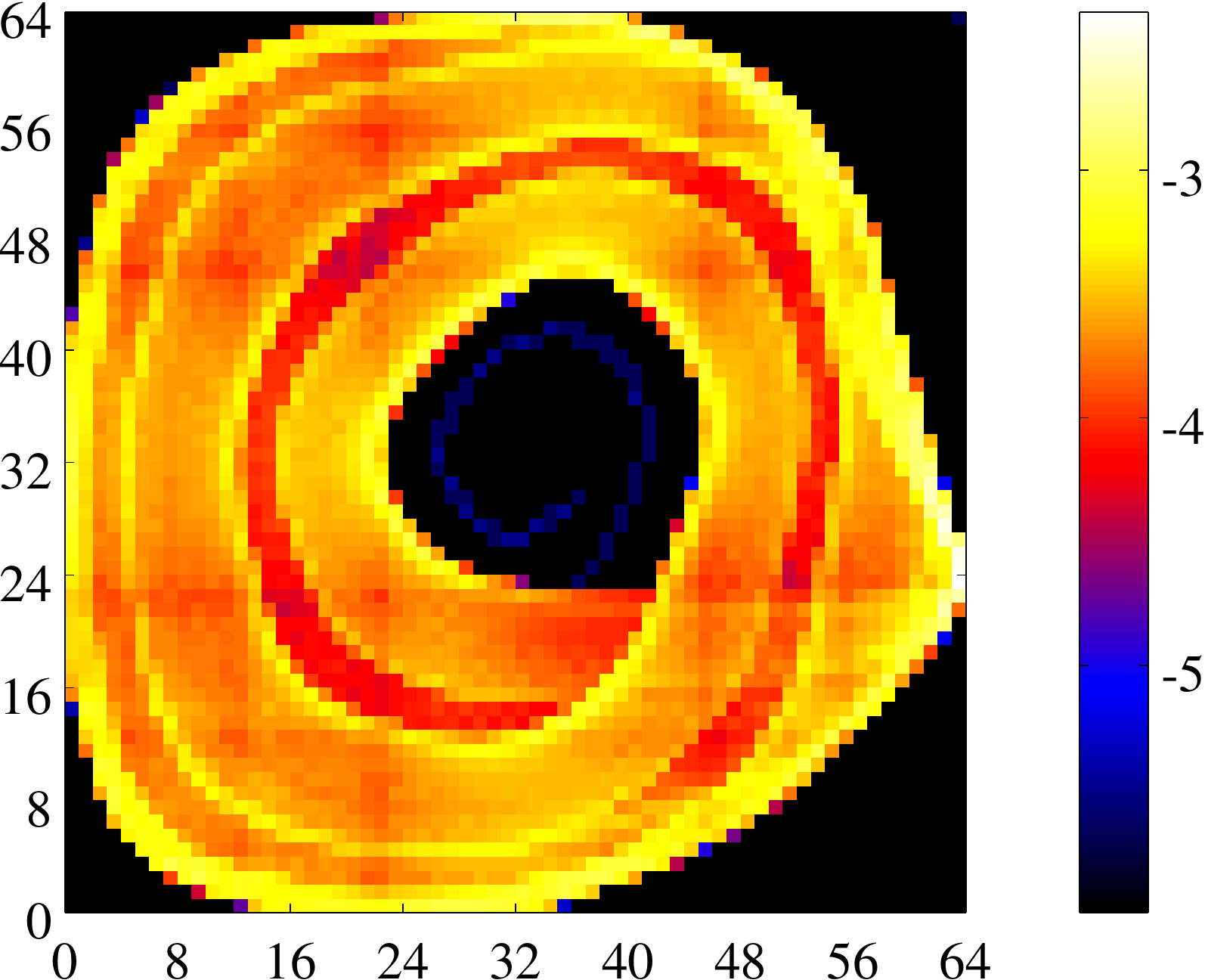}\hspace{16pt}%
  \includegraphics[width=0.3\textwidth]{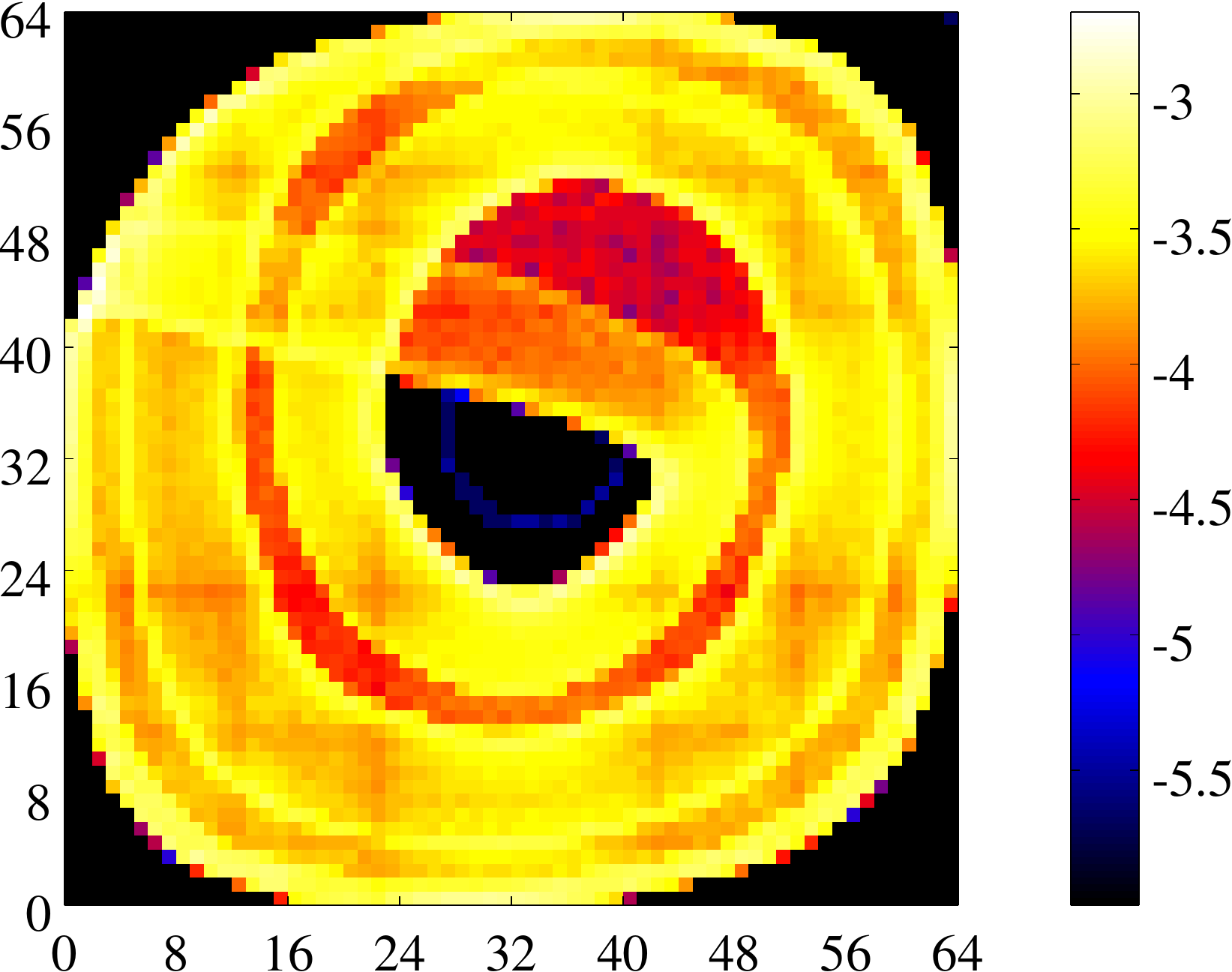}\hspace{16pt}%
  \includegraphics[width=0.3\textwidth]{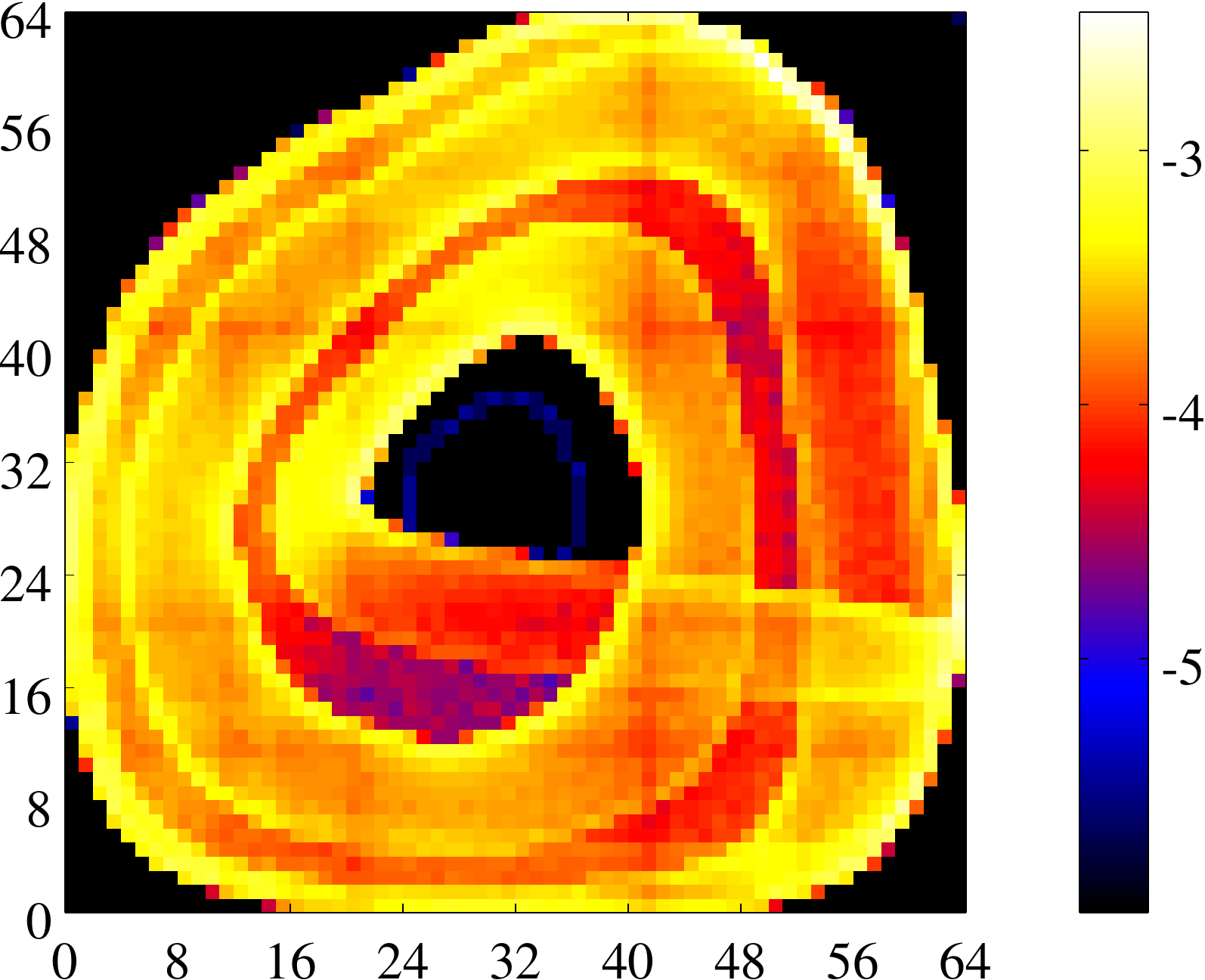}\\
  \includegraphics[width=0.3\textwidth]{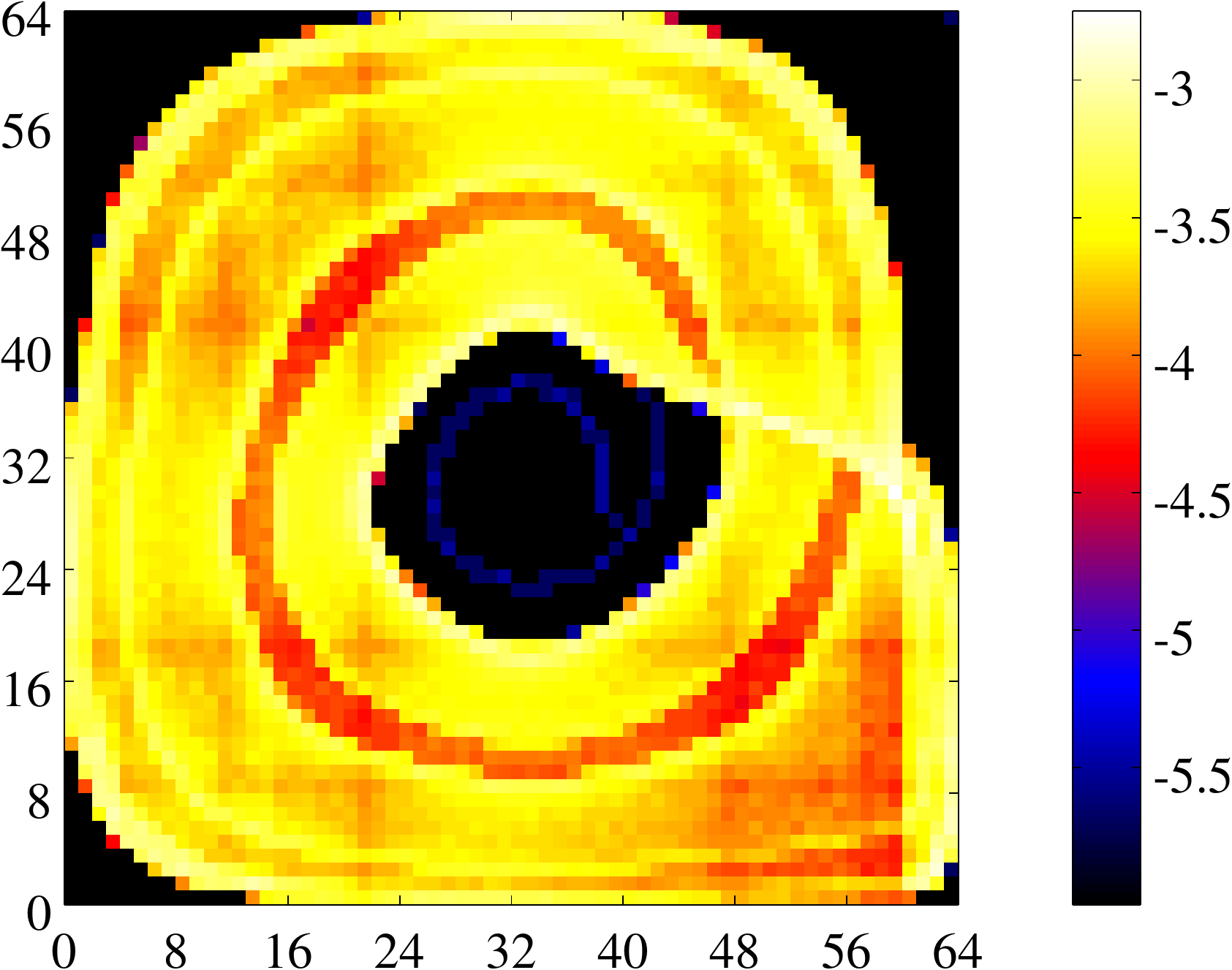}\hspace{16pt}%
  \includegraphics[width=0.3\textwidth]{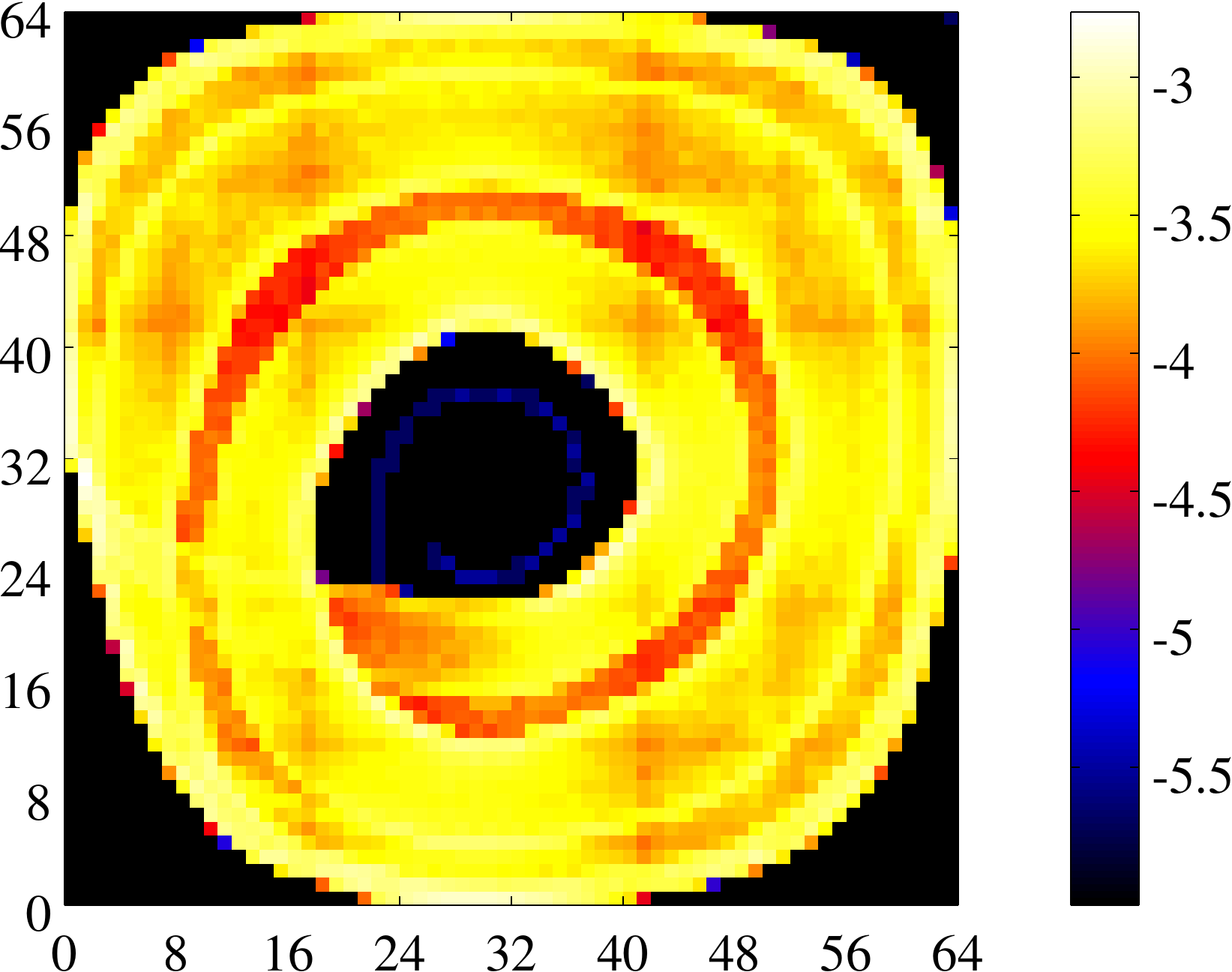}\hspace{16pt}%
  \includegraphics[width=0.3\textwidth]{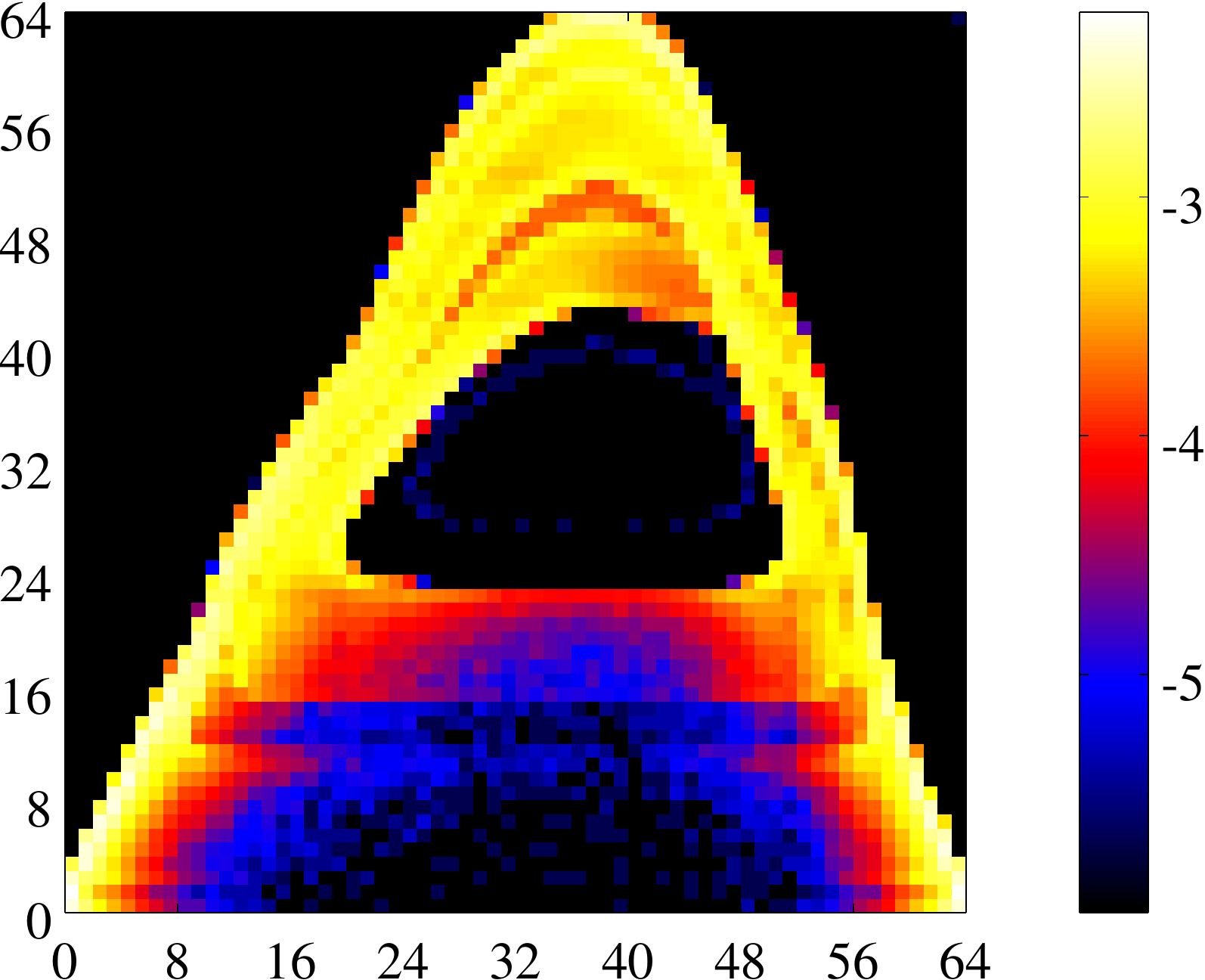}\\
  \includegraphics[width=0.3\textwidth]{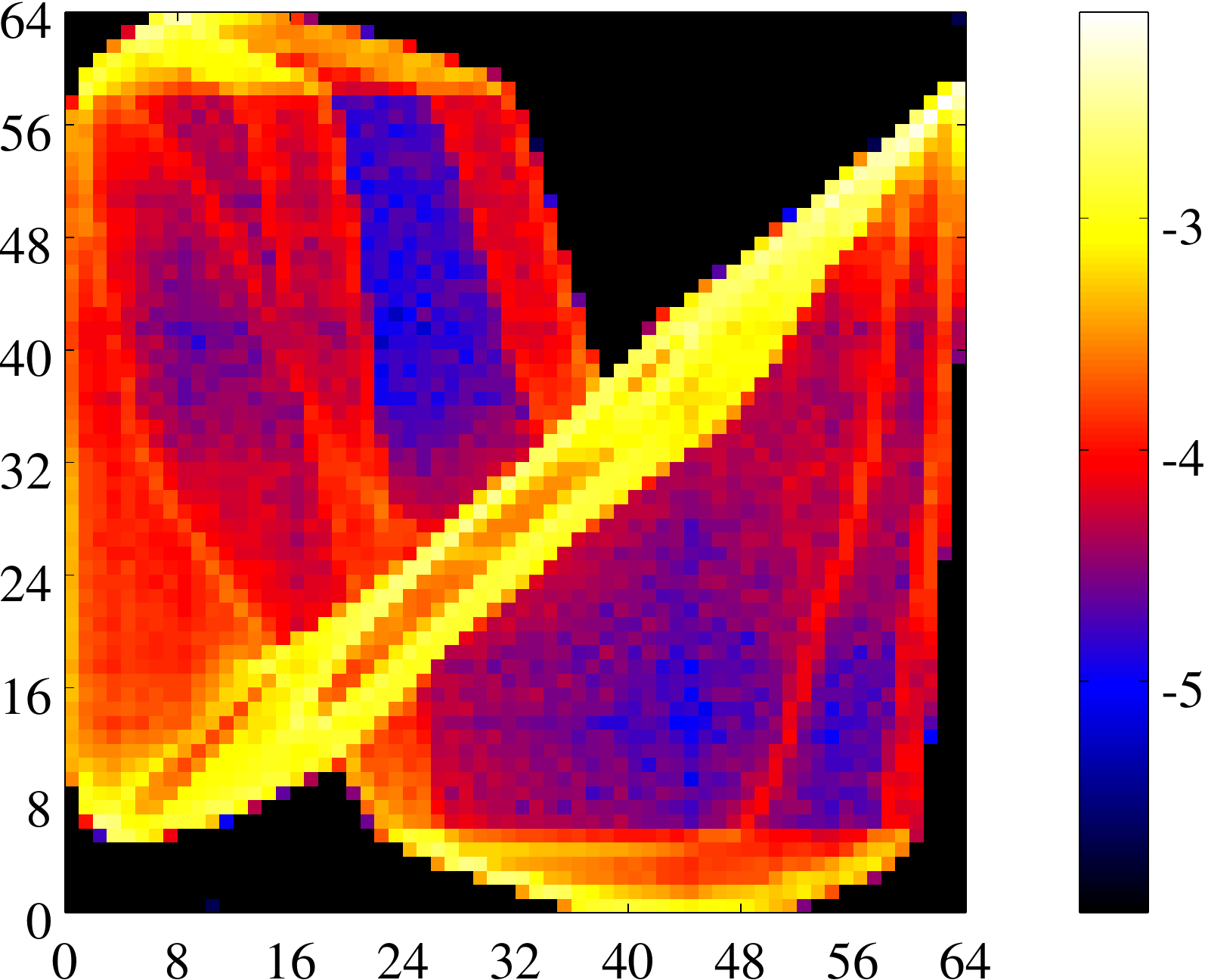}\hspace{16pt}%
  \includegraphics[width=0.3\textwidth]{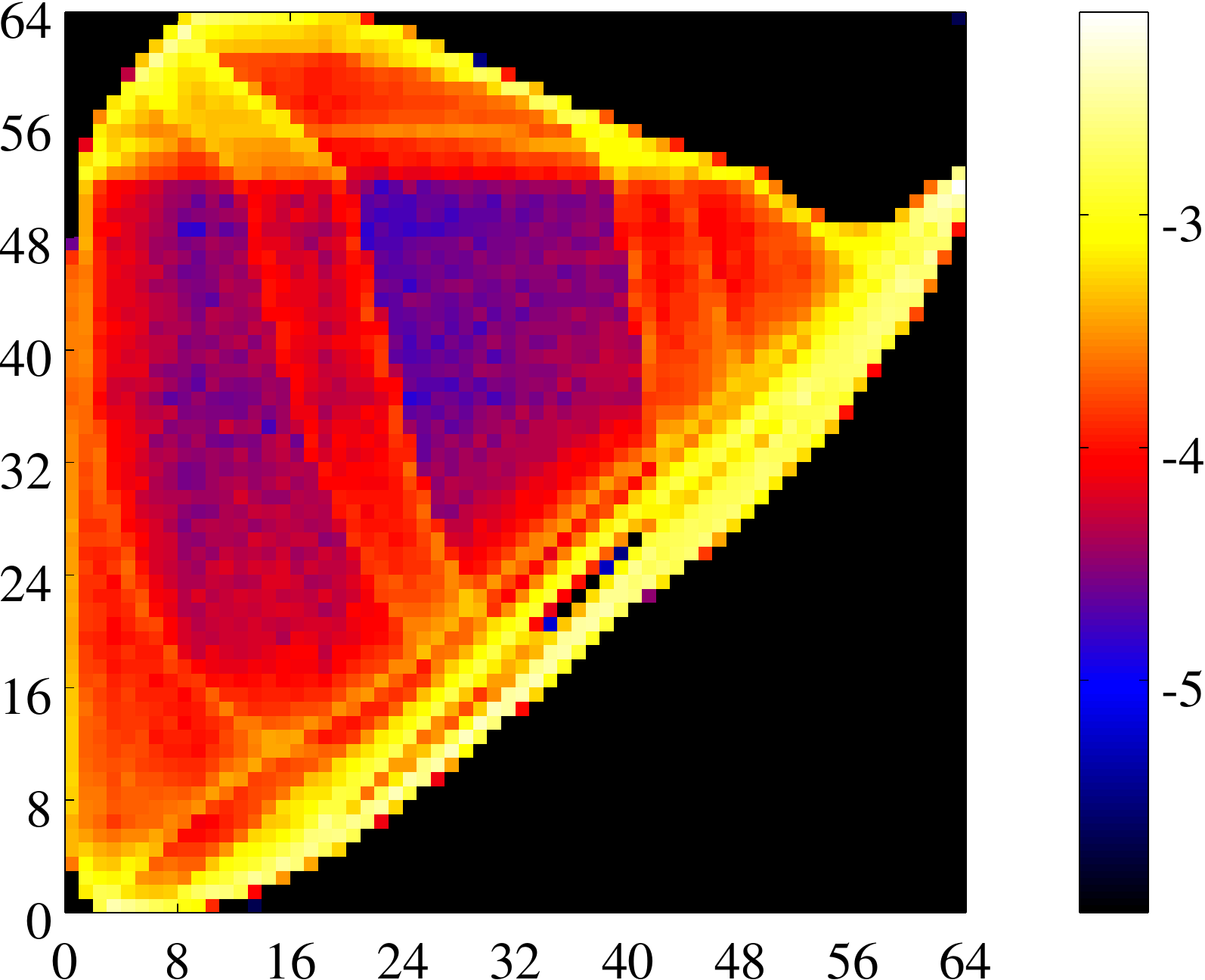}\hspace{16pt}%
  \includegraphics[width=0.3\textwidth]{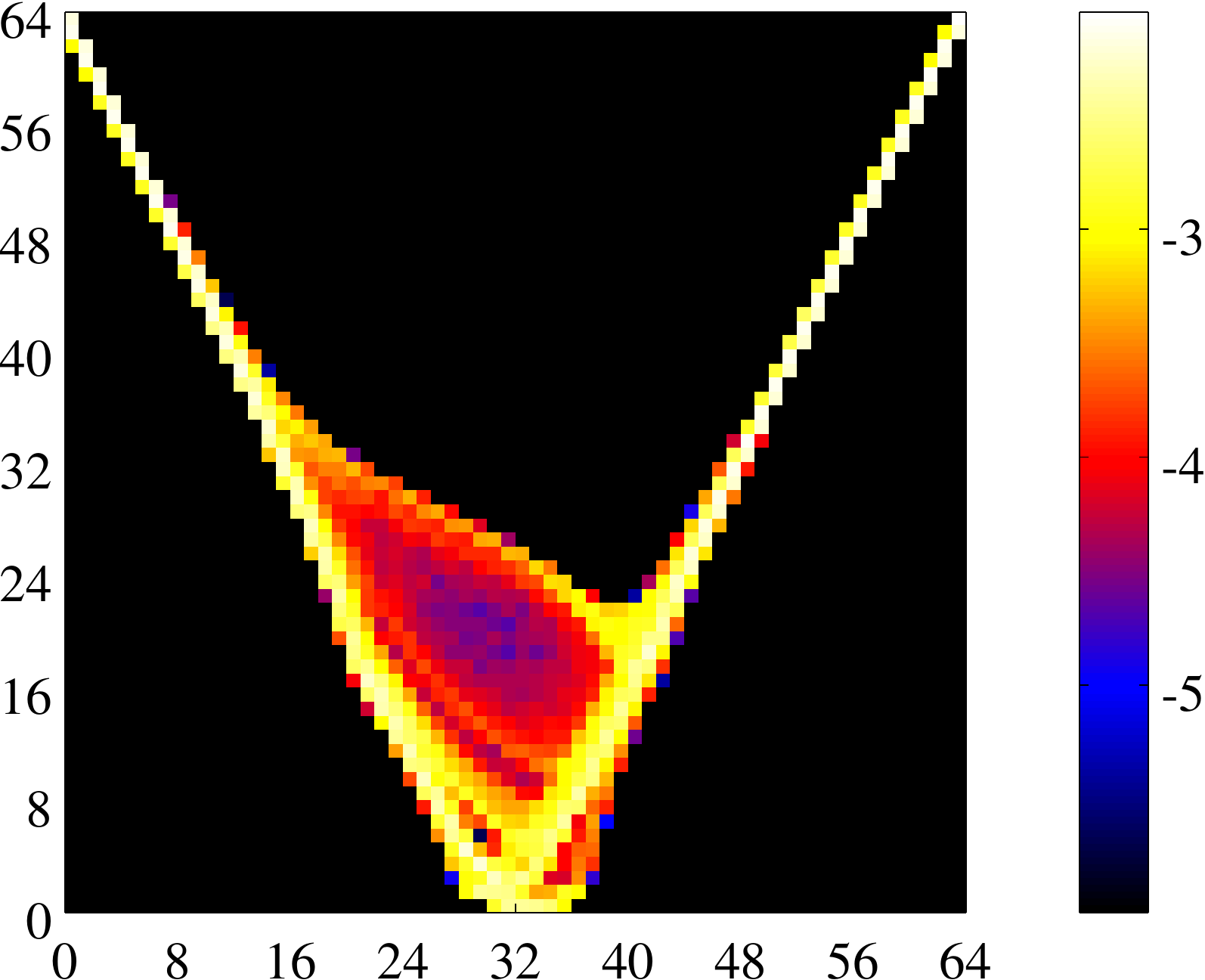}\\ 
  \caption{Significant phase portraits for R\"{o}ssler system using equiprobable binning. Portraits are arranged by state variable $x$ (left), $y$ (center), and $z$ (right).  The standard lag view of first local minima by variable are shown in the top row.  The middle row depicts the minimum lagged orthogonalized view. The bottom row shows the maximum mutual information orthogonalized view.}
  \label{Fig:ShadowsOfStraightRosslerEquibin}
\end{figure}

Figure \ref{Fig:ShadowsOfStraightRosslerEquibin} depicts the transformed phase portraits based on the equiprobable binning that compare to the equivalent criteria as those shown in Figure \ref{Fig:ShadowsOfStraightRossler}.  For the $x$ and $y$ coordinates, the views appear to be relatively minor transformation of the uniform bin views.  This is the result of the R\"{o}ssler system already populating those coordinates well.  

\begin{figure}
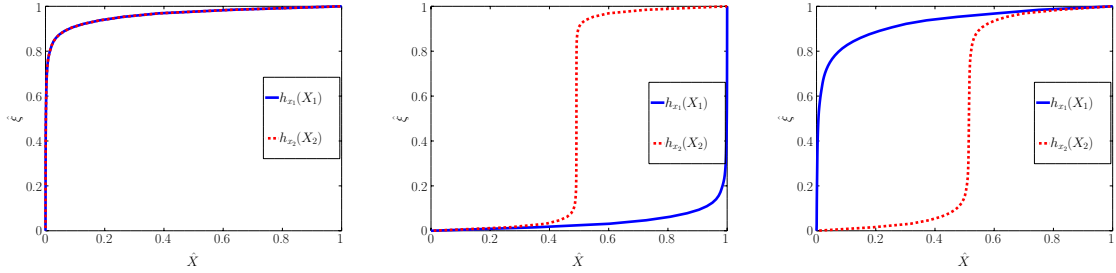

  \scalebox{0.25}{\include{ShadowEquibinTransform_14EquibinRosslerZ}}\hspace{0pt}%
  \scalebox{0.25}{\include{ShadowEigEquibinTransform_1EquibinRosslerZ}}\hspace{0pt}%
  \scalebox{0.25}{\include{ShadowEigEquibinTransform_32EquibinRosslerZ}}%
  \caption{Normalized mesh transformation functions for equiprobable bins for the R\"{o}ssler $z$ variable.  Left is the transformation applied to the time delay coordinates.  The middle figure is the transforms for the minimum lag orthogonalized view while the right is the transforms for the maximum mutual information view.  Note that the middle figure $h_1$ curve is mirrored over $\hat{X}=0.5$ and $\hat{\xi}=0.5$ due to the arbitrary sign of the eigenvectors which is similarly reflected in the up-down mirroring of the phase portrait compared to other orthogonalized views of $z$.}
  \label{Fig:TransformsRosslerEquibin}
\end{figure}

The views of the $z$ variable were considerably altered. The effect of equiprobable binning on the standard lagged view of the $z$ variable is most surprising in how much more similar it looks to the $x$ and $y$ variable views than any of the uniform binning results.  The fact that the equiprobable bin transformation so effectively converts the $z$-data into a view comparable to $x$ and $y$ makes it unsurprising that the evolution of mutual information with respect to lag are so tightly clustered in Figure \ref{Fig:EquibinRosslerMI}. Figure \ref{Fig:TransformsRosslerEquibin} compares the three mesh transformations for the $z$ variable.  This demonstrates the highly nonlinear mesh transformation that successfully converted the R\"{o}ssler system's $z$ variable into a similar phase portrait as $x$ and $y$.

The impacts of equiprobable binning on the $z$ portraits in the orthogonalized coordinates are considerably less dramatic.  The stretching does open the distribution in novel ways compared to the uniform bin version in particular by capturing the hole in the center of the distribution seen in $x$, $y$, and the lagged $z$ portraits.  The maximum mutual information view in the orthogonalized coordinates again appear to be potentially poor embedding views in that there is considerable ambiguity resulting from parallel but opposite direction phase space flows being compressed into small regions of the portrait.  Again, these are only two dimensional phase portraits of an intrinsically three dimensional system which enables such degenerate views when \MI is maximized.

\begin{figure}
  \includegraphics[width=0.3\textwidth]{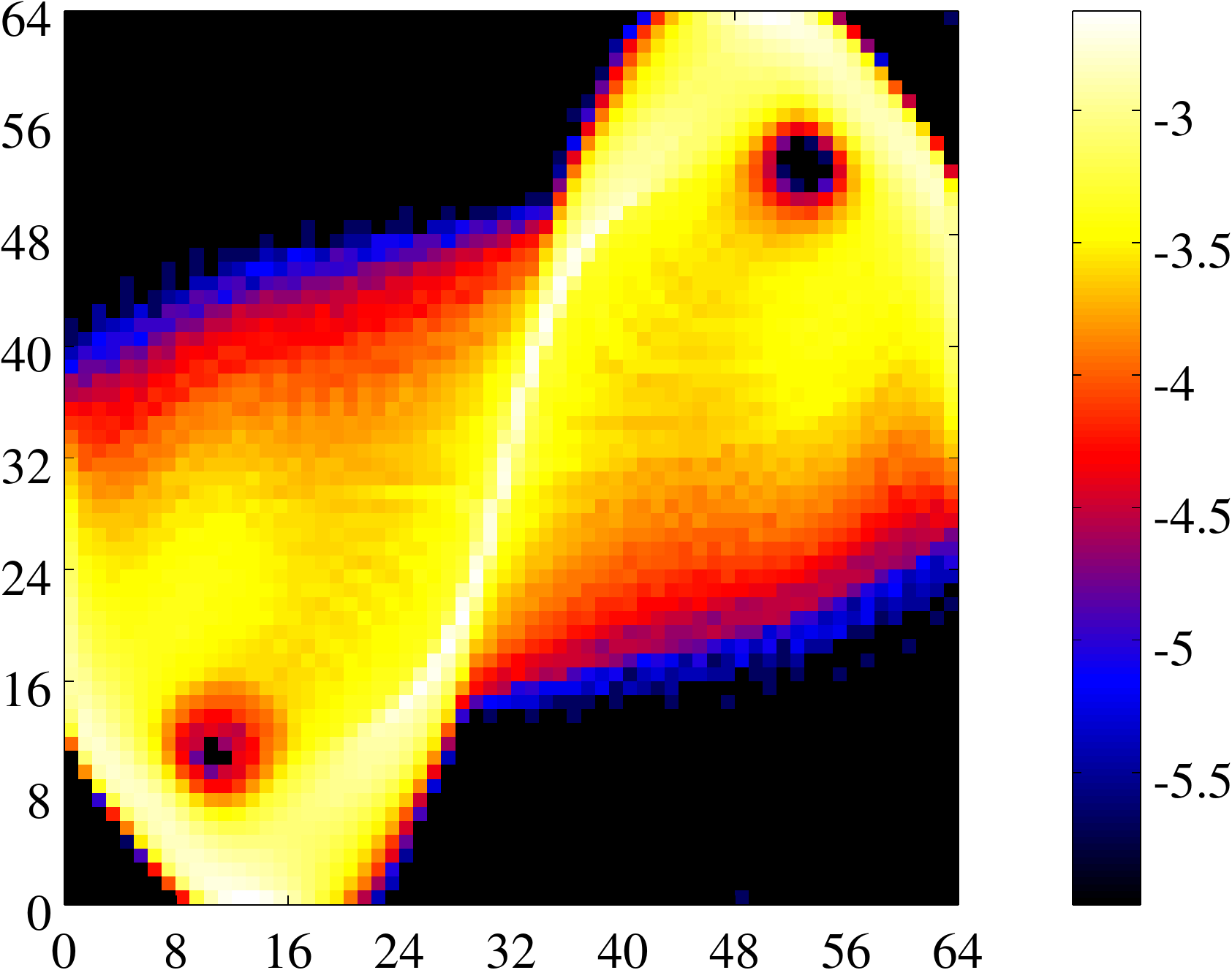}\hspace{16pt}%
  \includegraphics[width=0.3\textwidth]{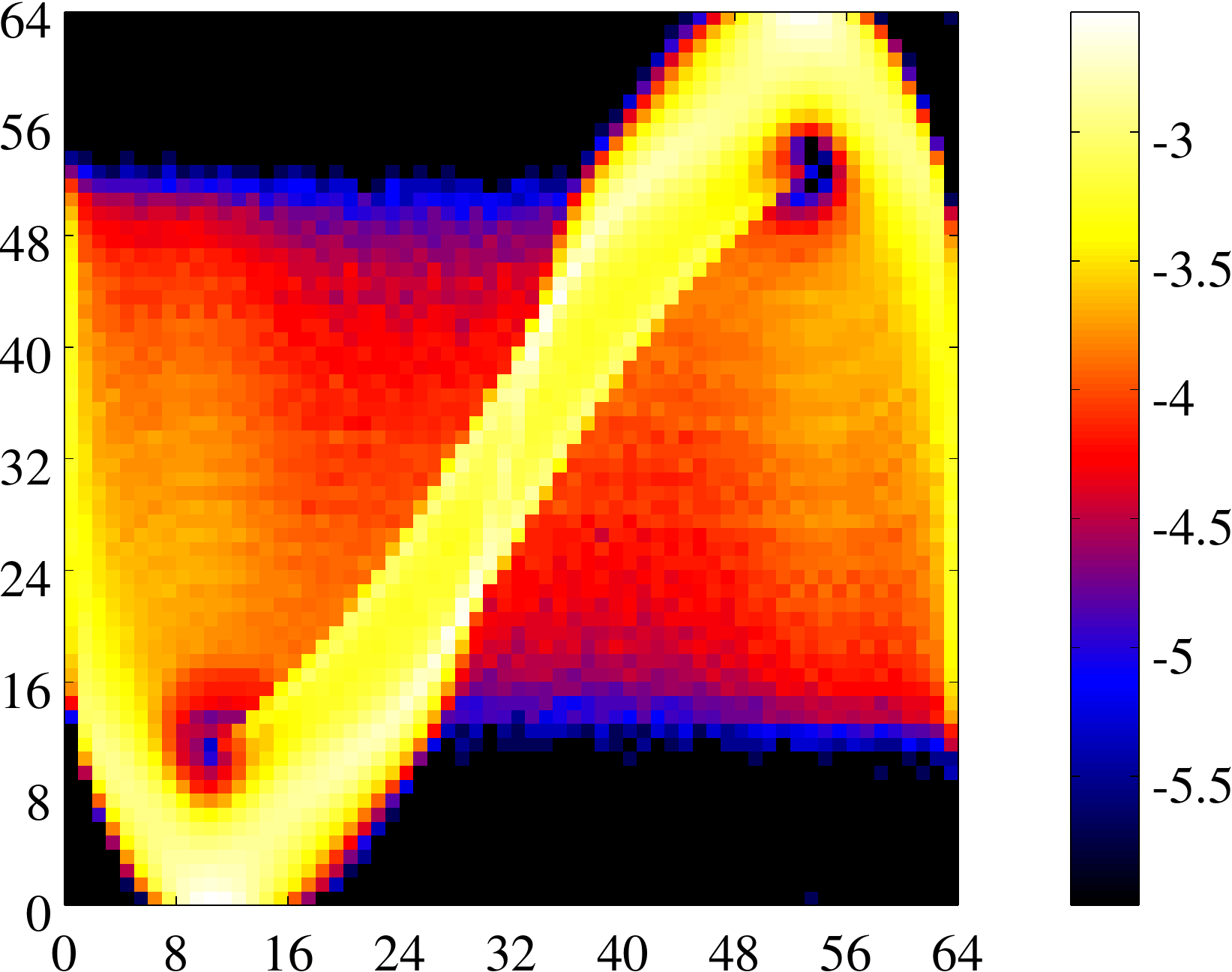}\hspace{16pt}%
  \includegraphics[width=0.3\textwidth]{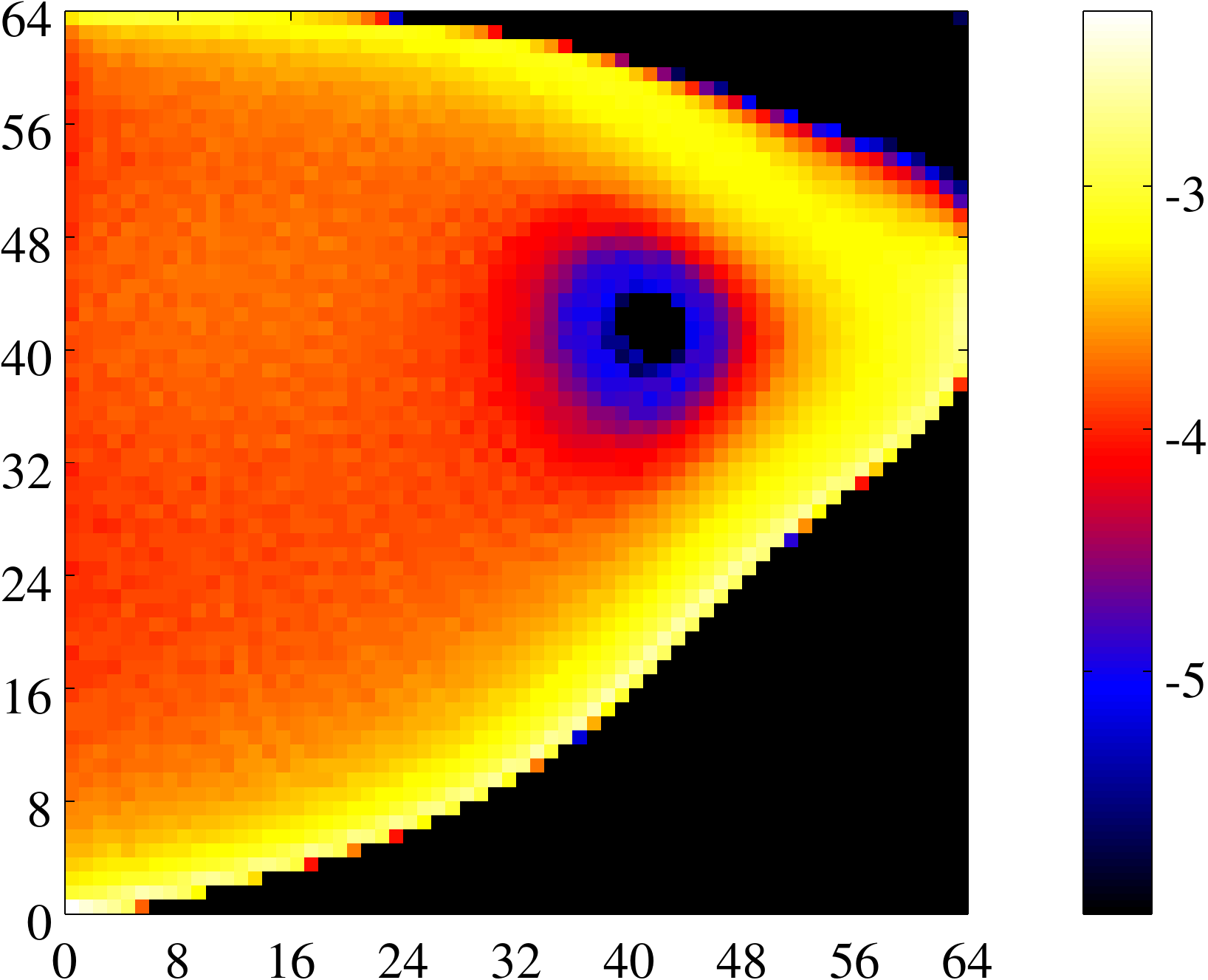}\\
  \includegraphics[width=0.3\textwidth]{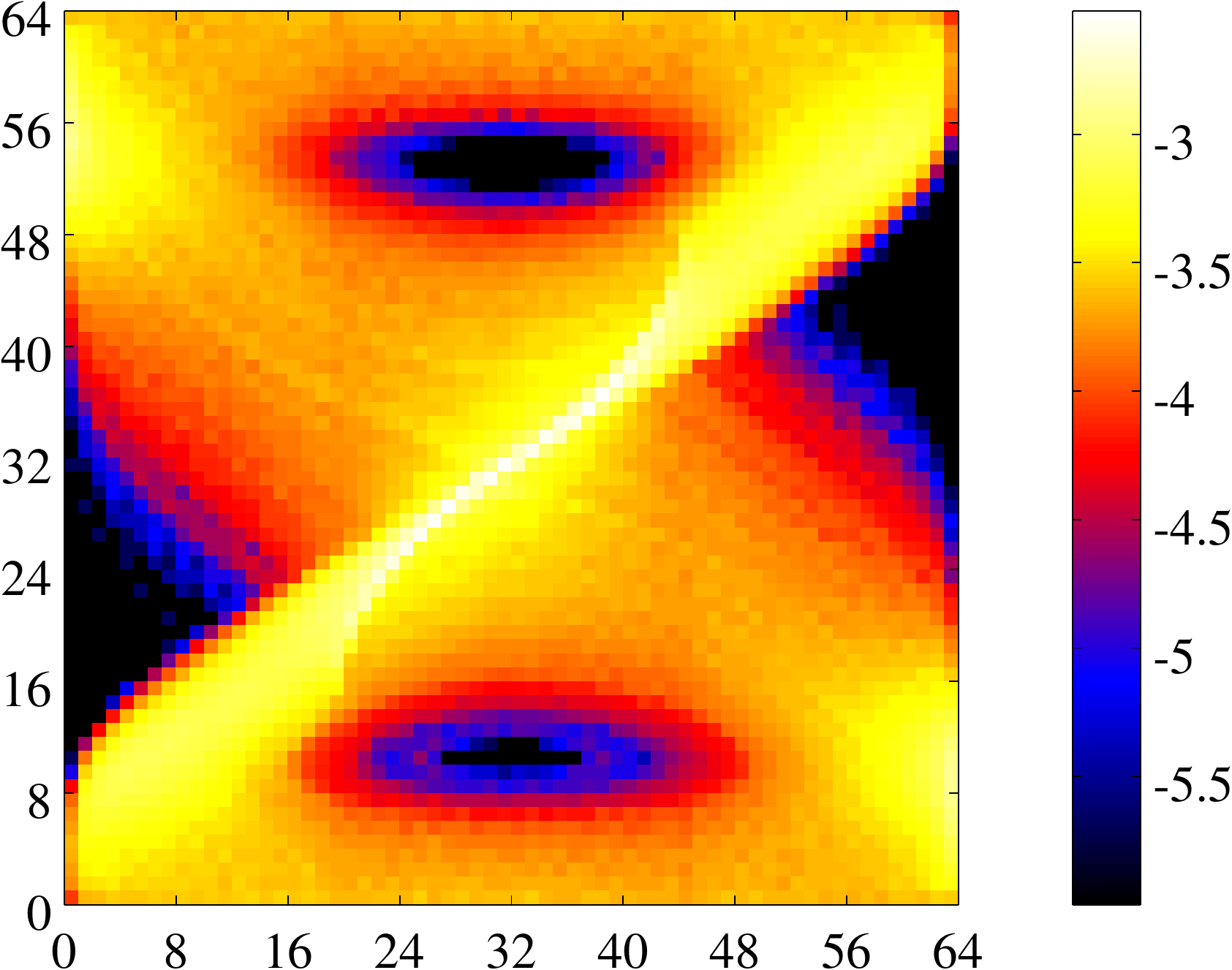}\hspace{16pt}%
  \includegraphics[width=0.3\textwidth]{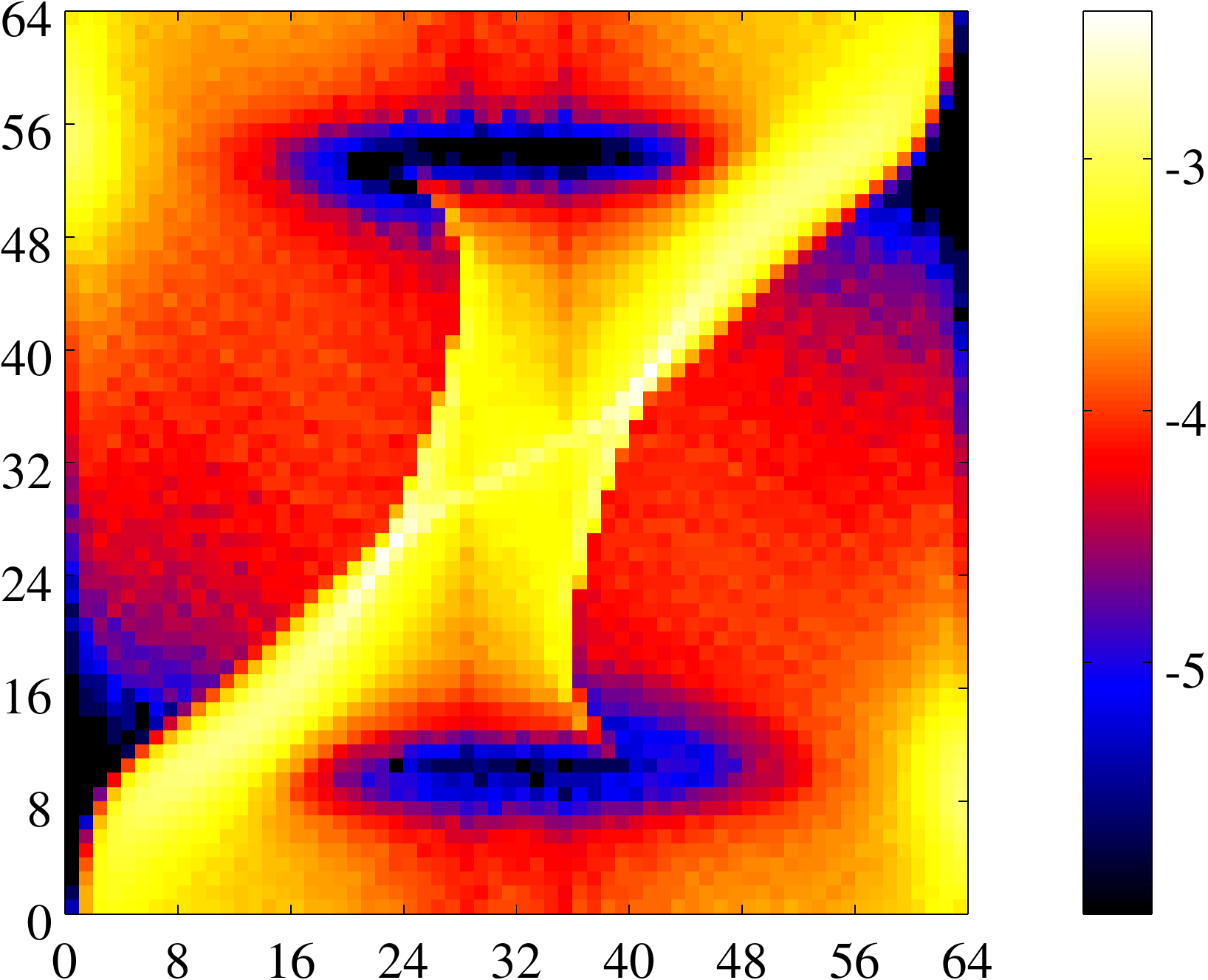}\hspace{16pt}%
  \includegraphics[width=0.3\textwidth]{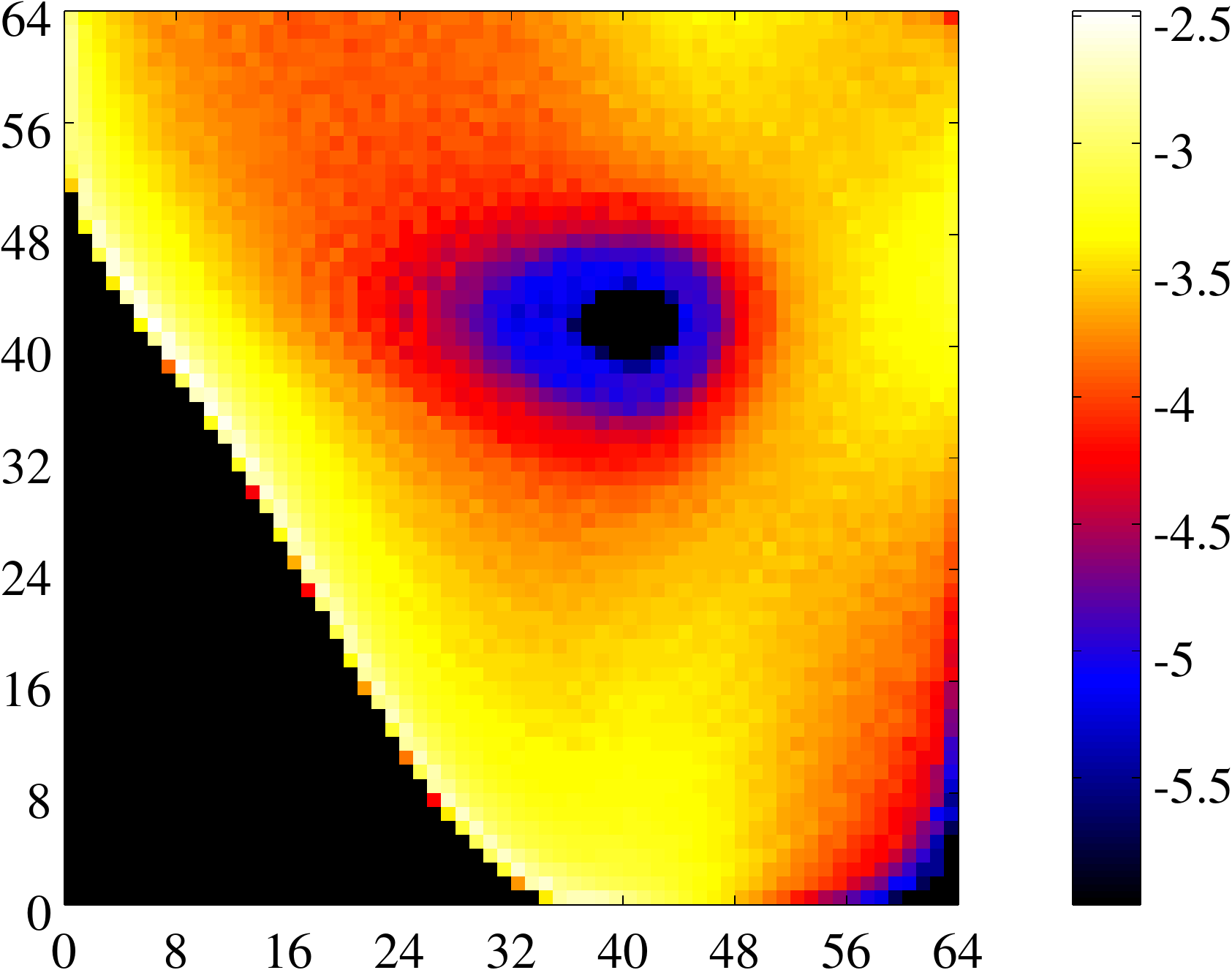}\\
  \includegraphics[width=0.3\textwidth]{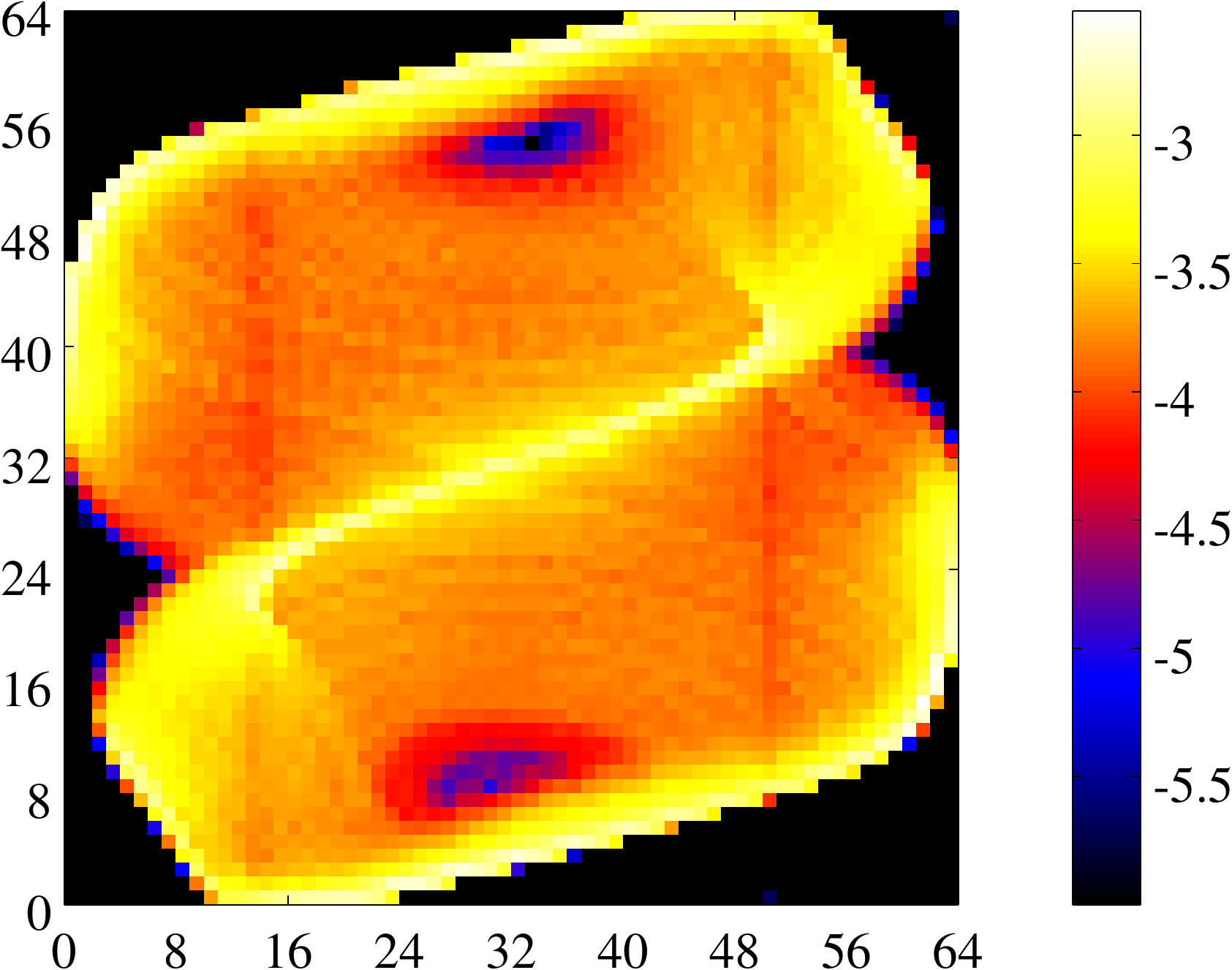}\hspace{16pt}%
  \includegraphics[width=0.3\textwidth]{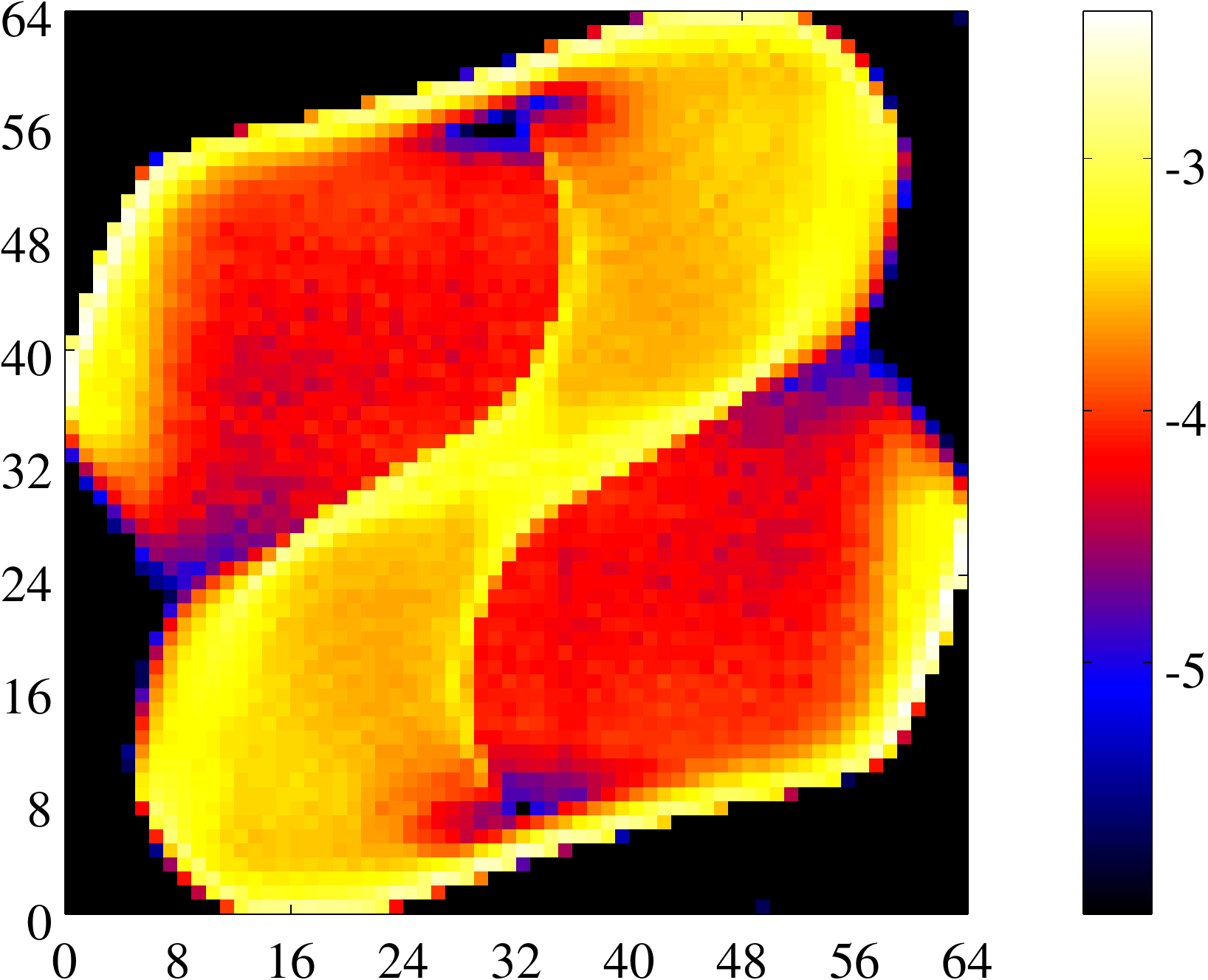}\hspace{16pt}%
  \includegraphics[width=0.3\textwidth]{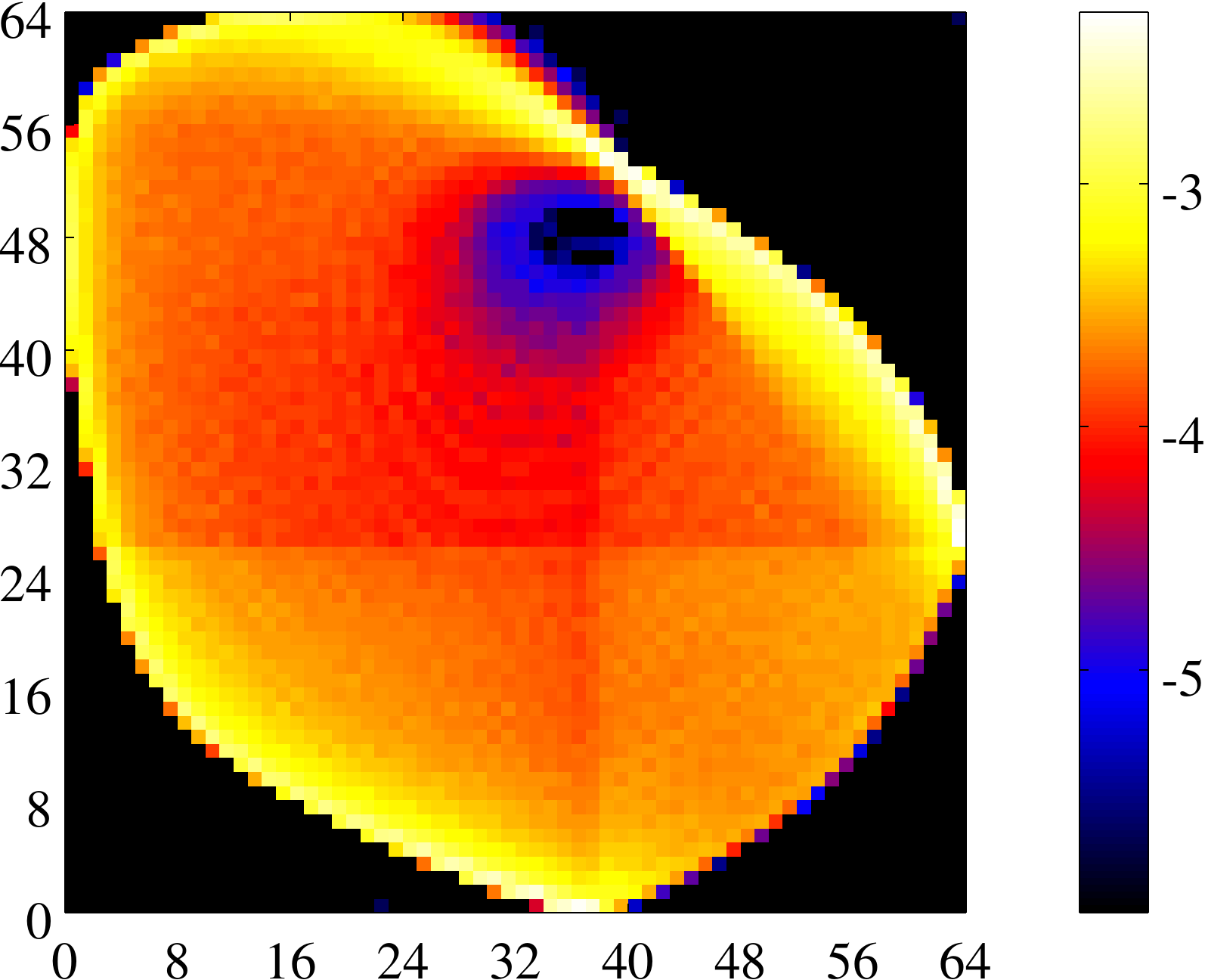}\\ 
  \caption{Significant phase portraits for Lorenz system using equiprobable binning. Portraits are arranged by state variable $x$ (left), $y$ (center), and $z$ (right).  The standard lag view of first local minima by variable are shown in the top row.  The middle row depicts the minimum lagged orthogonalized view. The bottom row shows the maximum mutual information orthogonalized view.}
  \label{Fig:ShadowsOfStraightLorenzEquibin}
\end{figure}

To demonstrate that the dramatic similarity of the R\"{o}ssler $z$ variable shadow to the $x$ and $y$ variable in the stretched lag view is principally a result specific to the R\"{o}ssler, the same array of views were calculated for the Lorenz system as shown in Figure \ref{Fig:ShadowsOfStraightLorenzEquibin}.  Note that the similarity between the lagged and orthogonalized first and maximum mutual information views of the $z$ variable are significantly more similar and consistent for the Lorenz system. The $z$ views are also more distinct from the $x$ and $y$ views for the Lorenz system in another critical way.  The $z$ view of the Lorenz system attractor has only one hole rather than two as a result of the $z$ dynamics symmetry. The dynamics are equivalent for $z$ given points mirrored across the $x$ and $y$ planes as $(x,y)\rightarrow (-x,-y)$.  This means that the $z$ time history can not be used to distinguish between points in phase space where the $x$ and $y$ dynamics have been mirrored.  The resulting shadows are then no longer one-to-one with the original phase space.  The failure of this projection to preserve the topology of the attractor is covered by one of the assumptions in the proof of Takens' theorem which requires that ``no two fixed points of $\phi$ are in the same level of $y$''. It is a consequence of the observation of the system being uniquely aligned with this symmetry of the system.  The number of such views is vanishingly small such that almost any projection, and particularly random projections, of the finite dimensional attractor would not suffer this degeneracy.

\section{Impact of Noise on Lag Selection}
\label{sec:noise}

It is well known that noise in data can corrupt the estimate for velocity and higher time derivatives as mentioned in References \cite{CASDAGLI199152,Letellier}. This is a principle motivation for the use of the time delay embedding instead of the difference embedding.  For a difference embedding, a wider finite difference stencil from a larger $\tau$ can be used as a low-pass filter to partially mitigate the noise of the derivative, but only at the expense of reducing the accuracy and potentially contracting the extrema of the derivative. The reason this occurs is that the magnitude of the signal increases with $\tau$ while the magnitude of the noise is independent of $\tau$, at least in the case of white noise.  However, in chaotic systems, the signal becomes decorrelated with itself for large values of $\tau$ which transforms the dynamics into noise of a different type.  The chaotic loss of the signal could be avoided by ensuring that the lag, $\tau$, is relatively short compared to the maximum Lyapunov exponent of the system. However, for general data coming from noisy observations where the underlying system dynamics are unknown, determining the Lyapunov exponents is a problem of equal magnitude as the original embedding problem that is the focus of this work.  In fact, creating a good reconstruction of phase space is a critical first step in attempting to determine the maximum Lyapunov exponent from experimental data as noted in Reference \cite{Abarbanel}.

The optimal choice of $\tau$ would balance denoising with randomness resulting from large lags, but this is a significantly challenging problem when both the signal and noise are broad spectrum. Distinguishing between the two requires defining a metric in which the competing sources of randomnesses can be compared in a meaningful manner.  In this section, the impact of noise on the \MI content is explored for both the Lorenz and R\"{o}ssler systems in the 2D time delay and orthogonal embedding.  The extension of the orthogonal embedding to higher dimensions is provided and additional investigation is performed into noise impact on the high dimensional embeddings for both canonical systems. 

\subsection{Noise in low dimensional embeddings}

 Figure \ref{Fig:NoisyLorenz} shows the impact of several signal-to-noise ratios (SNR) on the \MI content (using uniform bin sizes) of the canonical Lorenz system. Comparing Figure \ref{Fig:NoisyLorenz} to the mutual information curves from Figure \ref{Fig:ShadowStraight}, it is clear that the added noise decreases the initial mutual information for short time lags.  For the orthogonal embeddings, the shortest time lag has approximately zero mutual information whereas the standard lagged views still exhibits the short time redundancy at a reduced magnitude due to the additional noise.  Both results are consistent with expectations. For the lag views, the noise thickens the $i$=$j$ line for all short $\tau$ decreasing the maximum observed \MI. For the orthogonal views, the high frequency noise dominates the numerical derivative as $\tau\rightarrow 0$.  As this noise is random and independent of the value of $x$, the mutual information should be significantly decreased.  In fact, the only reason that the \MI for the 15dB SNR instance of the minimum time orthogonalized embedding is not exactly zero is that the magnitude of the signal for the finite small $dt$ of 0.01 remains significant compared to the noise.  In the limit as $t\rightarrow 0$, it would be identically zero.

\begin{figure}
  \scalebox{0.5}{\include{ShadowStraightVsEigGaussianNoise}}
  \caption{Mutual information with respect to lag using standard lagged (black) and orthogonalized (blue) coordinates for the Lorenz system's $x$ variable with additive Gaussian noise.  Line styles indicate differing signal to noise ratios.}
  \label{Fig:NoisyLorenz}
\end{figure}

\begin{figure}
  \includegraphics[width=0.3\textwidth]{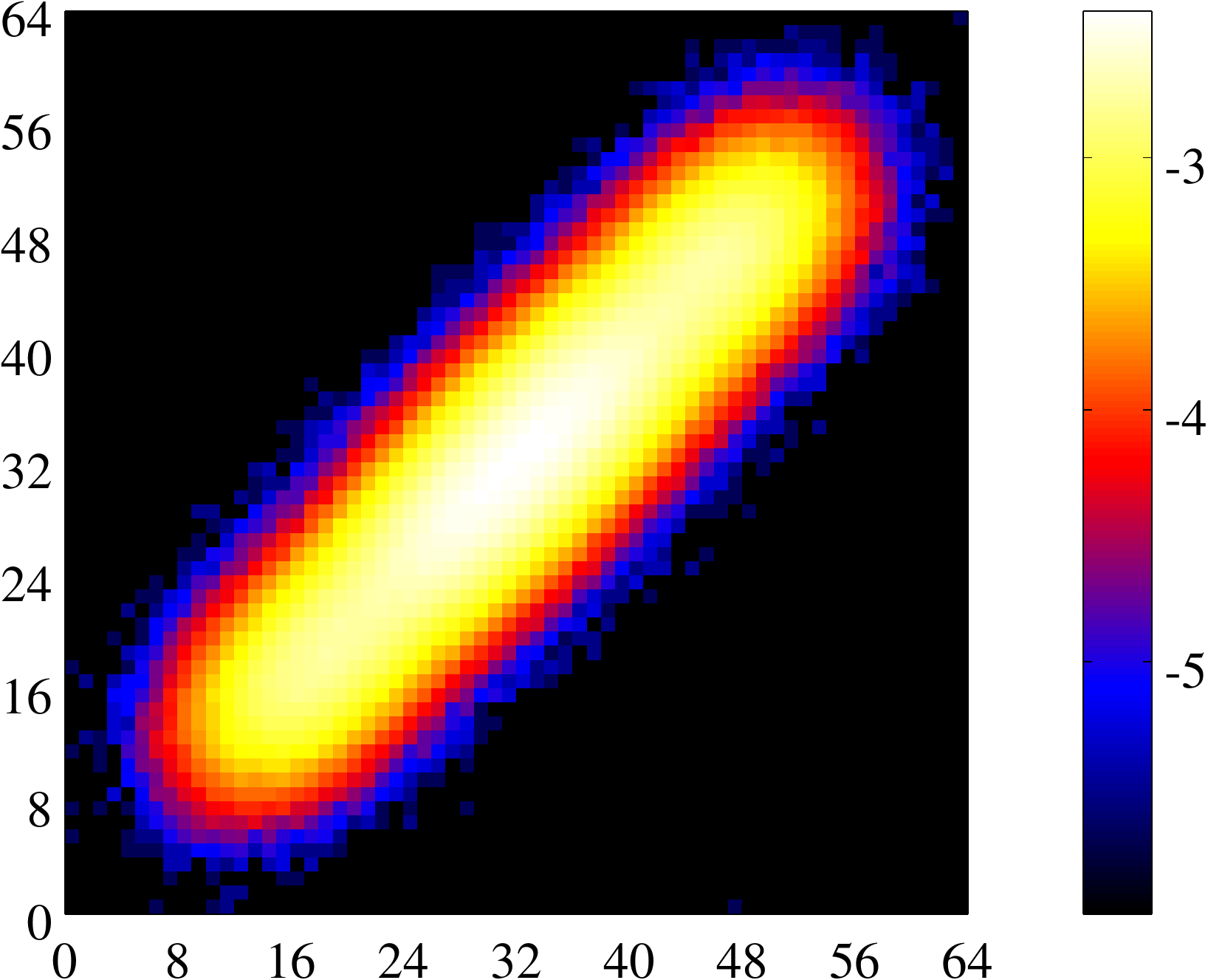}\hspace{16pt}%
  \includegraphics[width=0.3\textwidth]{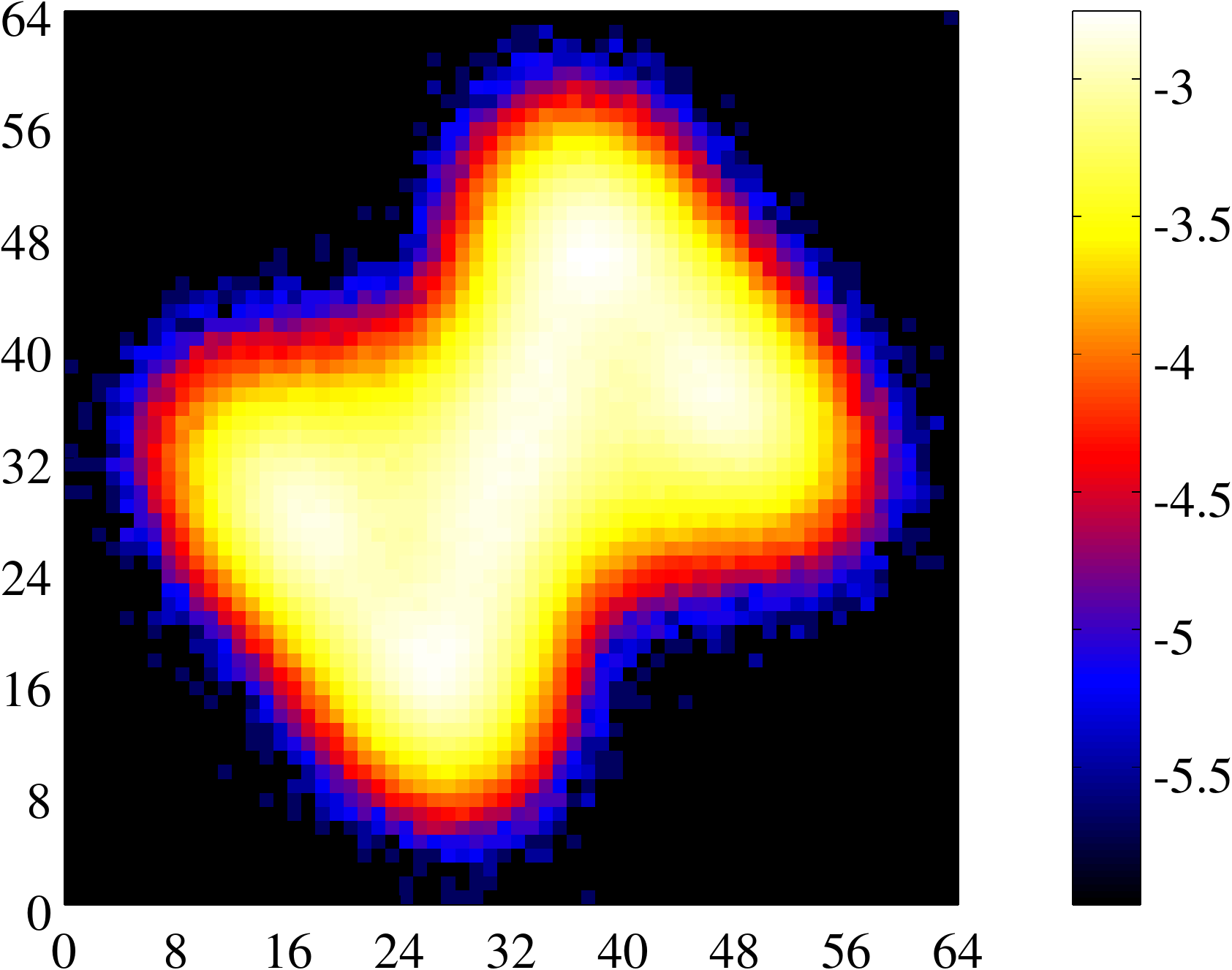}\hspace{16pt}%
  \includegraphics[width=0.3\textwidth]{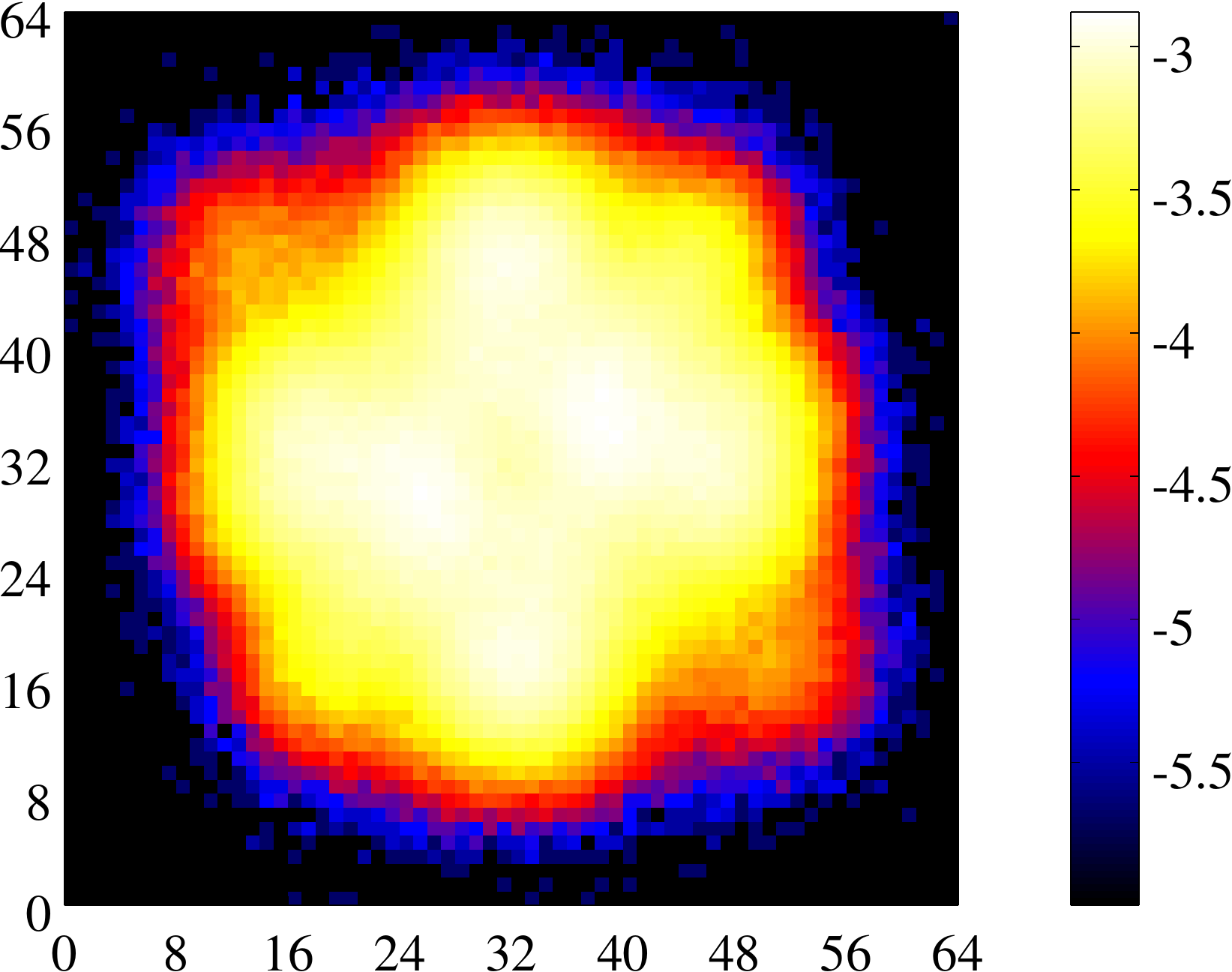}\\
  \includegraphics[width=0.3\textwidth]{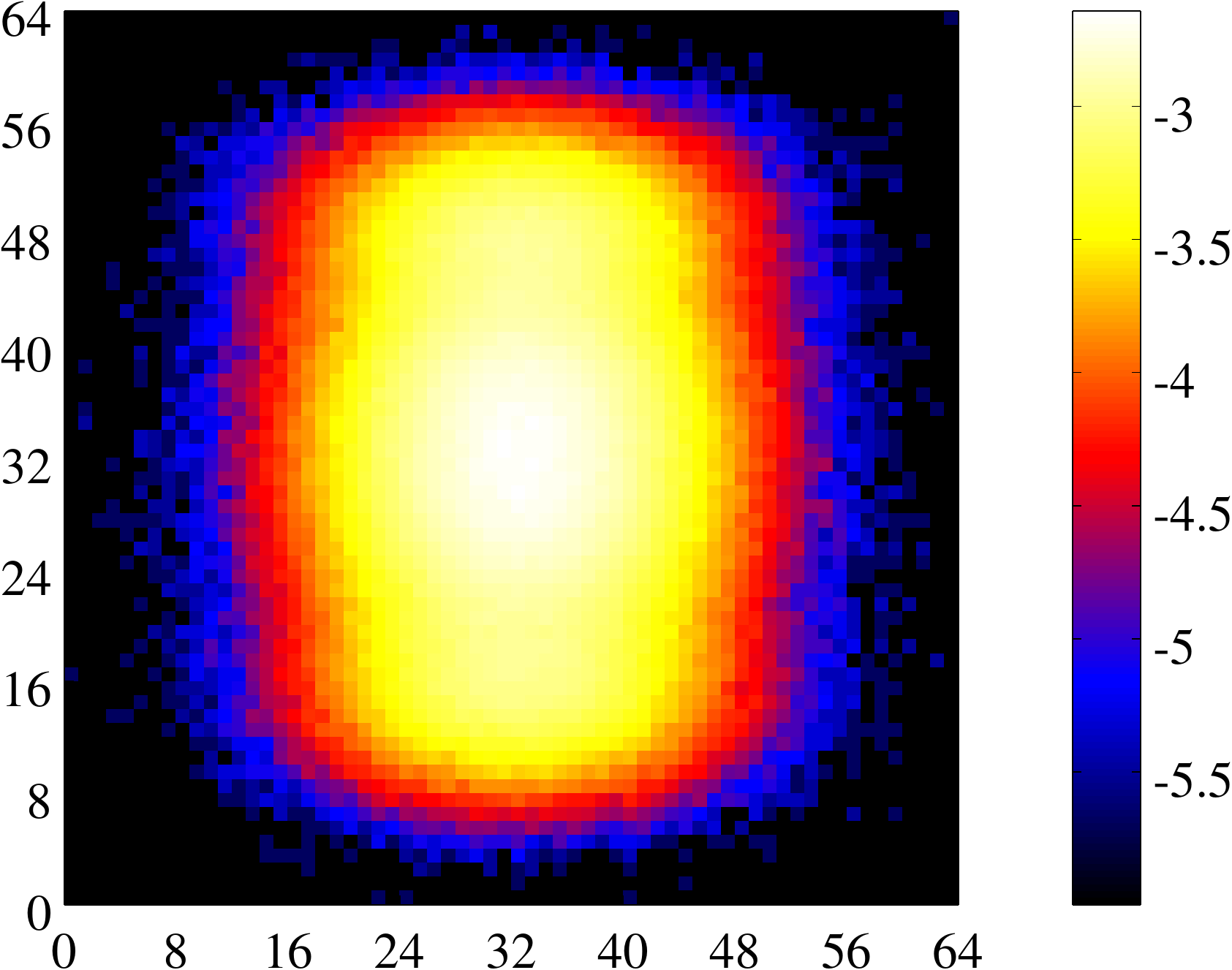}\hspace{16pt}%
  \includegraphics[width=0.3\textwidth]{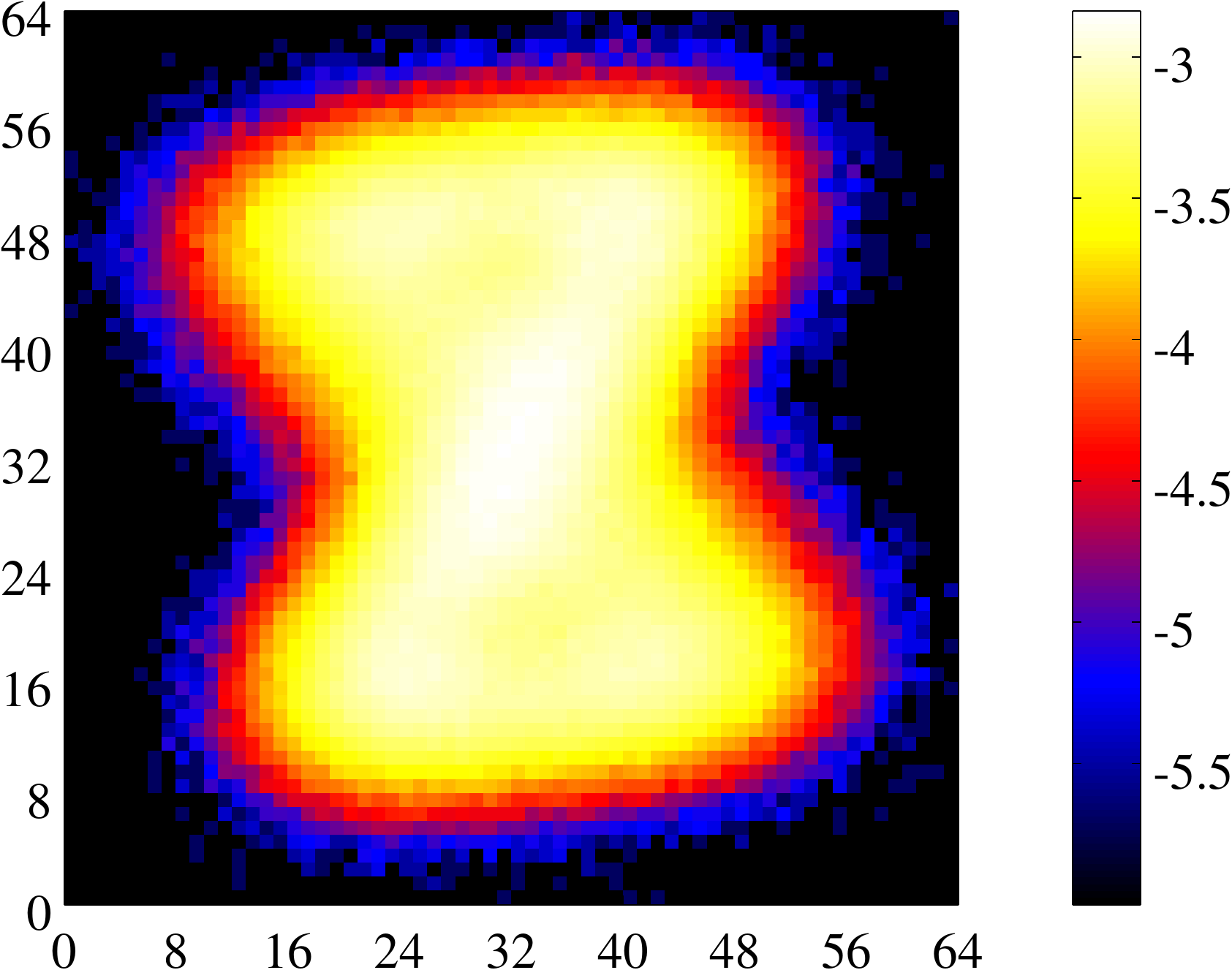}\hspace{16pt}%
  \includegraphics[width=0.3\textwidth]{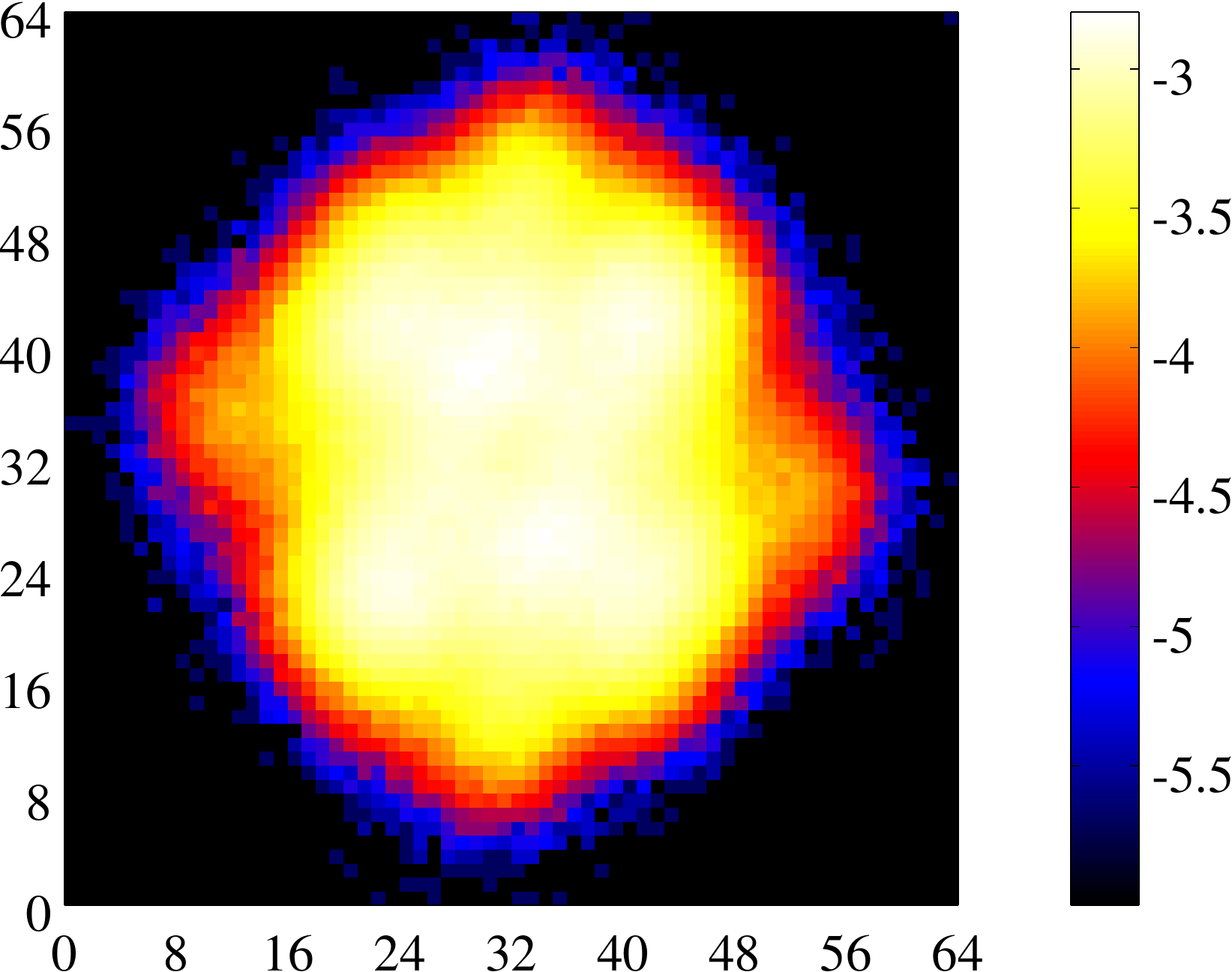}\\  
  \caption{Phase portraits from $x$ variable of lagged Lorenz data with SNR of 5dB additive Gaussian noise. Top row is original lagged coordinates and bottom row is in rotated orthogonal coordinates.  The left and right figures show minimal lag and longest lag portraits denoted by ``$\circ$'' and ``$*$'' symbols on the mutual information plot.  Middle figure shows views from ``$+$'' marked points due to the associated first minimum and global maximum criteria on mutual information.}
  \label{Fig:ShadowsOfStraightNoise}
\end{figure}

Figure \ref{Fig:ShadowsOfStraightNoise} depict the 5dB SNR noise level shadows commensurate with the clean shadows from Figure \ref{Fig:ShadowsOfStraight}. Note that the minimum lag views are from a lag of 1 rather than zero so that different realizations of the noise are applied. The figures would be nearly identical for $\tau\rightarrow 0$ so long as the noise on the original and lagged variables was not identical.  In the case that the noise is identical, the $\tau=0$ result reverts back to all the data landing on the $i=j$ diagonal which is only an artifact of finite $dt$ in the sampling of the noise.  As $\tau \rightarrow 0.19$ or a lag of 19 steps, corresponding roughly to the middle column of Figure \ref{Fig:ShadowsOfStraightNoise}, a blurred version of the shadow of the causal dynamics of the system reemerges. Beyond these mutual information based optimal lags, the distribution folds back on itself and the chaos of the system depletes the mutual information for larger lags.  It is interesting to note that in this system, the first minimum in mutual information for the lagged view and the global maximum in the orthogonalized view tend to generally agree reasonably.

\subsection{Noisy Lags in Higher Dimension}
\label{SubSec:Opt}
Pushing the embedding to three and higher dimensions provides additional insights into the behavior of the mutual information quantities and their ability to detect a lag representing a good view of the embedding.  This section addresses the impact of higher dimensionality on the views of the data.  To better understand this impact, issues relating to the noise floor due to the sparseness of data in high dimensions and additional detail on the orthogonalized coordinates in higher dimension must first be addressed.

\subsubsection{Accounting for Sparseness in High Dimension}
\label{SubSubSec:Floor}
In higher dimensions, the concept of a noise floor becomes more pressing.  The noise floor, as will be included in subsequent mutual information plots such as Figure \ref{Fig:ShadowStraightEquibin3-5DLorenz}, represents the expected apparent mutual information resulting from the deviation of bin density from a uniform value due to the finite length of the data. As the number of points, $n$, approaches infinity, the sum of samples in each bin approaches an expected average such that the bin counts $\lambda_i \rightarrow \bar{\lambda}=n/n_{bin}$. The probability of a point falling within the $i^{th}$-bin, $p_i$, then similarly approaches $p_i=\lambda_i/n \rightarrow 1/n_{bin}$ with the resulting mutual information of $0$.  This is only the limiting case of many samples per bin.  When the number of data points is on the order of or fewer than the number of bins, the randomly distributed points result in Poisson distributed bin probabilities and a non-uniform probability density. This nonuniformity results in an apparent mutual information governed by the deviation of the random noise from the expectation.  The expected entropy of the noise floor can then be calculated by estimating the expected number of bins with each possible bin count via the Poisson distribution and summing over all possible bin counts, $k$, for $n$ points in $n_{bin}$ bins as shown in Equation \ref{Eq:Noise}. Note that the $k=0$ term of the Poisson distribution has been dropped using $0\;log_2(0)=0$ as in the other entropy calculations. 

\begin{equation}
H^{noise}=-\sum_{k=1}^{n} n_{bin} \left( \frac{\bar{\lambda}^ke^-\bar{\lambda}}{k!} \right) \frac{k}{n}\; log_2\left(\frac{k}{n}\right) \label{Eq:Noise}
\end{equation}

\subsubsection{Higher Dimensional Orthogonalization}
\label{SubSubSec:Higher}
The $n^{th}$-order orthonormal discrete Legendre rotation matrices, $V^{(n)}$, are shown in Equation \ref{Eq:Legendre35} for $n$=3, 4, 5 below as taken from the Appendix of Reference \cite{GIBSON19921}.  A recurrence relation is included in that reference for generating higher order versions. Reference \cite{Aburdene}, provides a parallel algorithm for computing the same coefficients.

\begin{align}
  V^{(3)}&=  \begin{bmatrix}v^{(3)}_1\\v^{(3)}_2\\v^{(3)}_3\end{bmatrix}  =  \begin{bmatrix}\{1,1,1\}/\sqrt{3}\\ \{-1,0, 1\}/\sqrt{2} \\ \{1,-2, 1\}/\sqrt{6}\end{bmatrix} \nonumber\\
  V^{(4)}&=  \begin{bmatrix}v^{(4)}_1\\v^{(4)}_2\\v^{(4)}_3\\v^{(4)}_4\end{bmatrix}  =  \begin{bmatrix}\{1,1,1,1\}/2\\ \{-3,-1,1,3\}/\sqrt{20} \\ \{1,-1,-1,1\}/2\\ \{-1,3,-3,1\}\sqrt{20} \end{bmatrix}\label{Eq:Legendre35}\\
  V^{(5)}&=  \begin{bmatrix}v^{(5)}_1\\v^{(5)}_2\\v^{(5)}_3\\v^{(5)}_5\\v^{(5)}_5\end{bmatrix}  =  \begin{bmatrix}\{1,1,1,1,1\}/\sqrt{5}\\ \{-2,-1,0,1,2\}/\sqrt{10} \\ \{2,-1,-2,-1,2\}/\sqrt{14} \\ \{-1,2,0,-2,1\}/\sqrt{10} \\ \{1,-4,6,-4,1\}/\sqrt{70} \end{bmatrix}\nonumber
\end{align}

In three dimensions, the discrete Legendre orthonormal coordinates coincide with central average, centered first derivative, and centered second derivative in terms of finite difference equations.  This is a natural consequence of the orthogonality of the position, velocity, and acceleration unit vectors for a smooth parameterized equation.  In four and higher dimensions, however, though the highest order vector is the finite difference approximation for that order derivative, the other vectors blend towards shapes reflecting the continuous Legendre polynomials as described in the Appendix of Reference \cite{GIBSON19921}.  The deviation from classical finite difference approximations with the availability of more points in the stencil than required by the order of the derivative helps to filter the data to improve the SNR. It is interesting to reiterate that the Legendre basis described in Reference \cite{GIBSON19921}  was first observed numerically in this work as an emergent property of the eigendecomposition of the covariance matrix of data for the problems studied regardless of added noise.
This universality of the coordinate system makes it an interesting option for further study of any continuous time dynamical systems of this type.

\subsubsection{Impact of Higher Dimension Views}

\begin{figure}
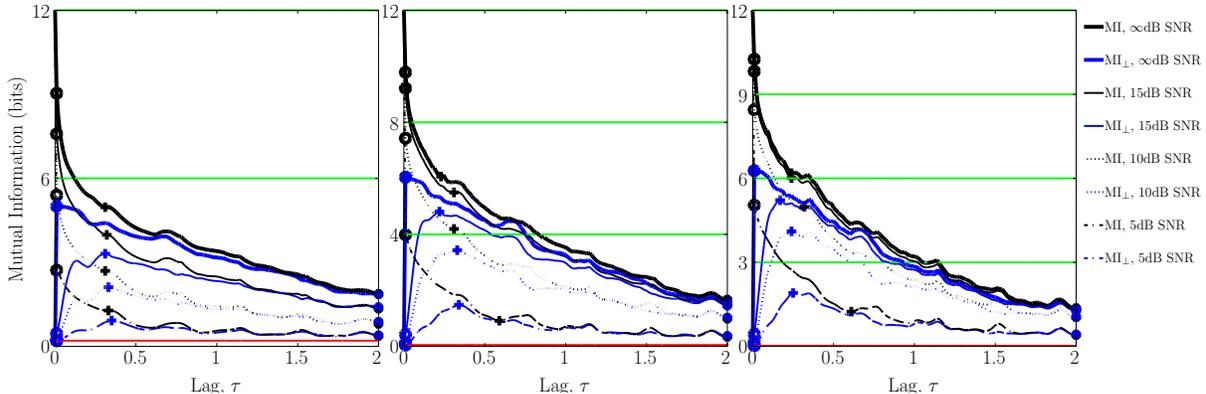

  \adjustbox{scale=0.36,trim=0pt 0pt 2.2in 0pt,clip}{
    \include{ShadowStraightVsEigLorenz6bXGaussianNoise3D}}%
  \adjustbox{scale=0.36,trim=.8in 0pt 2.2in 0pt,clip}{
    \include{ShadowStraightVsEigLorenz4bXGaussianNoise4D}}%
  \adjustbox{scale=0.36,trim=0.8in 0pt 0pt 0pt,clip}{
    \include{ShadowStraightVsEigLorenz3bXGaussianNoise5D}}%
  \caption{Mutual information for 3D-5D lags (left to right) of Lorenz-$x$ variable shadows on equiprobable bins in standard and orthogonalized coordinate systems.  The identified first minima and global maxima are marked with the ``$+$''-symbol for lagged and orthogonalized views respectively. Noise floor (red) and (d-1) lines representing $log_2(n_{edge})$ divisions (green) of the $n_{edge}^d$ bin hypercube are included for reference.}
  \label{Fig:ShadowStraightEquibin3-5DLorenz}
\end{figure}

Starting again with the Lorenz attractor, Figure \ref{Fig:ShadowStraightEquibin3-5DLorenz} depicts the evolution of mutual information with lag for lagged equiprobable bins and orthogonalized equiprobable bins in 3 to 5 dimensions.

\begin{figure} 
  \includegraphics[width=0.249\textwidth]{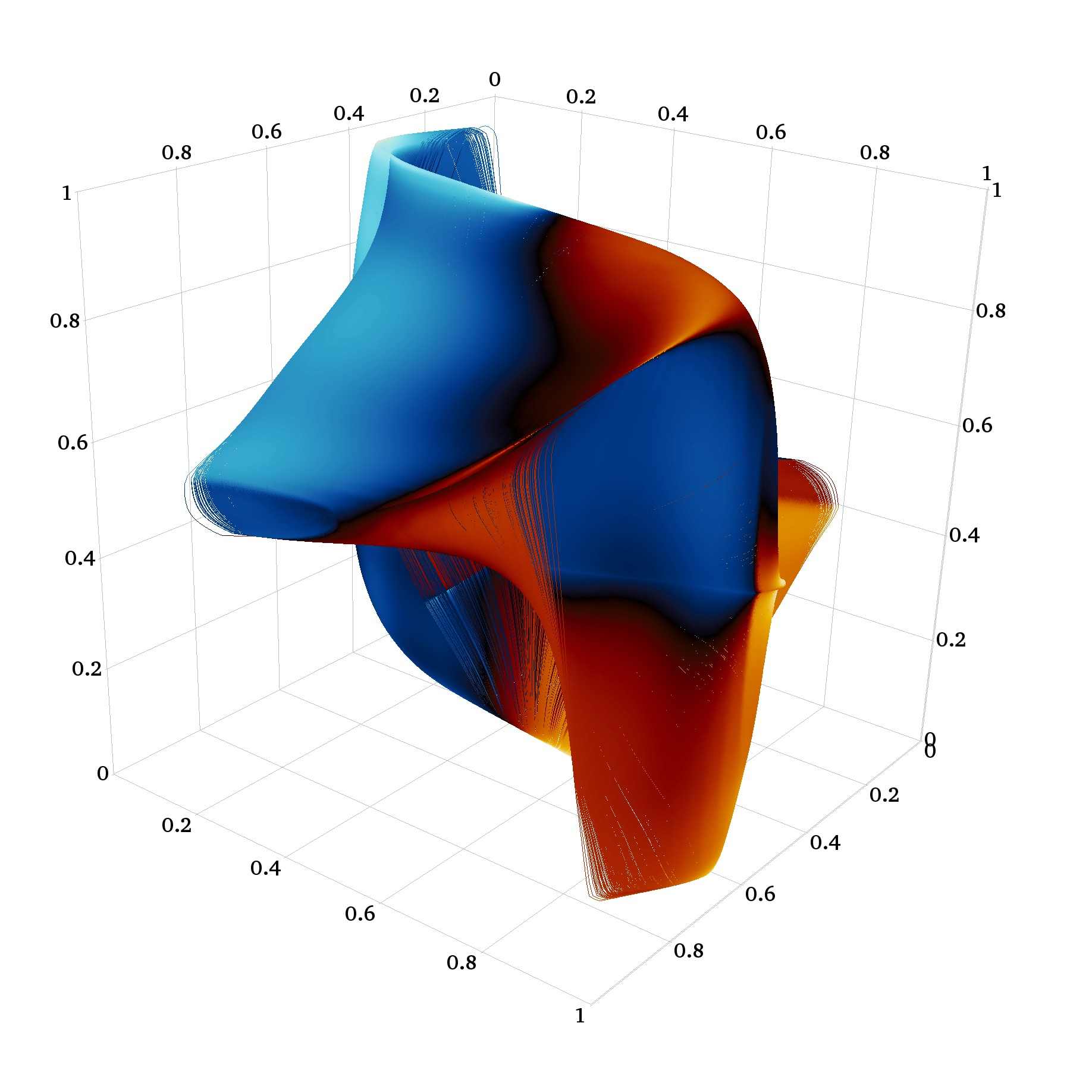}
  \includegraphics[width=0.249\textwidth]{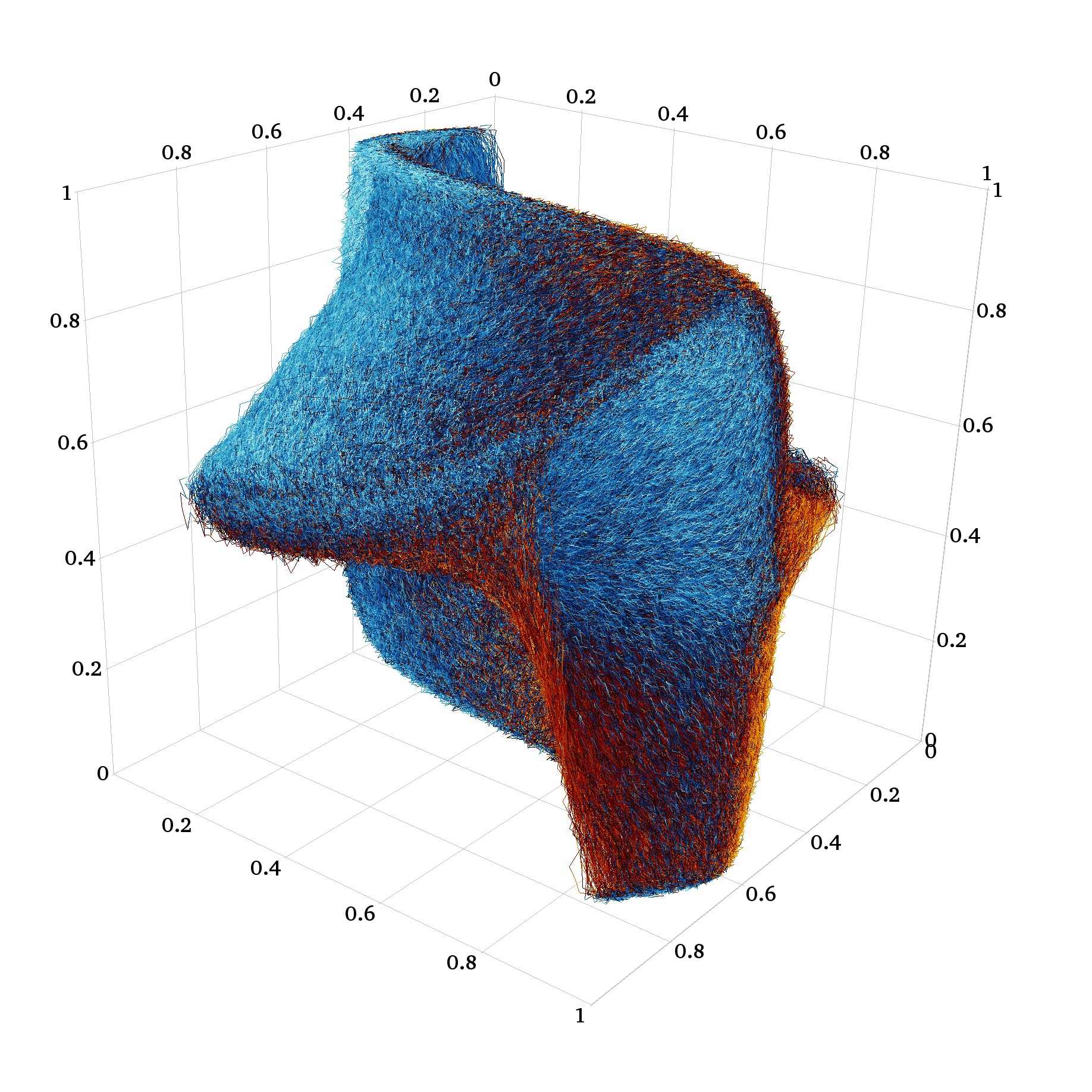}
  \includegraphics[width=0.249\textwidth]{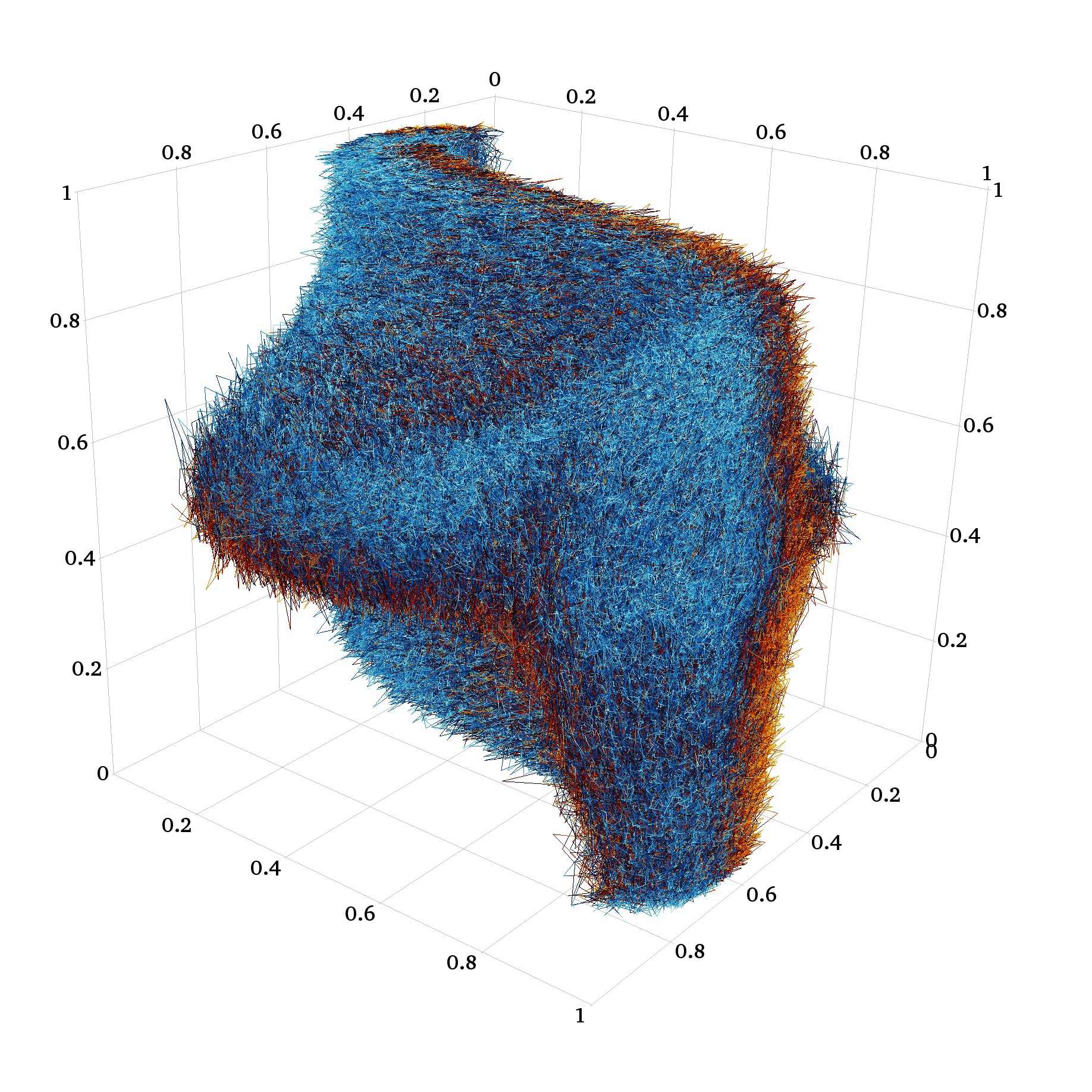}
  \includegraphics[width=0.249\textwidth]{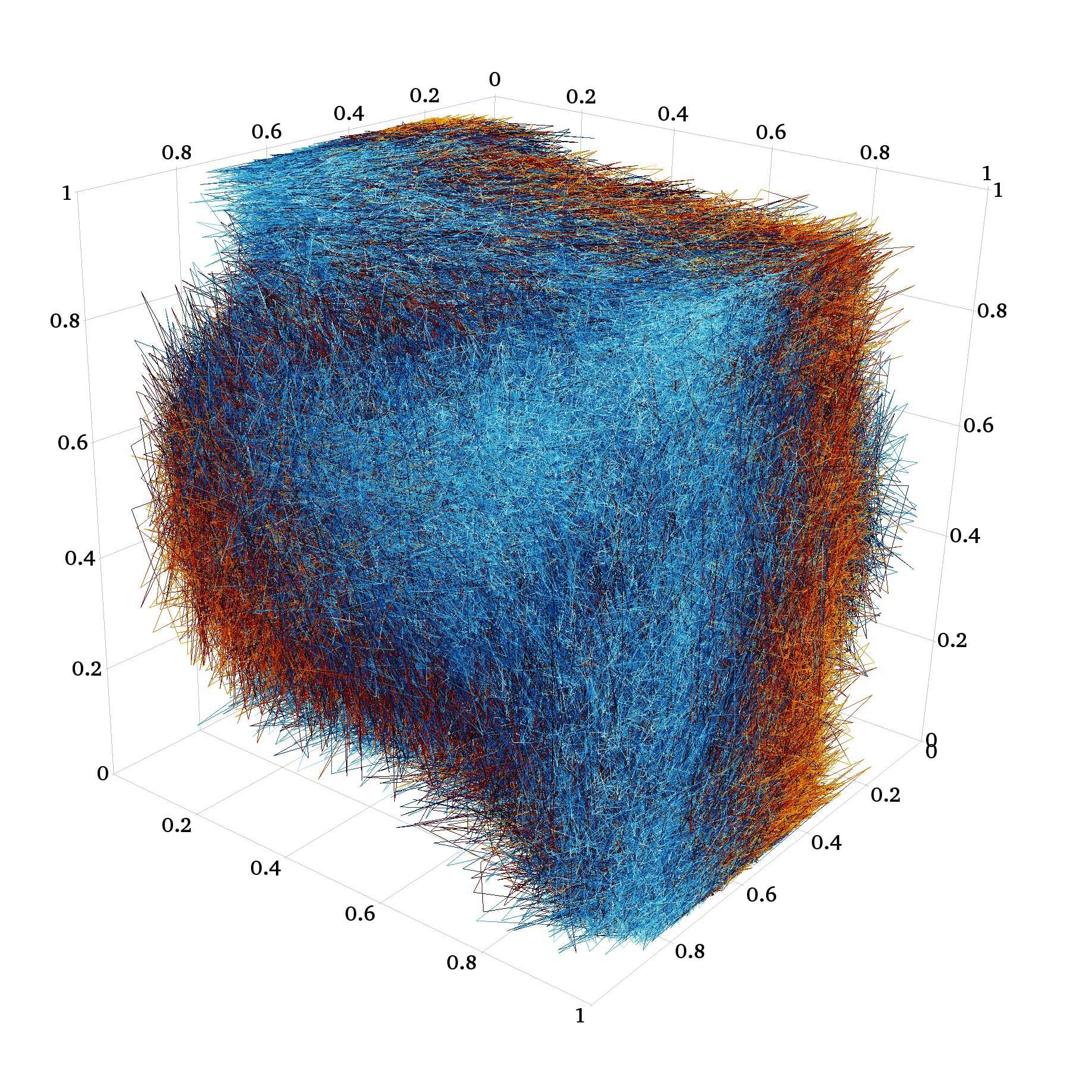}\\ 
  \includegraphics[width=0.249\textwidth]{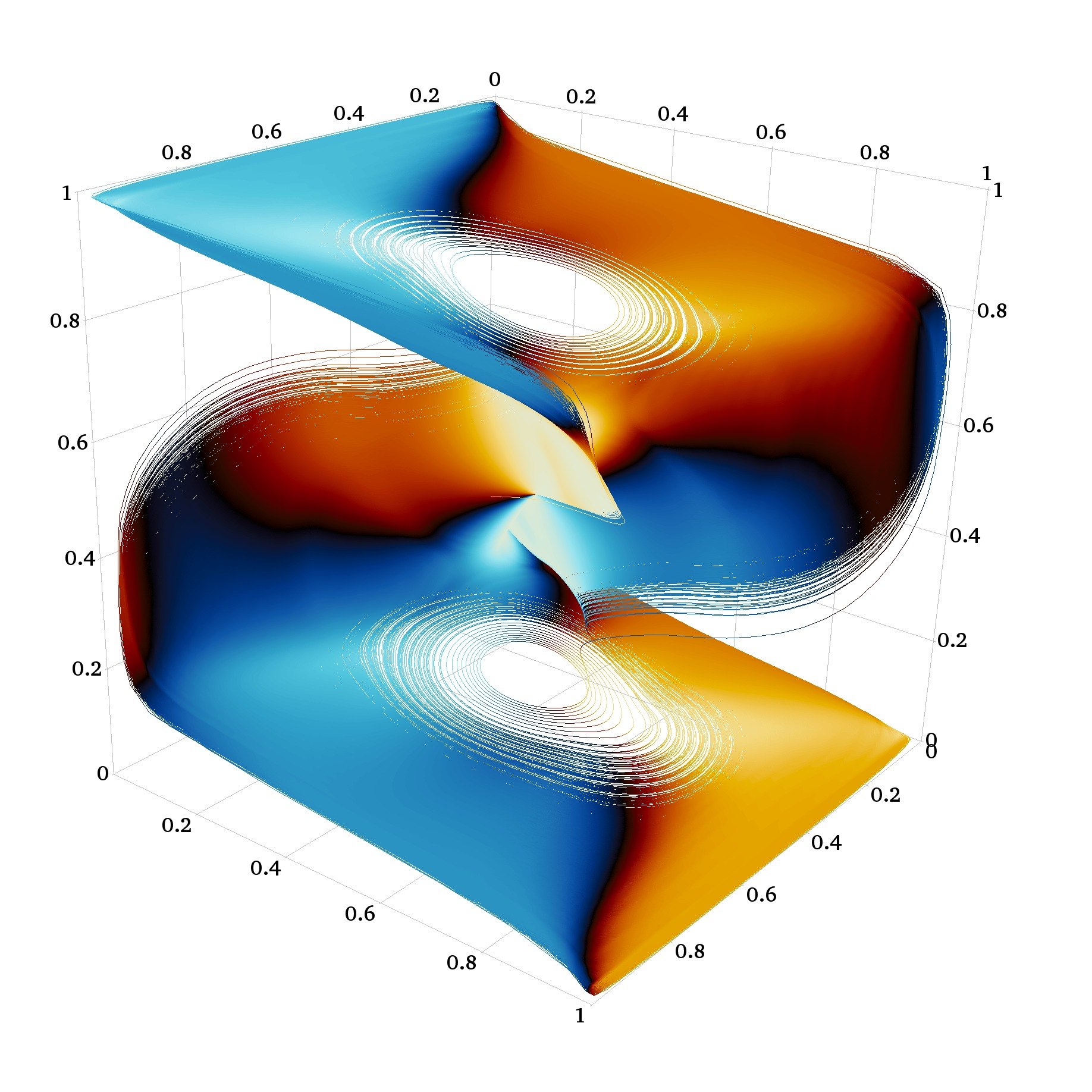}
  \includegraphics[width=0.249\textwidth]{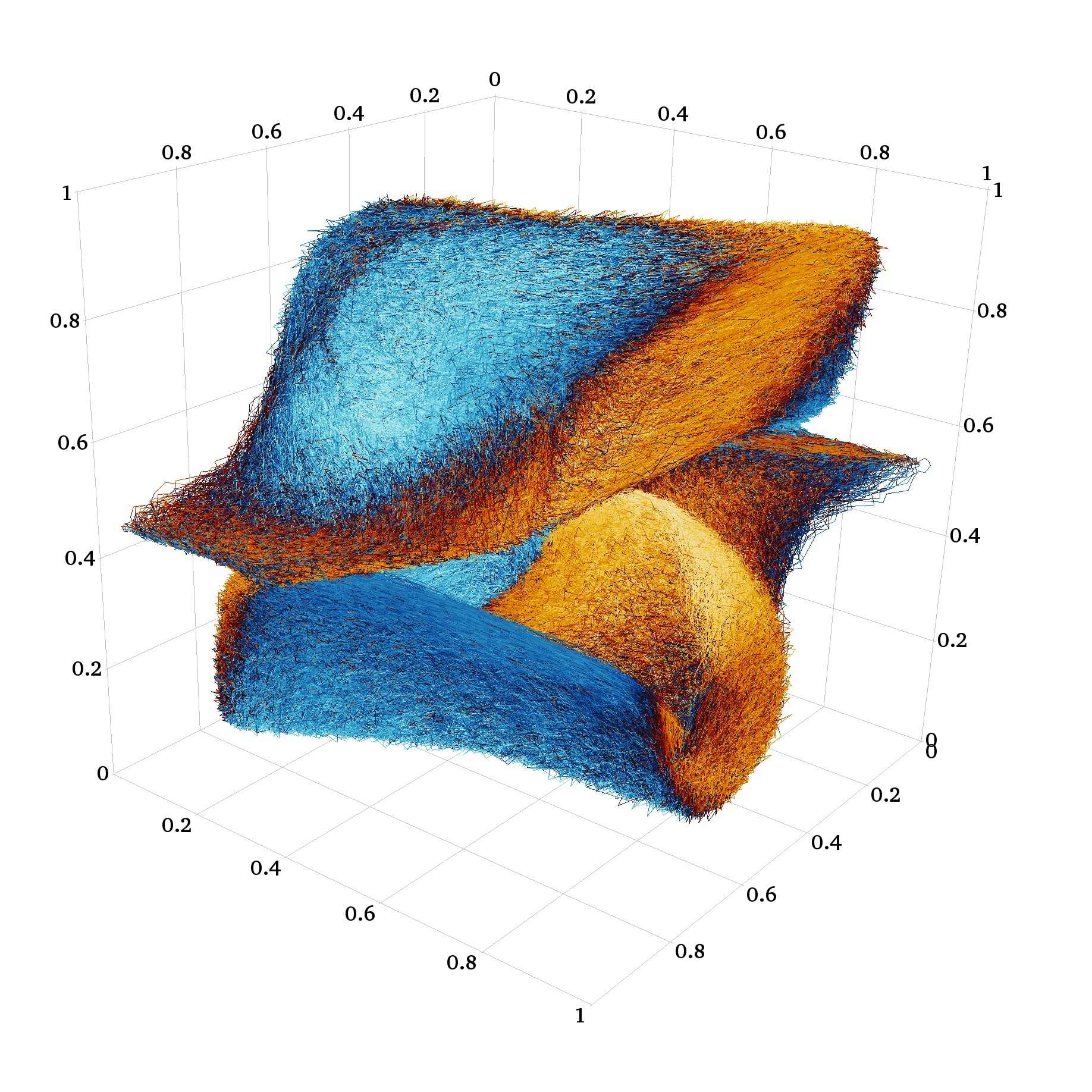}
  \includegraphics[width=0.249\textwidth]{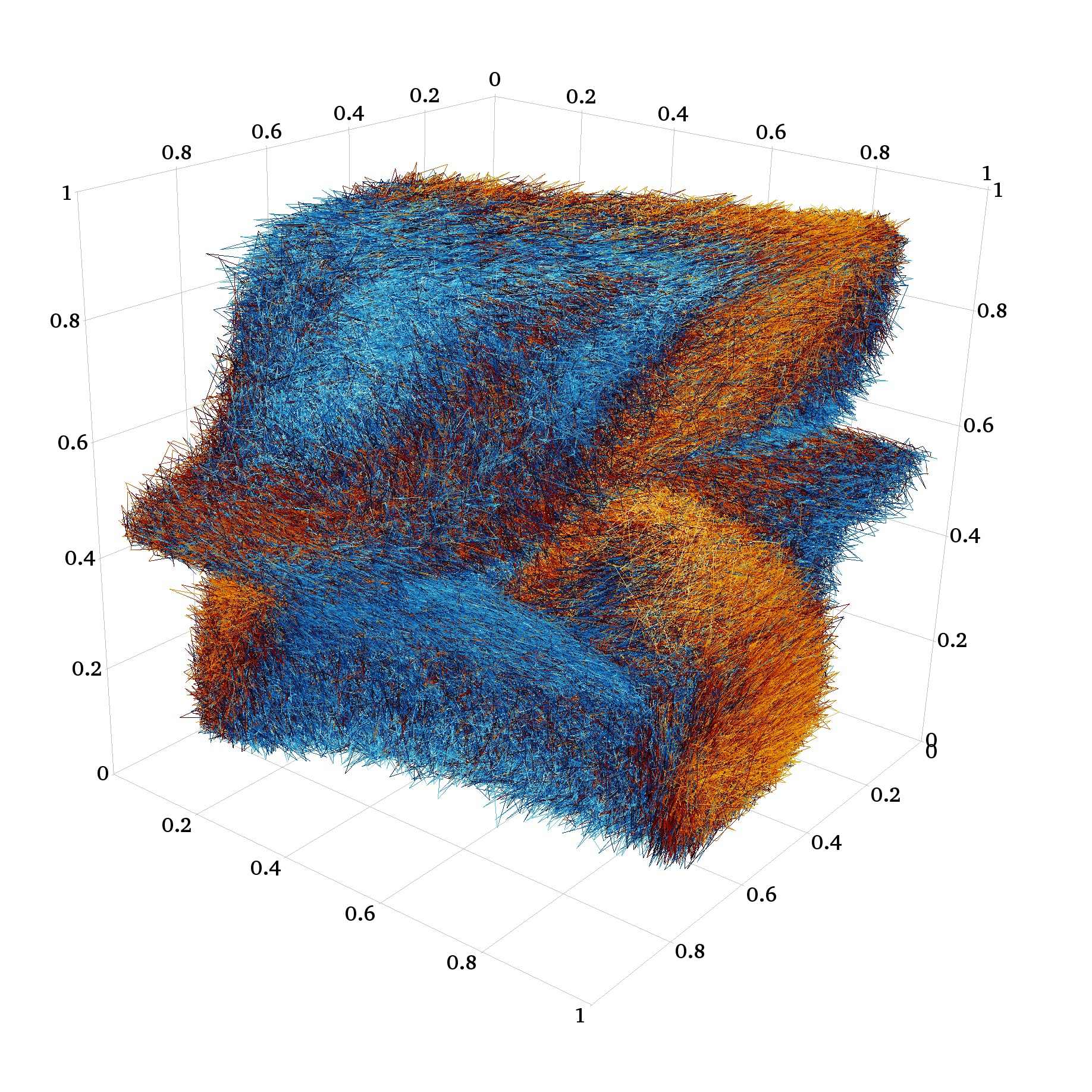}
  \includegraphics[width=0.249\textwidth]{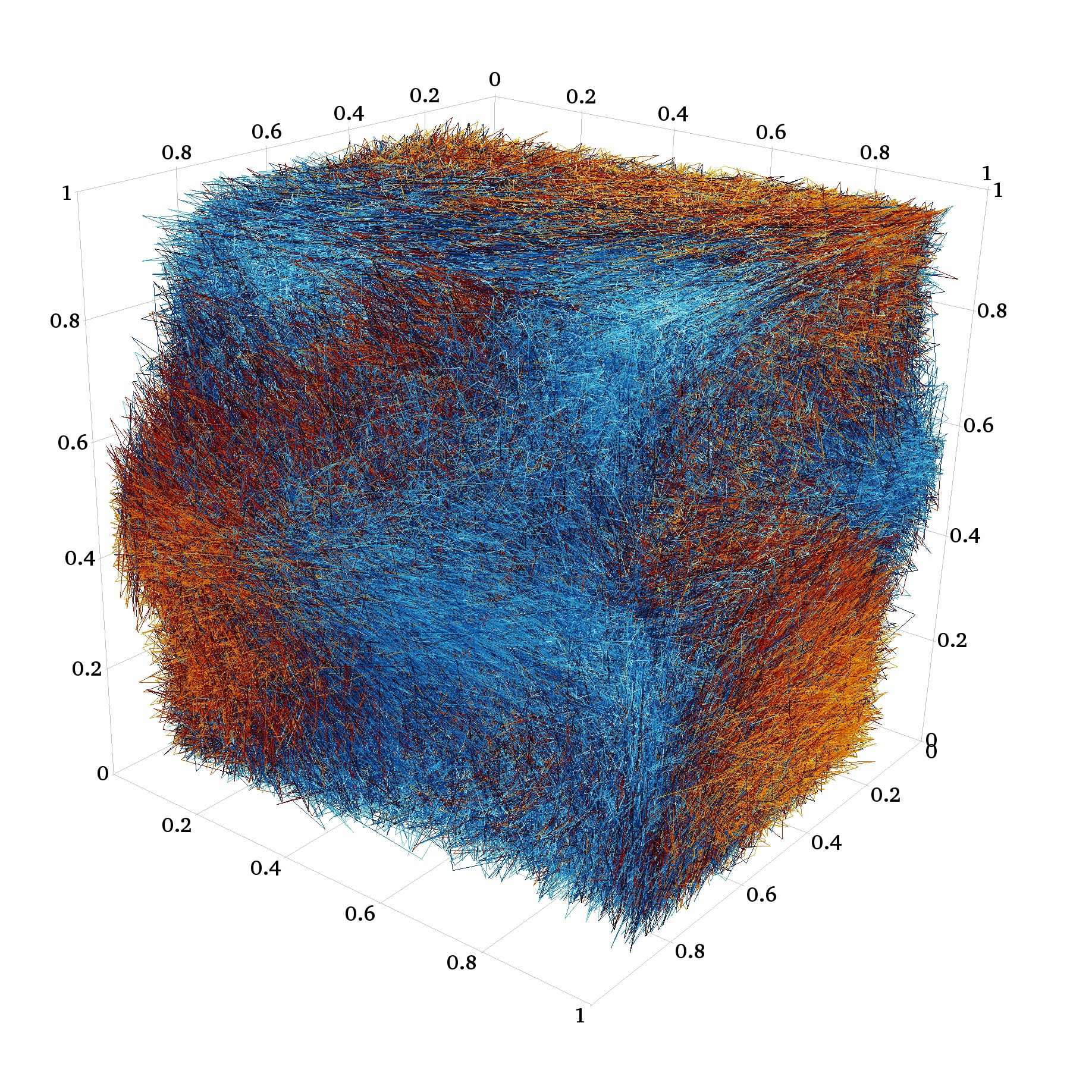}
  \caption{Three dimensional phase trajectory for Lorenz system from $x$ variable lags using equiprobable binning transformed coordinates. Portraits are arranged by signal to noise level with noiseless on the left and $5$dB SNR on the right.  The standard lag view of first local minima by variable are shown in the top row.  The bottom row shows the first global maxima of mutual information in the orthogonalized view. Segment are colored based off the angle between the segment and vector $\{1,2,\sqrt{3}\}$ to visually cue smoothness of attractor.}
  \label{Fig:ShadowsOf3D}
\end{figure}

Figure \ref{Fig:ShadowsOf3D} compares the resulting 3D manifold shadows in the stretched coordinates identified as optimal lags based off the local minima/global maxima criterions for all four noise levels from the 3D curves of Figure \ref{Fig:ShadowStraightEquibin3-5DLorenz}.

As can be visualized in the 5dB SNR curves of the 4D and 5D plots of Figure \ref{Fig:ShadowStraightEquibin3-5DLorenz}, the \MI from the traditional time delay view tends to show an inflection point near the orthogonal coordinates' maximum lag value even when the point is not selected as the first local minimum.   This identification of an inflection point rather than a local minimum in the \MI is due to a low or slightly fluctuating decay of redundancy resulting from the shape of the autocorrelation function.  Unfortunately, the logical outcome of this behavior is that the Fraser algorithm cannot reliably identify the point as a local minimum.  

\begin{figure}
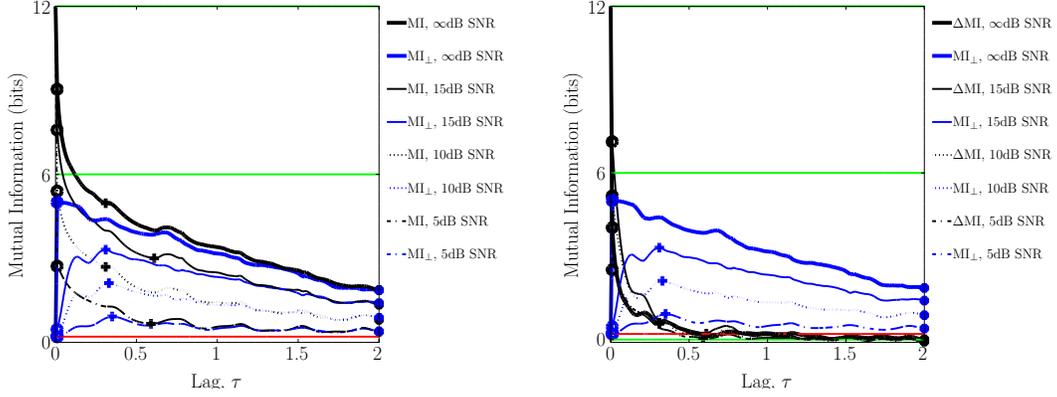

  \adjustbox{scale=0.36,trim=0in 0pt 0in 0pt,clip}{
    \include{ShadowStraightVsEigLorenz6bX-missMinimaGaussianNoise3D}}
  \adjustbox{scale=0.36,trim=0in 0pt 0in 0pt,clip}{
    \include{ShadowStraightVsEig-MissMinimaOnlyLorenz6bXGaussianNoise3D}}
  \caption{Mutual information for 3D lag of Lorenz-$x$ variable shadows on equiprobable bins in standard and orthogonalized coordinate systems. Figure depicts impact of alternative random number seed compared to 3D plot from Figure \ref{Fig:ShadowStraightEquibin3-5DLorenz} for both the standard lagged and orthogonalized mutual information dependence curves (left) and the quantity of removed mutual information, $\Delta$\MI=\MI-\MI$_\perp$, extracted from the natural lag coordinates via orthogonalization (right) for the same data.}
  \label{Fig:ShadowStraightEquibin3-missMinima}
\end{figure}

To demonstrate the issue more dramatically, the 3D curves from Figure \ref{Fig:ShadowStraightEquibin3-5DLorenz} can be rerun with different noise seeds which results in inconsistently identified local minima as shown in Figure \ref{Fig:ShadowStraightEquibin3-missMinima}. In this figure, minima are identified at approximately twice the lag for both the 5dB and 15dB SNR curves. This highlights the sensitivity of the \MI evaluation in identifying local minima.

\begin{figure} 
  \includegraphics[width=0.249\textwidth]{png/Attractor3DEquibin_Lorenz6bXGaussianNoise3D1_00032.jpg}
  \includegraphics[width=0.249\textwidth]{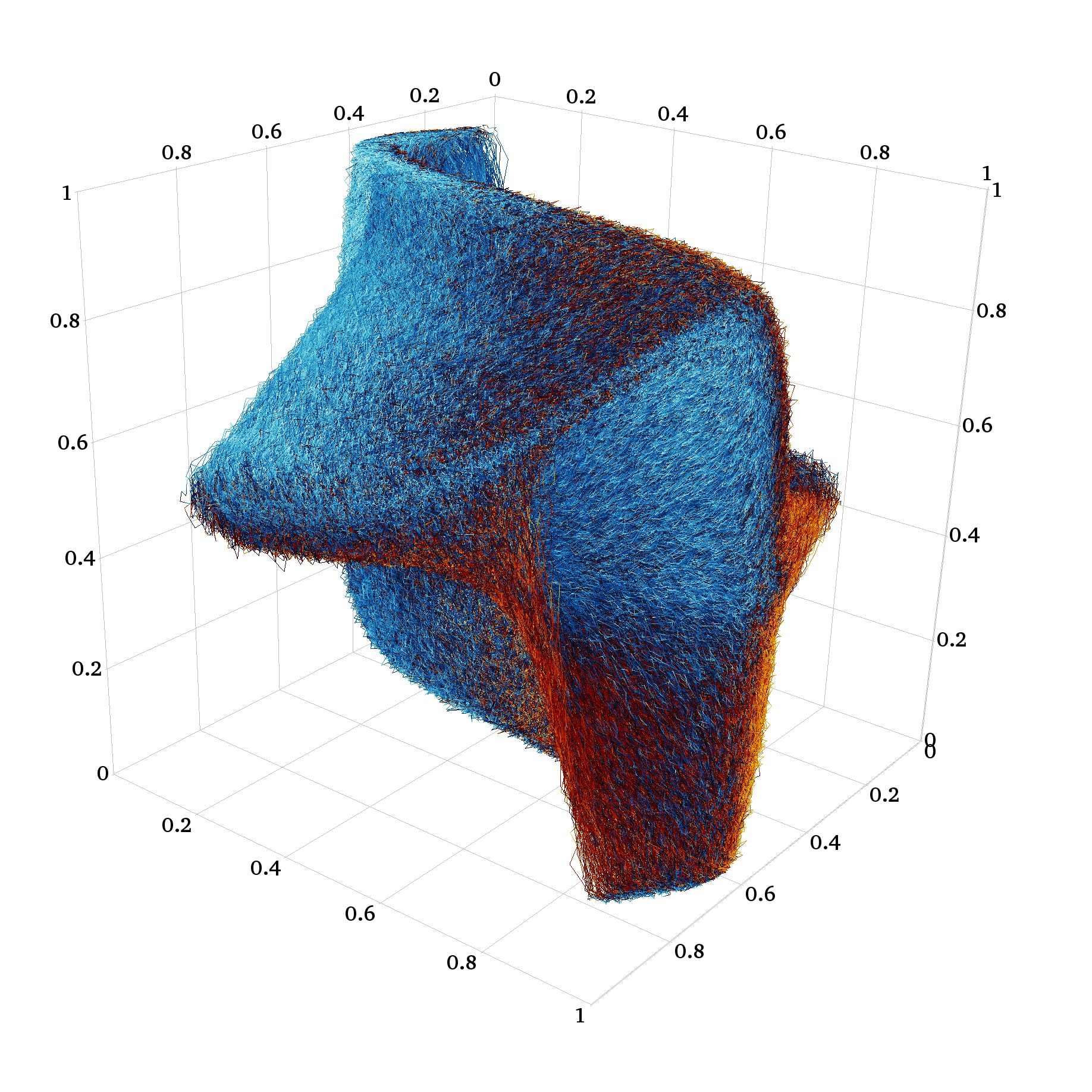}
  \caption{Comparison of nearly identical three dimensional phase trajectories for Lorenz system from $x$ variable lags using equiprobable binning transformed coordinates. Both portraits are the 15dB SNR level with different random seed for the noise. The left figure depicts the case where the local minimum is identified and the right depicts the same time lag when the minimum is not identified.  }
  \label{Fig:CompareMissedMinima}
\end{figure}

Figure \ref{Fig:CompareMissedMinima} depicts the lagged views of the two different random seeds of the 15dB shadows plotted with the lag identified by the first minimum in Figure \ref{Fig:ShadowStraightEquibin3-5DLorenz} rather than later time identified in Figure \ref{Fig:ShadowStraightEquibin3-missMinima}. In this case, the figure on the left is a local minimum of \MI whereas the figure on the right is not simply due to the difference in random seed.  The similarity of the two figures underscores the sensitivity of the first local minimum criteria for identification of useful time lags.

The right half of Figure \ref{Fig:ShadowStraightEquibin3-missMinima} is included to specifically highlight the behavior of the difference between the evaluation of \MI in the time lag and orthogonalized view, $\Delta$MI.  Note that $\Delta$\MI drops quite rapidly with lag and then oscillates around the expected noise floor for longer lags.  The quantity $\Delta$\MI that exceeds the noise floor represents the short time redundancy removed from the estimate by transforming to orthogonalized coordinates.  This redundant information is miss-identified as a quantity of mutual information in the lag coordinates due to the limitations of the numerical quadrature of the evaluation.

Assuming the redundant information, $\Delta$\MI, is a numerical artifact of the information estimation procedure, the quantity \MI$_\perp$ in the orthogonalized view is a closer approximation to the true \MI.  Having identified and estimated this artificial redundant information sheds some additional light on the first local minimum heuristic's emergence as an appropriate lag for smooth chaotic data masked by noise. The quantity of redundant information decreases rapidly with $\tau$ as $\tau$ increases from zero.  This causes the total estimated \MI of the time lag view to decrease from a maximum when the data fall on a diagonal line at $\tau$=0. At the same time, for data with a smooth continuous signal masked by noise, the mutual information increases from approximately $0$ at $\tau$=0 because the magnitude of the noise is approximately constant while $(\dot{x} \tau) \rightarrow 0$ as $\tau\rightarrow 0$.  If the redundant information was purely monotonically decreasing, it is clear that local minima in the time lag estimate of \MI could only occur at points prior to local maxima of the true mutual information when $d(\Delta$\MI$)/d\tau + d($\MI$)/d\tau = 0$.  This would make the first local minima of the time lagged \MI estimate the first point at which the growth of the true \MI out of the noise exceeds the decay rate of the redundant information.  This is often near a significant local maxima of \MI$_\perp$ at a $\tau$ sufficient to have left the rapid initial decay of redundant information.

In the examples in Figure \ref{Fig:ShadowStraightEquibin3-5DLorenz}, the first local minimum often coincides approximately with the global maximum of \MI$_\perp$ even if there is an earlier local maximum of \MI$_\perp$ that occurs while the decay of redundant information is still too significant to cause an inflection point.  Though often identifying similar lags, in some cases the growth of \MI$_\perp$ fails to exceed the decay of redundant information prior to the global maximum.  In these scenarios, the identified local maximum slips to some larger $\tau$ near where \MI$_\perp$ has another, lower local maximum and $\Delta$\MI is oscillating near the noise floor of the calculation. Because the decay and growth happen to be well matched in this example near the global maximum of \MI$_\perp$, this results in a shallow local minimum and strong sensitivity to noise in the identified local minimum as highlighted in Figure \ref{Fig:ShadowStraightEquibin3-missMinima}.  The global maximum in \MI$_\perp$ is much more robustly identified, and simultaneously the likely true target of the first local minimum criterion as it represents the point at which the data appears least random, highlighting the nonrandom structure hidden in the data. 

\begin{figure} 
  \includegraphics[width=0.249\textwidth]{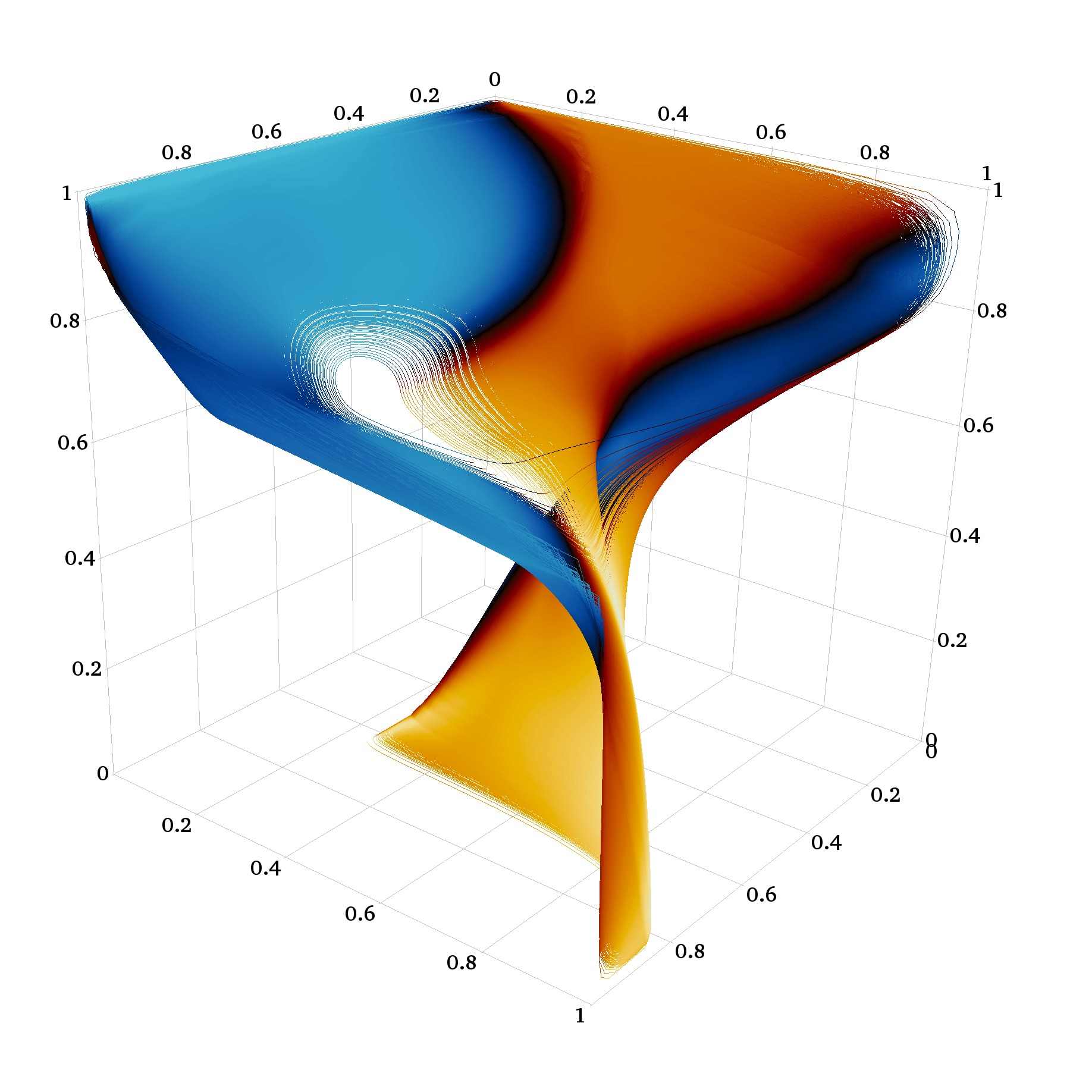}
  \includegraphics[width=0.249\textwidth]{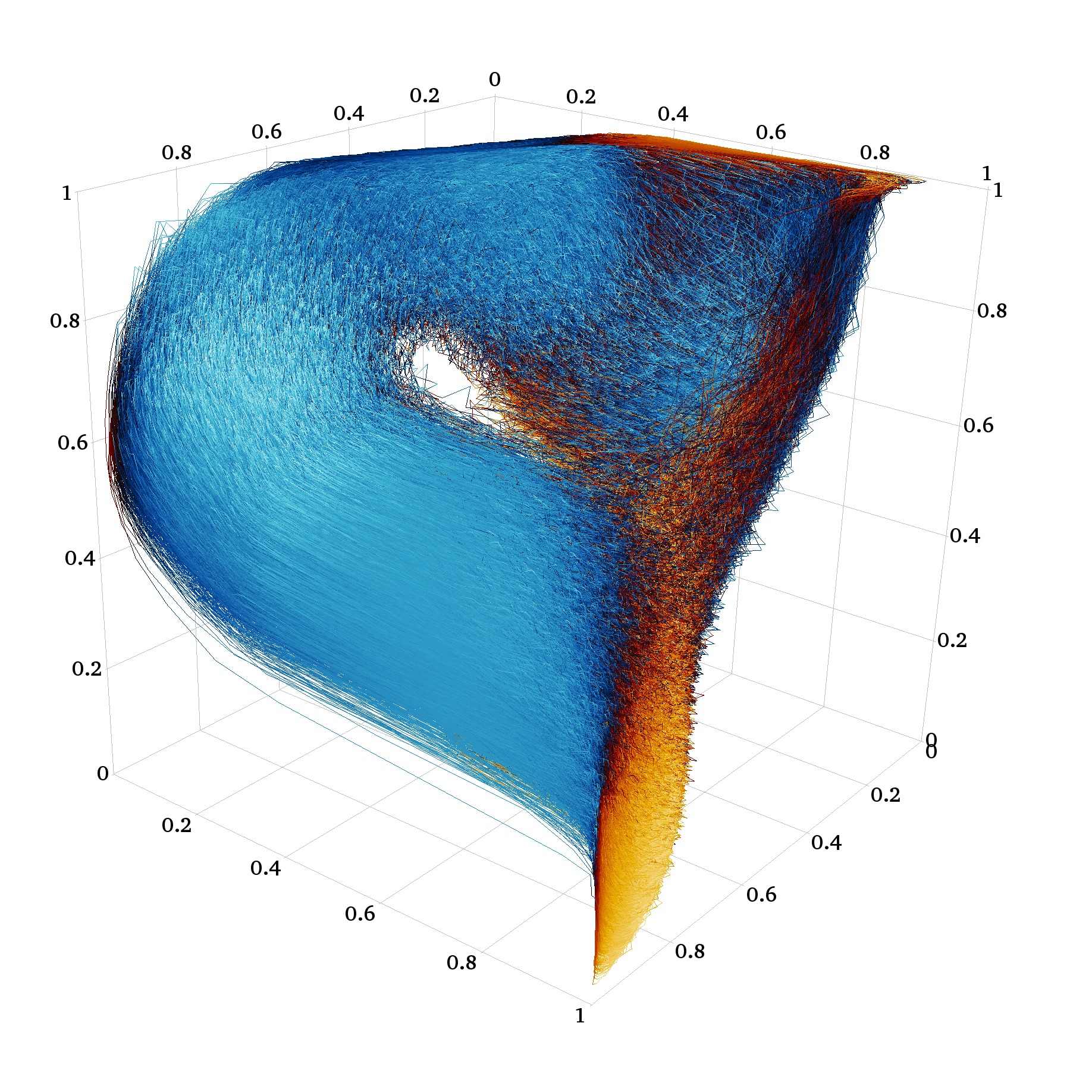}
  \includegraphics[width=0.249\textwidth]{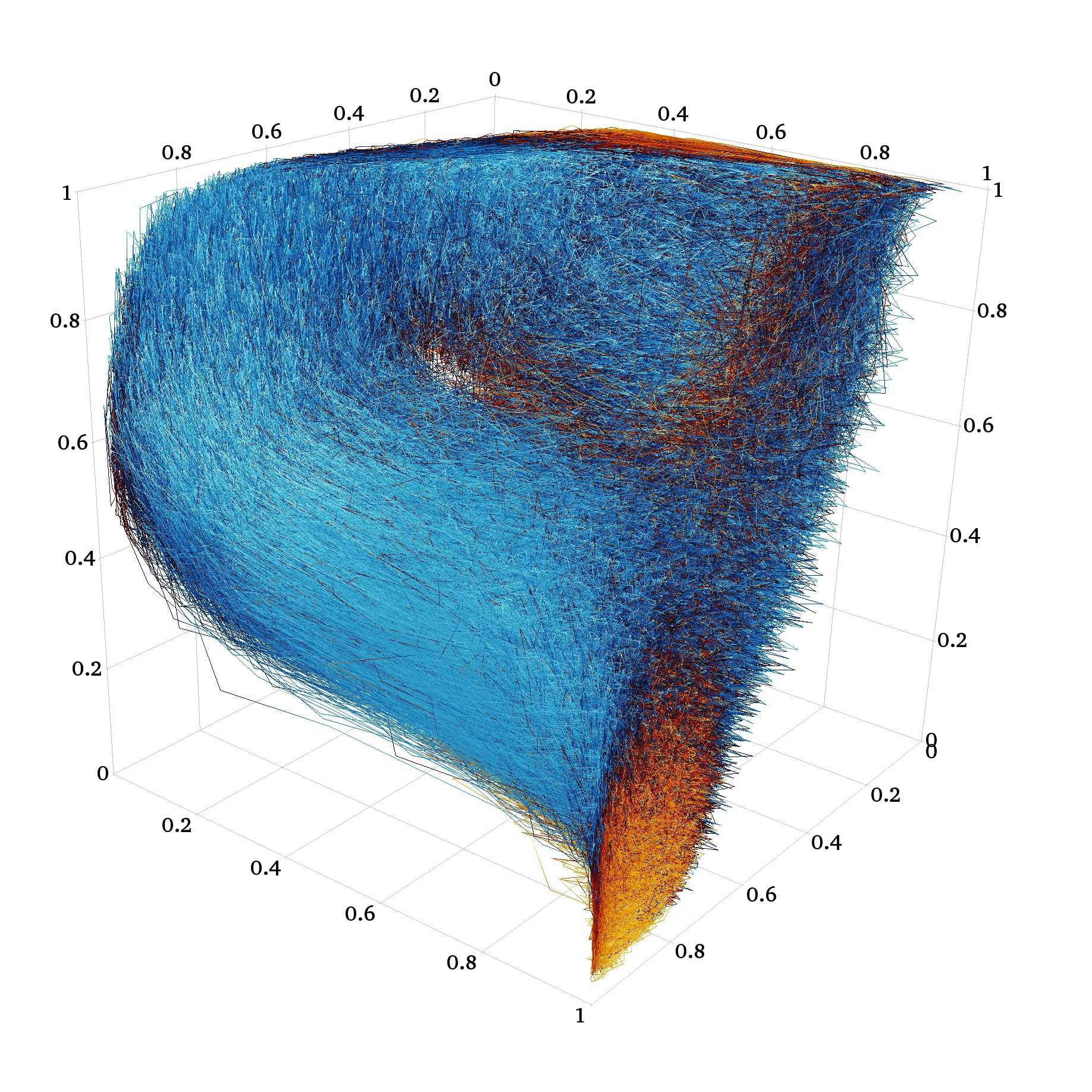}
  \includegraphics[width=0.249\textwidth]{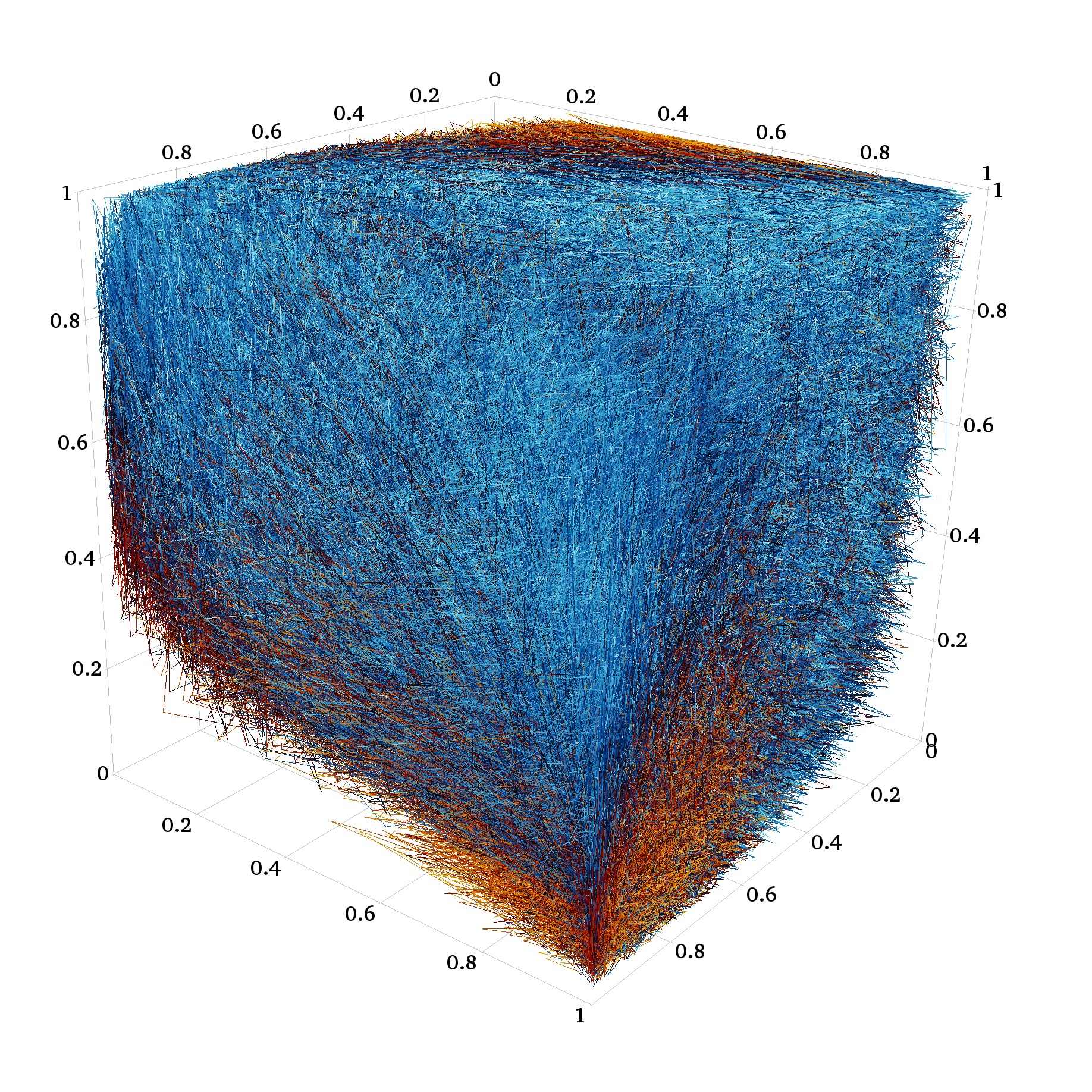}\\ 
  \includegraphics[width=0.249\textwidth]{png/Attractor3DEigEquibin_Lorenz6bXGaussianNoise3D0_00001.jpg}
  \includegraphics[width=0.249\textwidth]{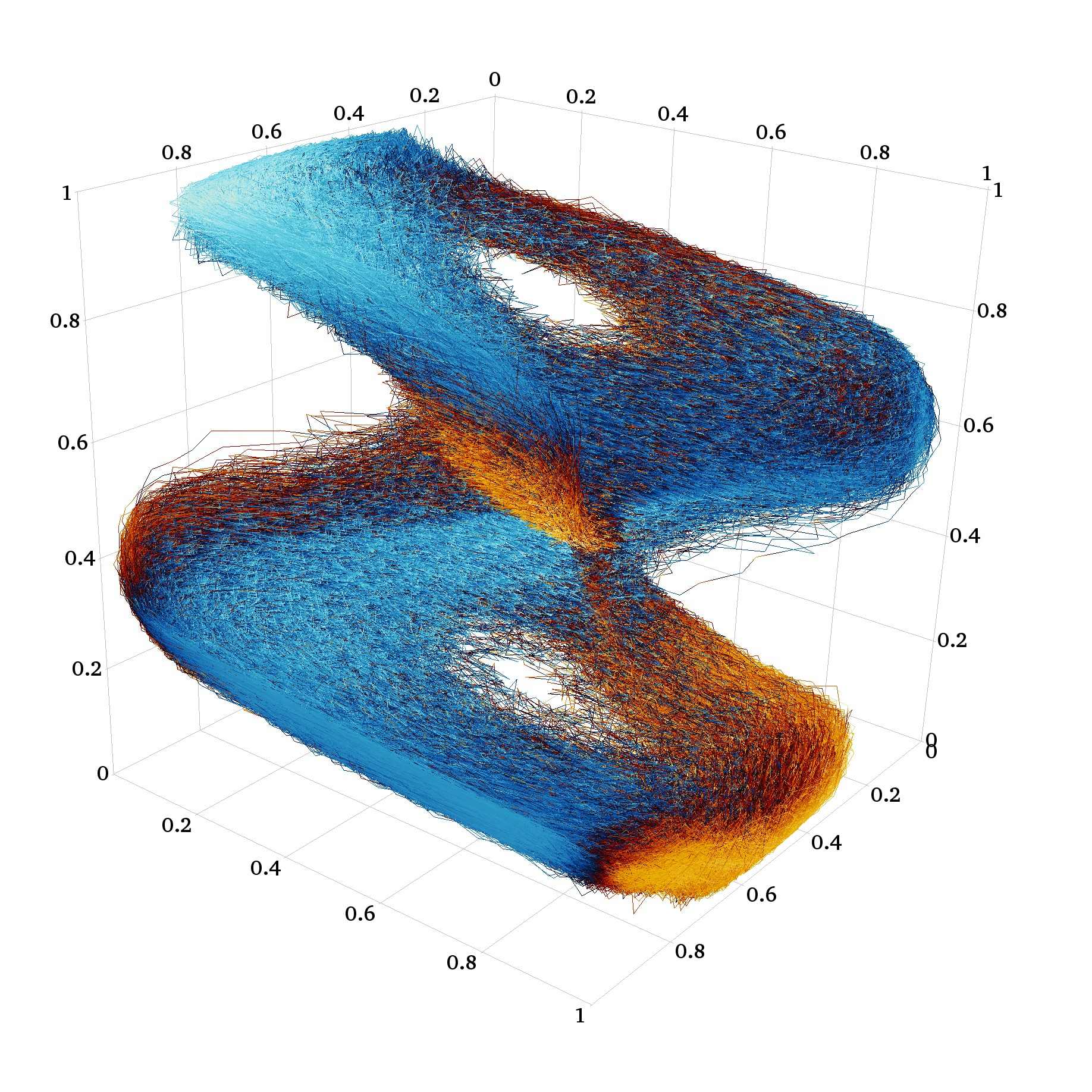}
  \includegraphics[width=0.249\textwidth]{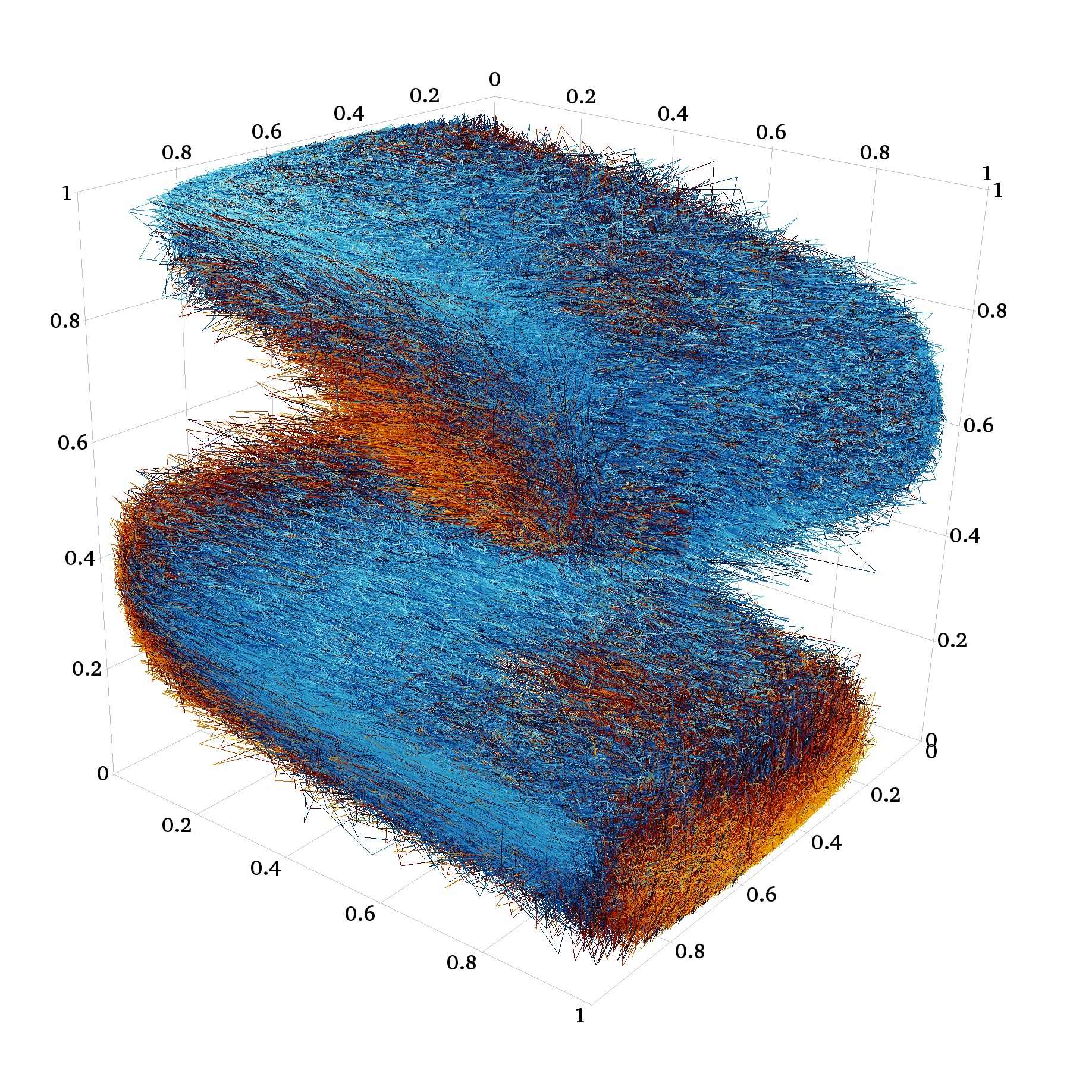}
  \includegraphics[width=0.249\textwidth]{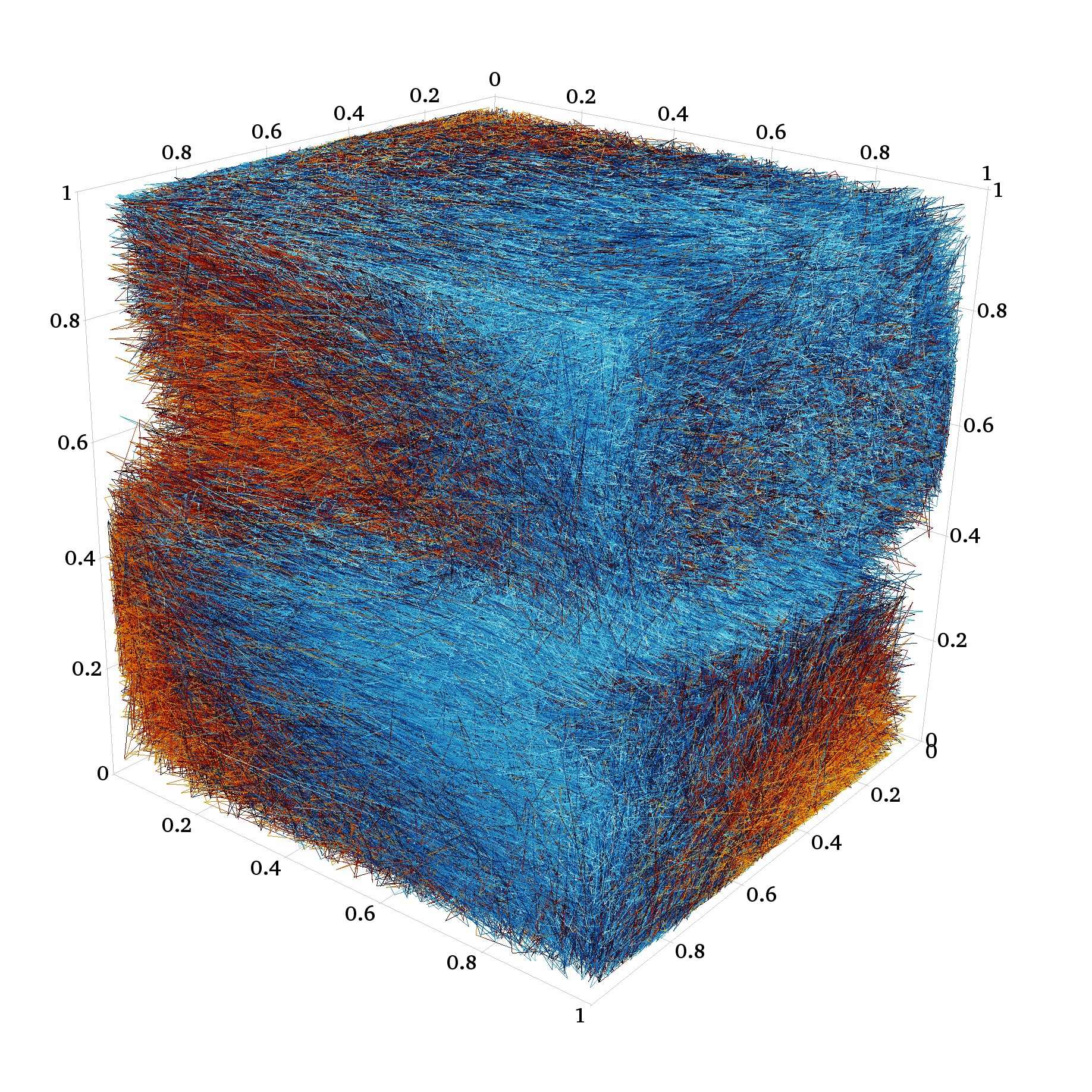}
  \caption{Three dimensional phase trajectory for Lorenz system from $z$ and $x$ variable lags using orthogonalized equiprobable binning transformed coordinates. Portraits are arranged by signal to noise level with noiseless on the left and $5$dB SNR on the right.  The lag of $z$ views by global maxima of mutual information in $z$ are shown in the top row.  The bottom row shows the orthogonalized view of $x$ using the lags from $z$.}
  \label{Fig:ShadowsOf3D-ZX}
\end{figure}

With regards to comparing the first local maximum to the global maximum as a prescription for the choice of lag, it is interesting to note that the $z$ variable of the Lorenz system identifies a lag that is approximately half that of $x$ and $y$.  Figure \ref{Fig:ShadowStraightEquibin3-5DLorenzZ} shows the mutual information curves for Lorenz-$z$ which can be compared to those of Figure \ref{Fig:ShadowStraightEquibin3-5DLorenz}. Notably, the $z$ \MI$_\perp$ curves' first local maxima are also the global maxima rather than being exceeded by the second peak as seen in the $x$ curves. This is the result of a symmetry in the Lorenz system that makes both halves of the butterfly attractor collapse onto each other exactly in the $z$ variable as discussed in more detail in Subsection \ref{SubSec:LorenzImpact}.  Figure \ref{Fig:ShadowsOf3D-ZX} shows the phase portraits based off the Lorenz $z$ global maximum mutual information time lags across noise level for both the $z$ and $x$ data.  It is interesting to note that at this shorter time lag identified for the noisy data, the $x$ shadows more closely resemble the the distinct two hole butterfly shape of the one-step lag orthogonalized portrait from the noiseless data. 

\begin{figure}
  \adjustbox{scale=0.36,trim=0in 0pt 2.2in 0pt,clip}{
    \include{ShadowStraightVsEigLorenz6bZGaussianNoise3D}}%
  \adjustbox{scale=0.36,trim=.8in 0pt 2.2in 0pt,clip}{
    \include{ShadowStraightVsEigLorenz4bZGaussianNoise4D}}%
  \adjustbox{scale=0.36,trim=.8in 0pt 0in 0pt,clip}{
    \include{ShadowStraightVsEigLorenz3bZGaussianNoise5D}}%
  \caption{Mutual information for 3D-5D lags (left to right) of Lorenz-$z$ variable shadows on equiprobable bins in standard and orthogonalized coordinate systems.  Note that the first extrema, minima or maxima, is marked with the ``$+$''-symbol for lagged and orthogonalized view respectively. Noise floor (red) and (d-1) lines representing $log_2(n_{edge})$ divisions (green) of the $n_{edge}^d$ bin hypercube  are included for reference.}
  \label{Fig:ShadowStraightEquibin3-5DLorenzZ}
\end{figure}

Upon initiation of the line of research, the portraits from the identified $z$ lags were actually the more expected result than those of Figure \ref{Fig:ShadowsOf3D-ZX}. It was expected that the noiseless short lag orthogonal coordinate results which immediately emphasizes the classic two hole butterfly attractor topology would carry through to the noisy versions.  This suggests that a first local maximum in \MI$_\perp$ criterion may represent yet another potentially attractive alternative prescription for lag selection.  However, it is important to remember that despite the appearance in the plot, the volume of phase space occupied by the longer lag version is actually a smaller volume of phase space (i.e. has higher mutual information). Though the underlying two dimensional surface structure has folded back towards itself more significantly in the longer lag identified in $x$ and $y$, it has not in fact crossed itself inducing the long time incoherence.  The thickness of the surface resulting from the the noise is smaller in the longer lag version such that the mutual information is larger even though the 2D image projection of the 3D view fails to make that apparent.

\section{Impact of Lag Choice on Cross Mapping Performance}
\label{Sec:Impact}

While the previous two sections focused on relationships between different embedding views in the presence of noise, the primary motivation of this work is to provide a mechanism for automatic selection of an optimal embedding lag to remove one free parameter in the CCM process.  While selection criteria based on \MI provide a window into the coherence of embeddings of individual time series data extracted from nonlinear coupled systems, the ultimate justification for a particular choice of lag depends on its ability to maximize the power of cross mapping correlations detected between different observations of the system.

One way to maximize the cross correlation strength for finite, noisy systems is to simply attempt reconstructions for every pair of variables with every possible time lag and select the maximum. This approach significantly compounds the computational cost of performing CCM calculations, particularly over a large selection of signal pairs.  The extent to which a choice of lag approximates this maximum provides a metric for assessing the quality of the lag choice in this context.  In Section \ref{SubSec:LorenzImpact}, three criteria for selection of the optimal lag are evaluated with respect to the swept lag results when performing CCM on the Lorenz system.  These criteria are the original first local minimum of \MI in an equiprobable time delay embedding from Fraser, the global maximum of \MI in an equiprobable orthogonal embedding, and the first local maximum of \MI in the equiprobable orthogonal embedding.   The ultimate test of these selection criteria is then conducted with application of the CCM method for real, noisy experimental data in Section \ref{SubSec:ImpactExpCCM}.  Finally, to further emphasize the importance of a robust lag selection criterion given finite, noisy experimental data, the cross mapping effectiveness for one particular case study pair of signals from the experimental data set which were found to be particularly sensitive to lag choice are shown in further detail in Section \ref{SubSec:DemoExp}.  

\subsection{Impact on Lorenz CCM}
\label{SubSec:LorenzImpact}
The first test of the impact of selecting a lag from the criteria for the mutual information comes from constructing cross mapping estimates for the ``$x$'' and ``$z$'' variables of Lorenz system.  The data was first split between a model and test set where the first half of the data set was used to generate model three-dimensional state space vectors and the second half was used to test the models. The state space reconstruction for variables ``$x$'' and ``$z$'' proceeded as follows.  For a given lag, training vectors for some fraction of the model data of ``$x$'' and ``$z$'' were constructed via time delay embedding. These vectors are provided to the kd-tree routines of the ANN library \cite{ANNLibrary}.  Reconstructions are then generated using this model data for approximately the first half of the test data (25k points) as follows.  For a given state vector of test data ``$x$'', the time index and distance of the five nearest neighbor point from the $x$-model data is determined.  A reconstructed estimate $\tilde{z}$ of $z$ is then generated using weighted radial basis functions of the $z$-model data using the weights $w_i=e^{-d_i/d_{min}}/\sum_j e^{-d_j/d_{min}}$. The correlation coefficient $\rho_{\tilde{z}|x}$ is then computed as $\rho_{\tilde{z}|x}=\mathtt{z} \cdot \tilde{\mathtt{z}} / \sqrt{ (\mathtt{z}\cdot \mathtt{z})(\tilde{\mathtt{z}}\cdot \tilde{\mathtt{z}})}$ where $\mathtt{z}$ and $\tilde{\mathtt{z}}$ are the original test $z$ series and the reconstruction of $z$ from $x$ with the mean of the vector subtracted such that $\mathtt{z}=z-\bar{z}$.  Similarly,  $\rho_{\tilde{x}|z}$ is the correlation of the original test $x$ with the reconstruction of $x$ from $z$.  These correlations can naturally range between $-1$ and $1$.  The correlation coefficient is the cosine of the angle between the high dimensional unit vectors where $1$ represents a perfect correlation, $0$ represents a complete failure of the reconstruction, and a hypothetical $-1$ would represent perfectly anti-correlated test and reconstruction vectors which only occurs if the training data trends are exactly the opposite sign of the test data. This can occur with very short training data on a nonlinear causal system such as encountered by the ``mirage'' correlations depicted in the coupled two-species nonlinear logistic difference system used to motivate the development of CCM \cite{Sugihara1227079}.  Note that negative correlation coefficients are likely only to occur for short model and test data. An example encountered in this work occurred with the Lorenz $x$ and $z$ data where the sign of correlation between $x$ and $z$ flipped between the training data and test data. The training data was short enough to only sample one lobe of a Lorenz attractor while the test data only came from the other lobe.  In practice on datasets sufficiently long to capture the full extent of the attractor if even sparsely, only slight negative correlation are observed due to an imbalance in the fraction of time spent in different lobes. The correlation oscillates with magnitude decreasing towards zero with additional data. 

\begin{figure}
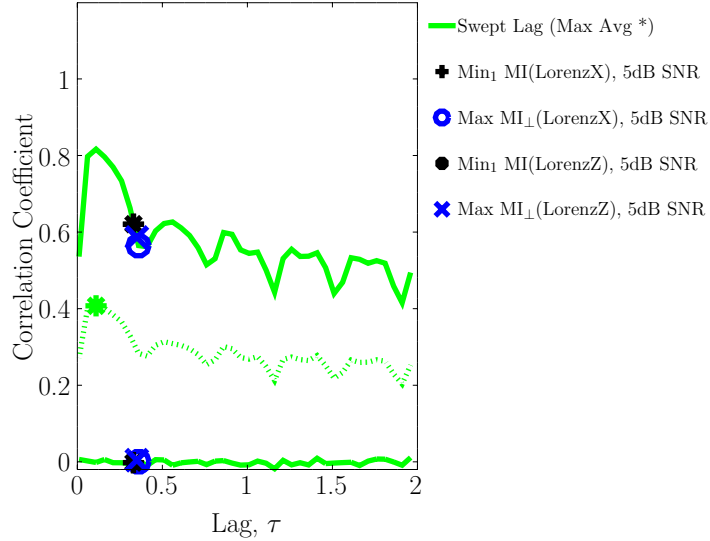

\scalebox{0.5}{\include{CCM_CorrelationVsLag_LorenzX-LorenzZ5dB}}
  \caption{Cross mapping correlation as a function of swept time lag for Lorenz ``$x$'' and ``$z$'' data set with 5dB SNR in stretched version of both lag and orthogonalized coordinates. The upper curves correspond to $\rho_{\tilde{z}|x}$ and lower curve $\rho_{\tilde{x}|z}$.  The dashed line is the average bidirectional correlation with the maximum value marked.  The first local minimum and global maxima points are marked with symbols for comparison with respect to the swept curves.}
  \label{Fig:CCM_CorrelationVsLag_LorenzX-LorenzZ5dB}
\end{figure}

Figure \ref{Fig:CCM_CorrelationVsLag_LorenzX-LorenzZ5dB} depicts how the cross correlation coefficients vary as a function of time lag for the Lorenz ``$x$'' and ``$z$'' pair case with a signal to noise ratio of only 5dB.  Note that for this particular combination, though lags of $x$ generate good reconstruction of $z$, the inverse is not true. This is not surprising considering the significant topological difference between the $x$ and $z$ phase portraits shown in Figure \ref{Fig:ShadowsOfStraightLorenzEquibin}.  This topological change, i.e. two holes to one, is not an artifact of the low dimensional phase portrait.  Once the system dynamics have been projected onto the $z$ axis, the 3D  manifold of the true dynamics has has suffered a noninvertible 2:1 projection.  Subsequent manifold points $(x(t_i),y(t_i),z(t_i))$ and $(x(t_i+\tau),y(t_i+\tau),z(t_i+\tau))$ map to shadow point $(z(t_i),z(t_i+\tau))$, but the mirrored pair of points $(-x(t_i),-y(t_i),z(t_i))$ and $(-x(t_i+\tau),-y(t_i+\tau),z(t_i+\tau))$ also map to shadow point $(z(t_i),z(t_i+\tau))$.  Regardless of the number of lags taken, this 2:1 mapping cannot be undone.  This means that, while the near neighbors of $x$ or $y$ lagged points can reconstruct the $z$ data, the $z$ lags cannot be used to reconstruct $x$ or $y$.  The nearest neighbor points in $z$-lag shadows result from two potential manifold points of opposite sign.  If the reconstruction uses only the nearest point, the reconstruction jumps erratically depending on which piece of phase trajectory is closest.  If radial basis functions are used for the reconstruction, the reconstruction is a random weighted average of the opposite sign points.   

When different mutual information curves indicate different choices of best lag, the shorter of the two lags is used in this work.  The idea behind this choice is that, though there may be excessive redundancy in the variable with longer indicated lag using the shorter lag, this is superior to ambiguity and irrelevance resulting from the shorter lag variable folding back upon itself due to the use of a longer lag. In practice, it was observed that the shorter of the two lags more frequently approximated the swept optimal value. For systems with a single dominant time scale, one would hope that the various views would indicate the same time lag. For observations involving mixtures of independent subsystems operating at different timescales, this may be a more questionable assumption. Exploration of the impact of multiple scales on these choices will have to be addressed in future work.

\begin{figure}
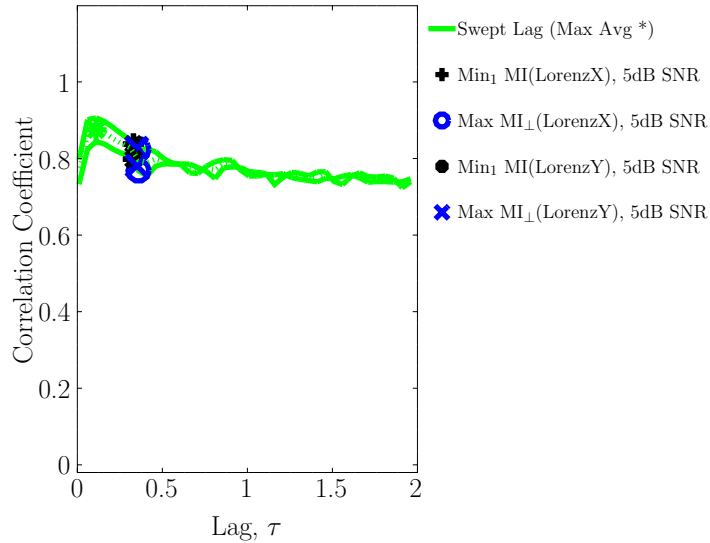

\scalebox{0.5}{\include{CCM_CorrelationVsLag_LorenzX-LorenzY5dB}}
  \caption{Cross mapping correlation as a function of swept time lag for Lorenz ``$x$'' and ``$y$'' data set with 5dB SNR in stretched version of both lag and orthogonalized coordinates. The upper curves correspond to $\rho_{\tilde{y}|x}$ and lower curve $\rho_{\tilde{x}|y}$.  The dashed line is the average bidirectional correlation with the maximum value marked.  The first local minima and global maxima points are marked with symbols for comparison with respect to the swept curves.}
  \label{Fig:CCM_CorrelationVsLag_LorenzX-LorenzY5dB}
\end{figure}

In the case of the Lorenz $x$-$z$ pair, this means that the shorter lag $z$ dynamics time scale would be the marginally preferred choice for the application of CCM for the pair as indicated by the 'x' markers in Figure \ref{Fig:CCM_CorrelationVsLag_LorenzX-LorenzZ5dB}.  Exploring this further, the same plot for the Lorenz $x$-$y$ pair is shown in Figure \ref{Fig:CCM_CorrelationVsLag_LorenzX-LorenzY5dB}.

\begin{figure}
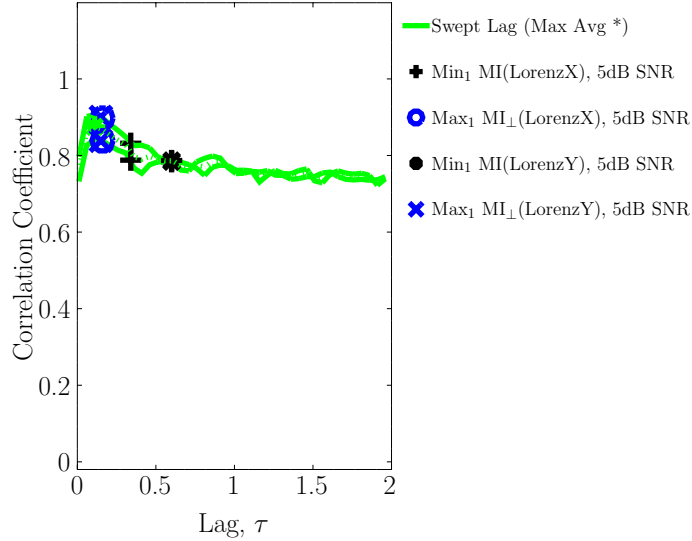

\scalebox{0.5}{\include{CCMFirstMax_CorrelationVsLag_LorenzX-LorenzY5dB}}
  \caption{Cross mapping correlation as a function of swept time lag for Lorenz ``$x$'' and ``$y$'' data set with 5dB SNR in stretched version of both lag and orthogonalized coordinates. The upper curves correspond to $\rho_{\tilde{y}|x}$ and lower curve $\rho_{\tilde{x}|y}$.  The dashed line is the average bidirectional correlation with the maximum value marked.  The first local minimum and first local maxima points are marked with symbols for comparison with respect to the swept curves.}
  \label{Fig:CCMFirstMax_CorrelationVsLag_LorenzX-LorenzY5dB}
\end{figure}

This figure indicates that the choice of shorter lag from the Lorenz $x$-$z$ data may have been a serendipitous choice in that the $z$ lag happened to be a good choice of lag for high CCM correlation. If instead the first local maximum in the orthogonalized coordinates is chosen, the resulting lag choice is much closer to corresponding to the peak of the CCM correlation coefficients as seen in Figure \ref{Fig:CCMFirstMax_CorrelationVsLag_LorenzX-LorenzY5dB}.

This lends credence to the idea that the first local maximum in the orthogonalized coordinates may in fact be a superior choice to either the global maximum in those coordinates or the first local minimum in the lag coordinates.  Figure \ref{Fig:CCMFirstMax_CorrelationVsLag_LorenzX-LorenzZ5dB} revisits the $x$-$z$ pair using the first local maximum criteria. As in Figure \ref{Fig:CCM_CorrelationVsLag_LorenzX-LorenzZ5dB}, the lag set by the $z$ data is a good choice for high CCM correlation at least in the non-ambiguous direction.  More interesting is how consistent the first local maximum criteria is between $x$ and $z$.  This consistency is again an indication that the first local maximum may be a superior criteria for CCM.

\begin{figure}
\scalebox{0.5}{\include{CCMFirstMax_CorrelationVsLag_LorenzX-LorenzZ5dB}}
  \caption{Cross mapping correlation as a function of swept time lag for Lorenz ``$x$'' and ``$z$'' data set with 5dB SNR in stretched version of both lag and orthogonalized coordinates. The upper curves correspond to $\rho_{\tilde{z}|x}$ and lower curve $\rho_{\tilde{x}|z}$.  The dashed line is the average bidirectional correlation with the maximum value marked.  The first local minimum and first local maxima points are marked with symbols for comparison with respect to the swept curves.}
  \label{Fig:CCMFirstMax_CorrelationVsLag_LorenzX-LorenzZ5dB}
\end{figure}
  
\subsection{Impact on Experimental CCM}
\label{SubSec:ImpactExpCCM}
To assess the real world impact of the different choices of embedding lags, a collection of data from an array of high frequency observations of a Hall effect thruster operating within an isolated instrumented confinement cage was used.  A portion of this data was previously used to test the performance of shadow manifold based interpolation compared to prior state-of-the-art FFT-based cross mapping in Reference \cite{eckhardtspatio}.  In that work, brute force sweeping was used to select a lag that provided a good cross mapping performance.  As part of the Air Force Research Laboratory's ``Electric Propulsion Test \& Evaluation Methodologies for Plasma in the Environments of Space and Testing'' (EP TEMPEST) program \cite{EP-TEMPEST}, the use of CCM to detect causation within this coupled nonlinear system is the focus of a concurrent investigation as described in Reference \cite{Eptempest2018}. Additional details of the physical experimental setup are left to description in Reference \cite{Eptempest2018}.  For the purposes of this work, the cross-mapping of a set of ten fundamental current measurements as well as two derived current measurements which were all sampled at 25 MHz for 5ms were used. A time lag, $\tau$, was swept between 40ns-40$\mu$s for a four-dimensional embedding.  It was observed that the thruster dynamics were dominated by an approximately one-dimensional ``breathing-mode'' limit cycle for these particular operating conditions suggesting that a three-dimensional embedding should be sufficient to satisfy Takens' theorem.  However, results were observed consistent between three- and four-dimensions as expected for an embedding of a dimensionality sufficient that the number of false near neighbors has been minimized.  Note, however, that the concept of nearest neighbor is somewhat obfuscated in the presence of noise as noisy representations of unique causal states may overlap in phase space\footnote{In Shannon's framework for information theory, the vast majority of possible symbol volume lays on the surface of a hypersphere such that the number of symbols that can be encoded on the surface without error depends on the symbol separation distance and signal to noise ratio.  This controls the channel bandwidth.  Similarly, states on the attractor manifold of causal dynamics overlap with finite noise. Ambiguity of near neighbors with noise can form an upper limit on the fidelity of a reconstruction. Exploration of this topic is left to future investigation.}.

To compare the impact of lag choice, cross mapping correlation was calculated for the 11x11 array of signal using an adaptation of the method described in the supplemental material of Reference \cite{Sugihara1227079}.  Of the approximately 125k samples of each data series, half was used for training and half for testing the model. The amount of data used for model construction was swept up to the maximum available to ensure that the correlation coefficients were asymptotically approaching constant values with increased data.

\begin{figure}
  \includegraphics[height=1.5in]{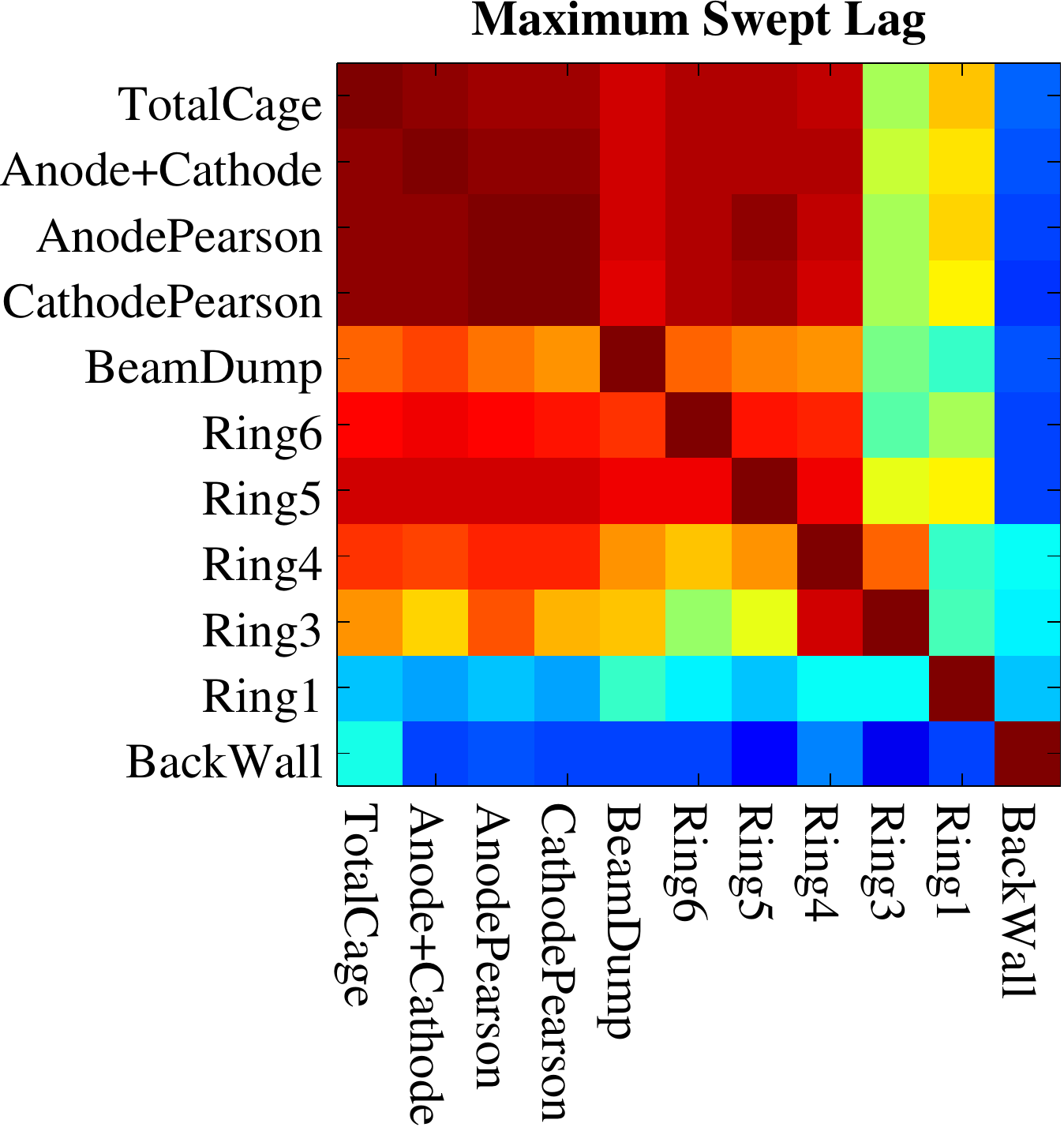}\\ 
  \includegraphics[height=1.5in]{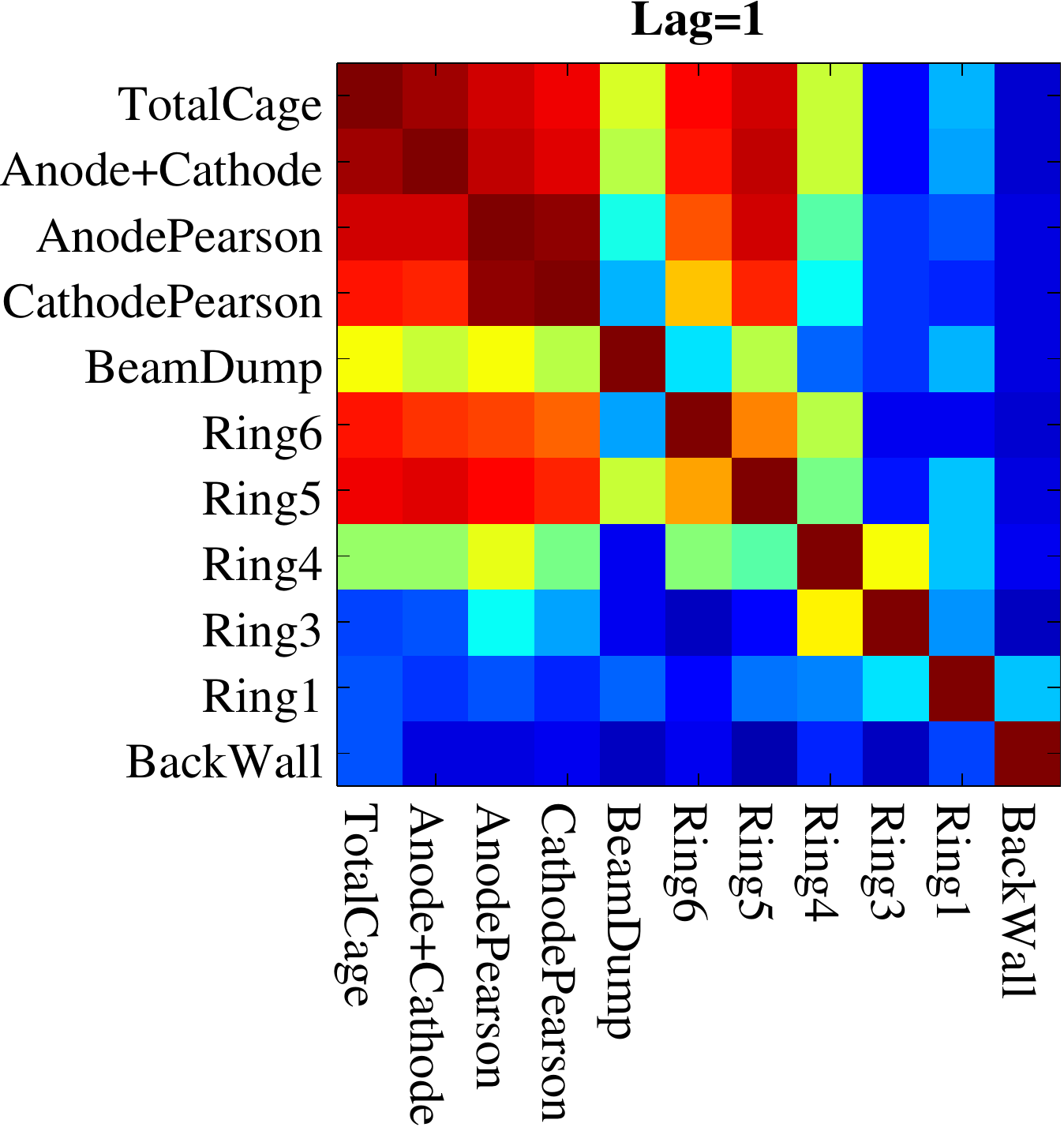}\;%
  \includegraphics[height=1.5in]{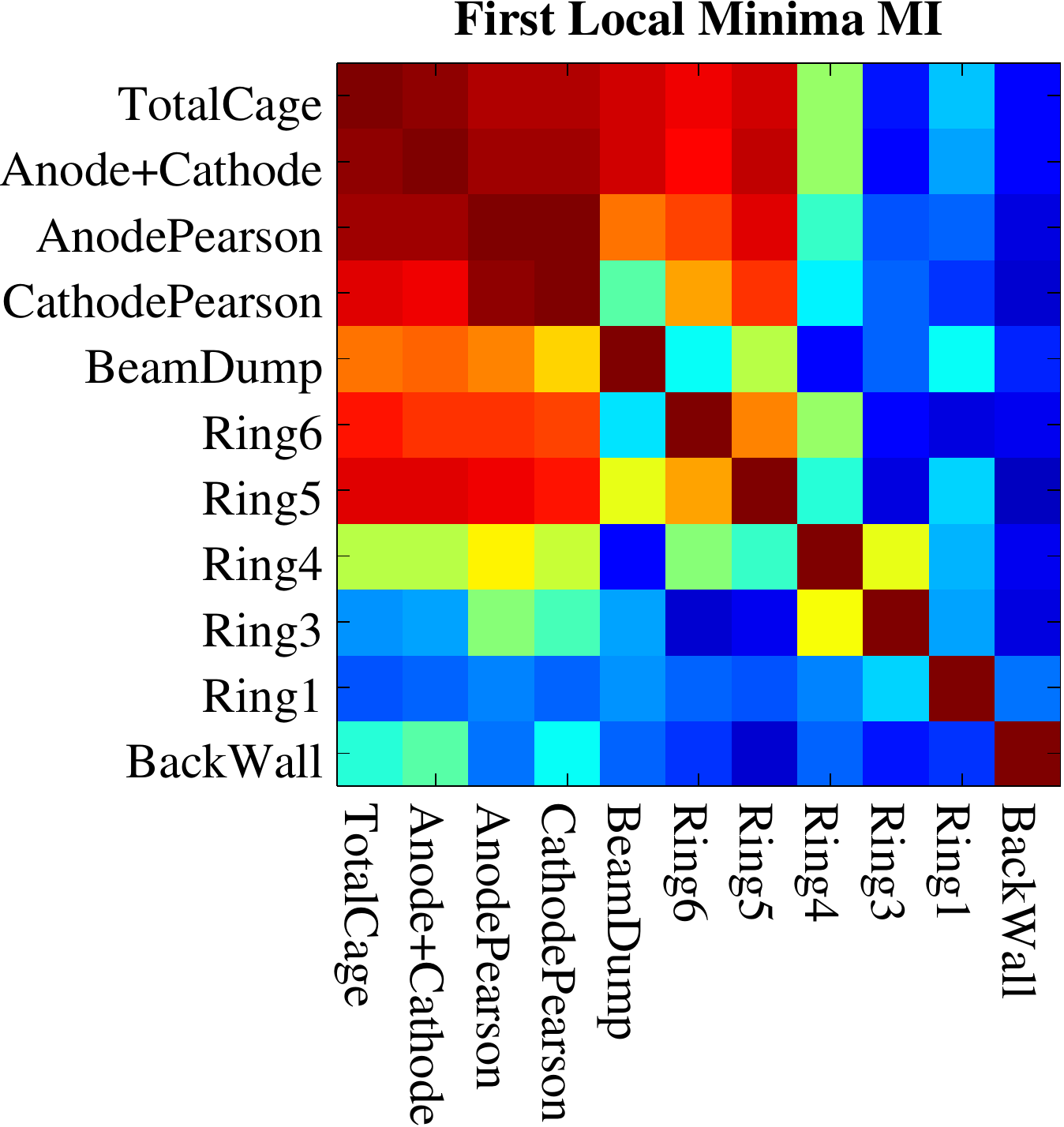}\;%
  \includegraphics[height=1.5in]{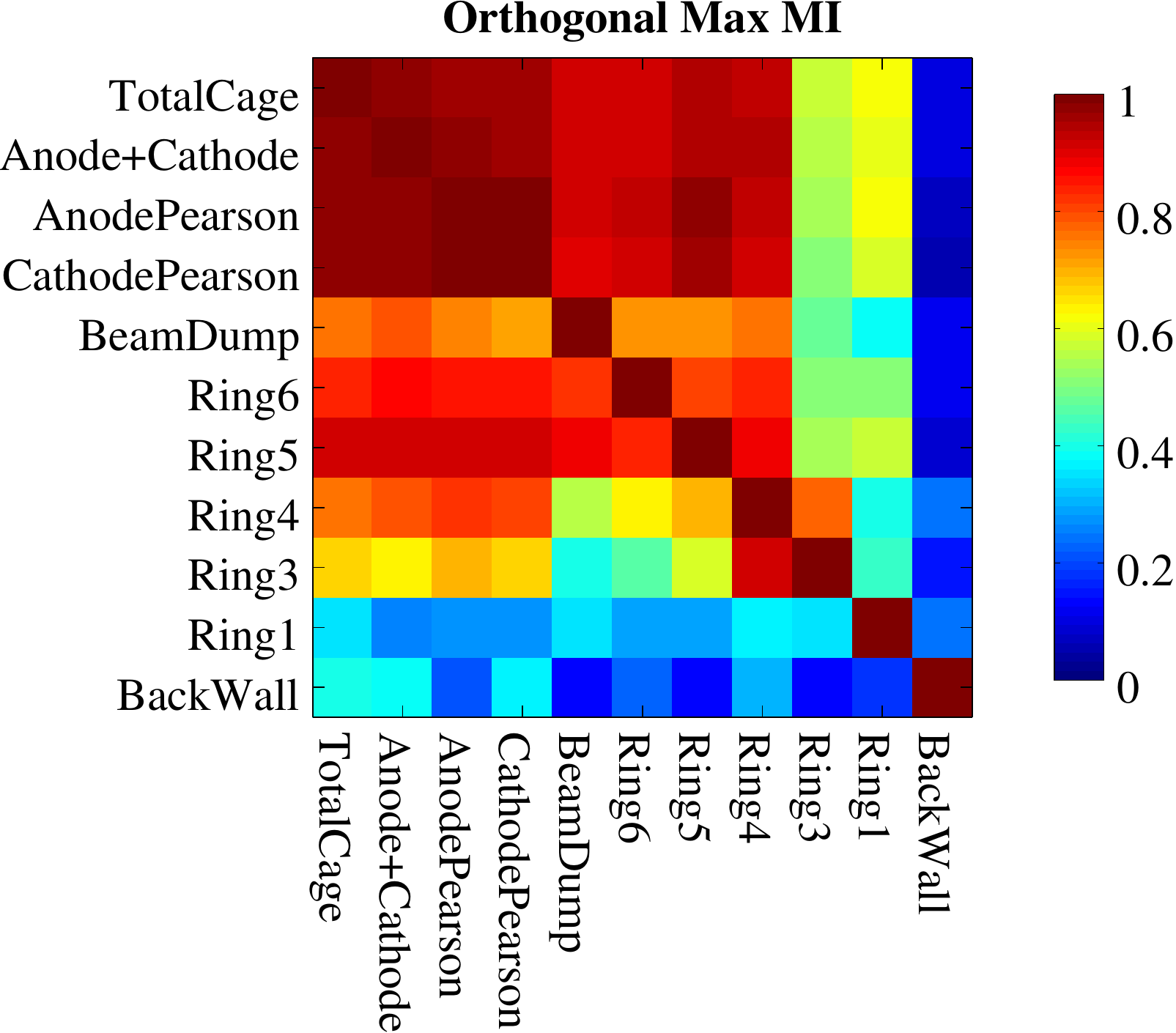}\;%
  \includegraphics[height=1.5in]{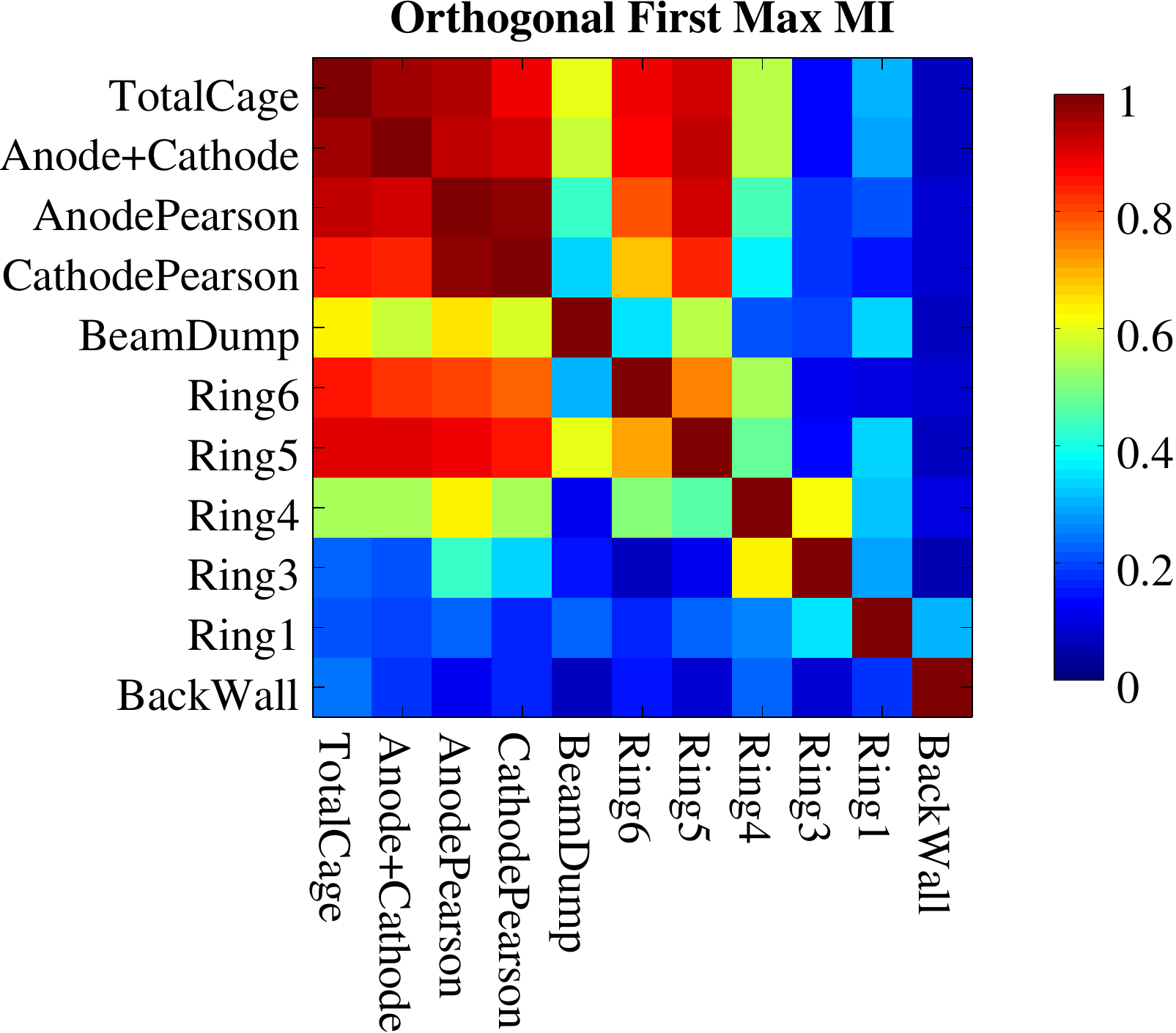}\\ 
  \hspace{1.37in}
  \includegraphics[height=1.5in]{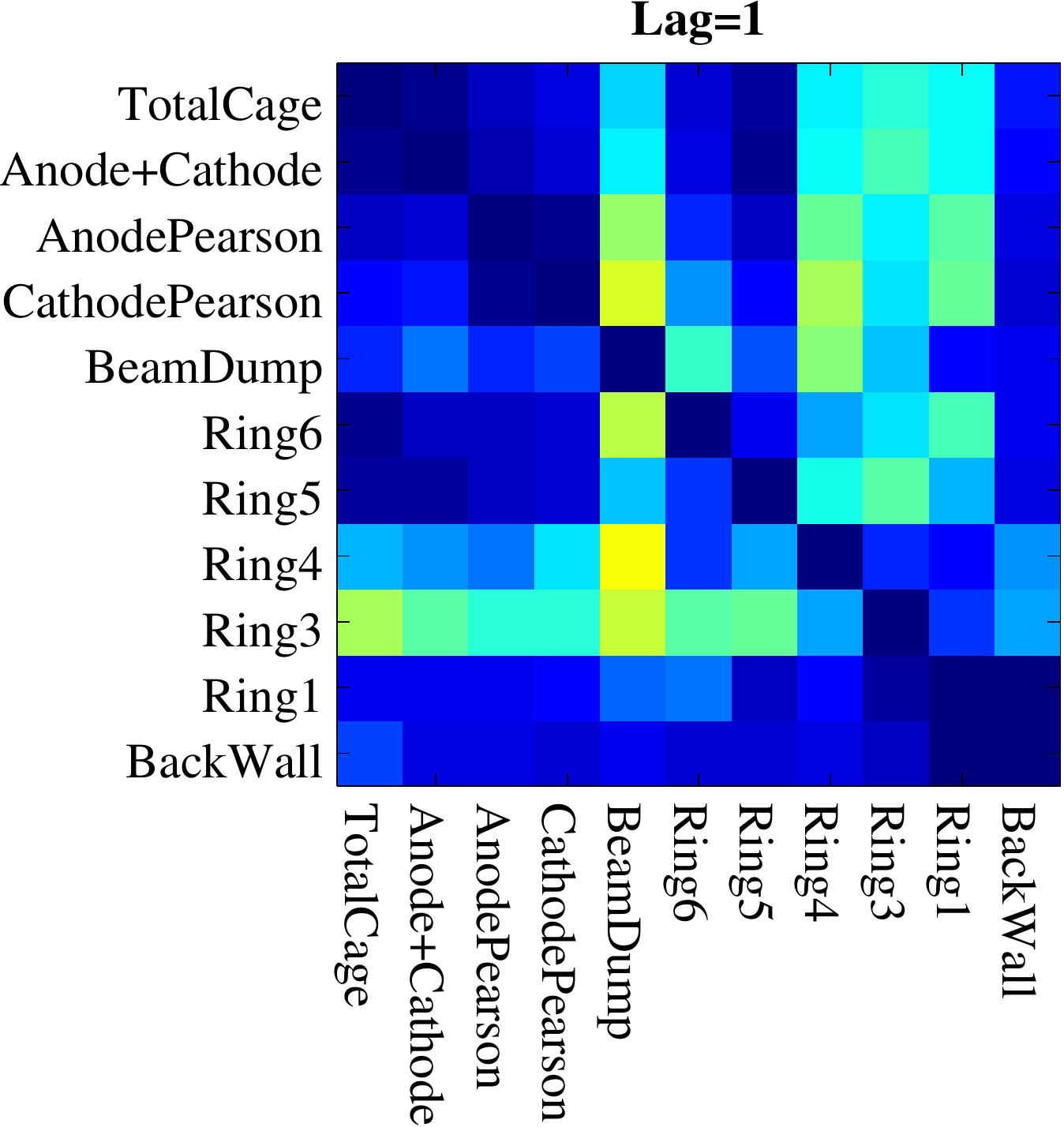}\;%
  \includegraphics[height=1.5in]{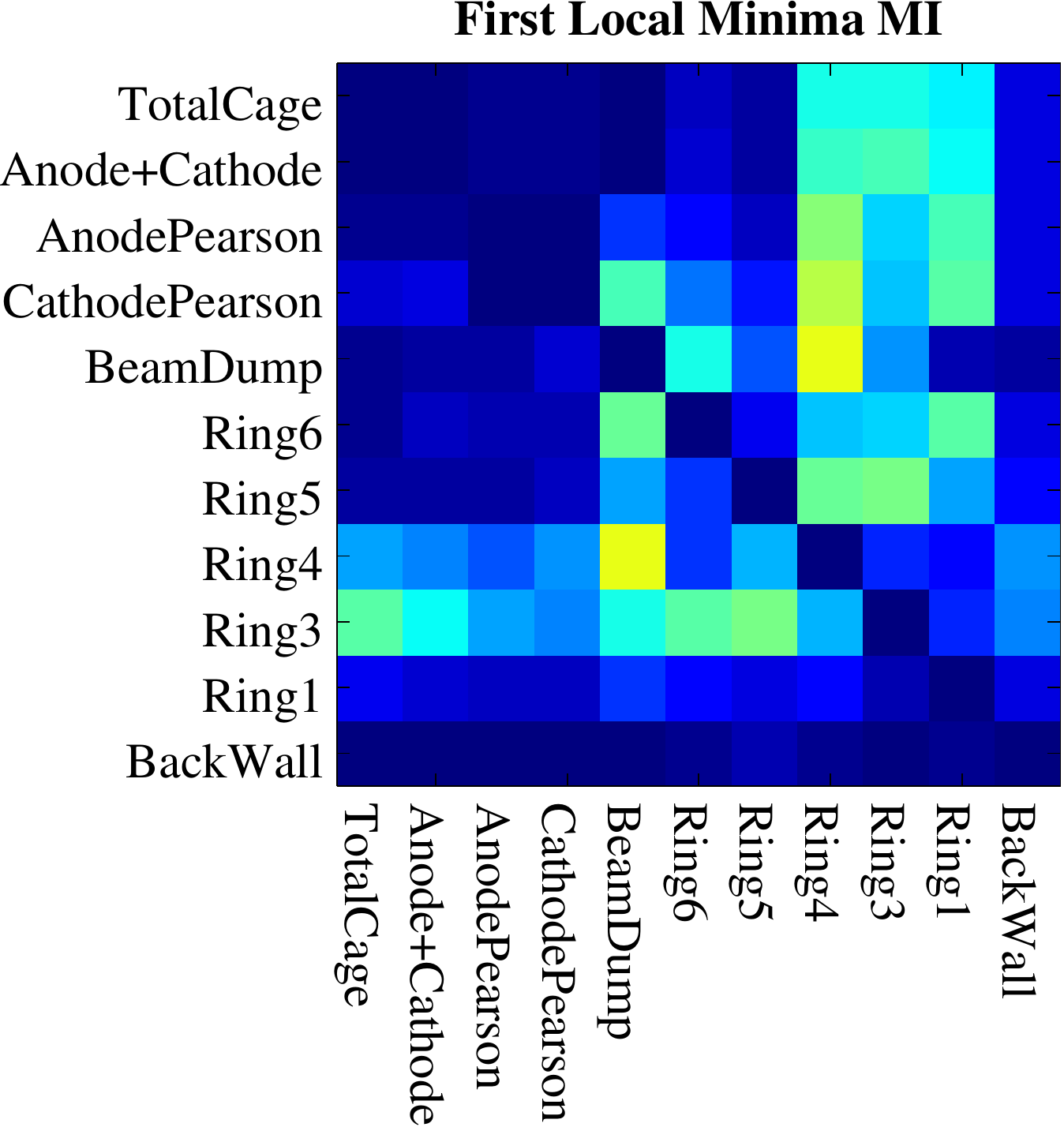}\;%
  \includegraphics[height=1.5in]{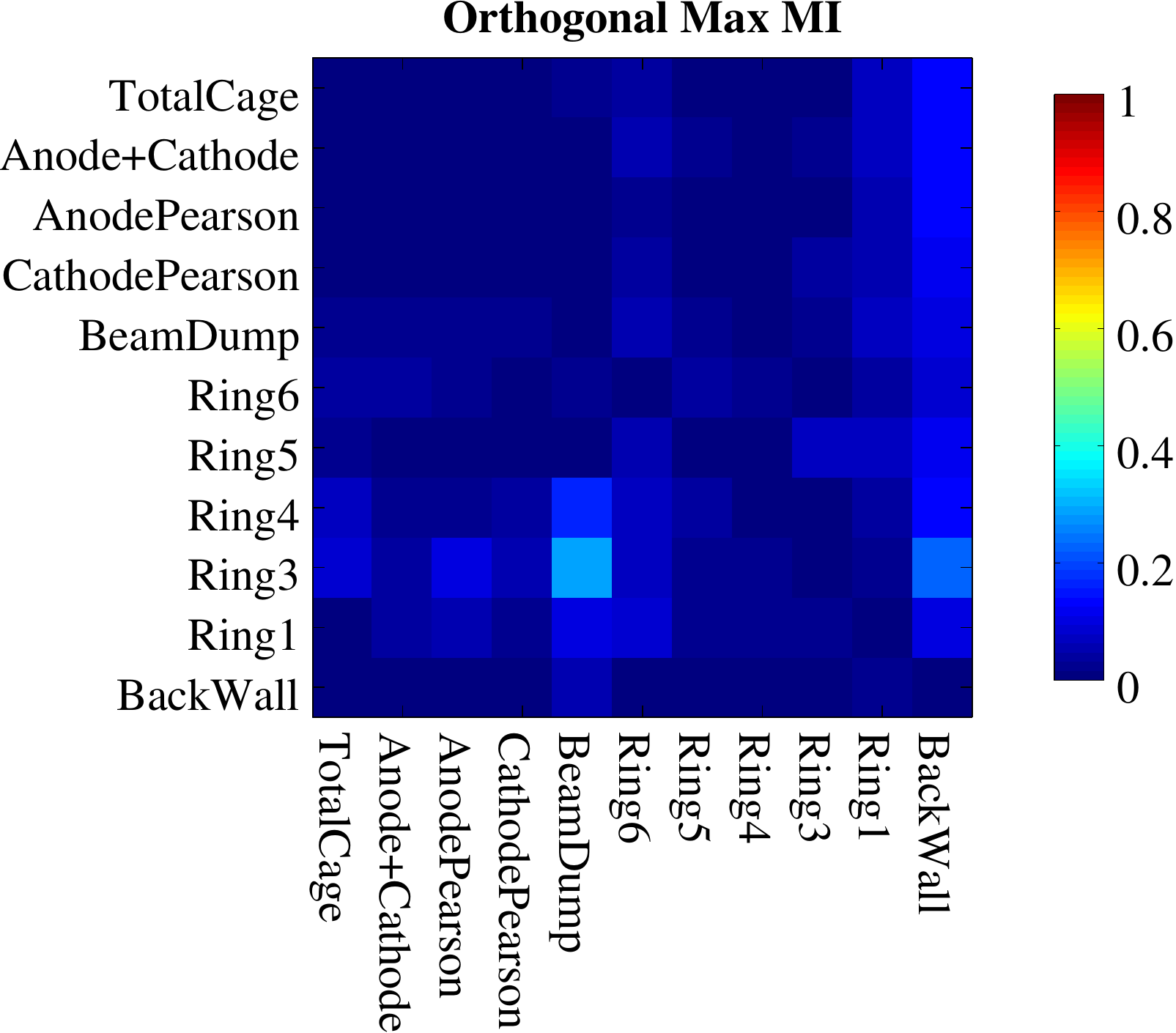}\;%
  \includegraphics[height=1.5in]{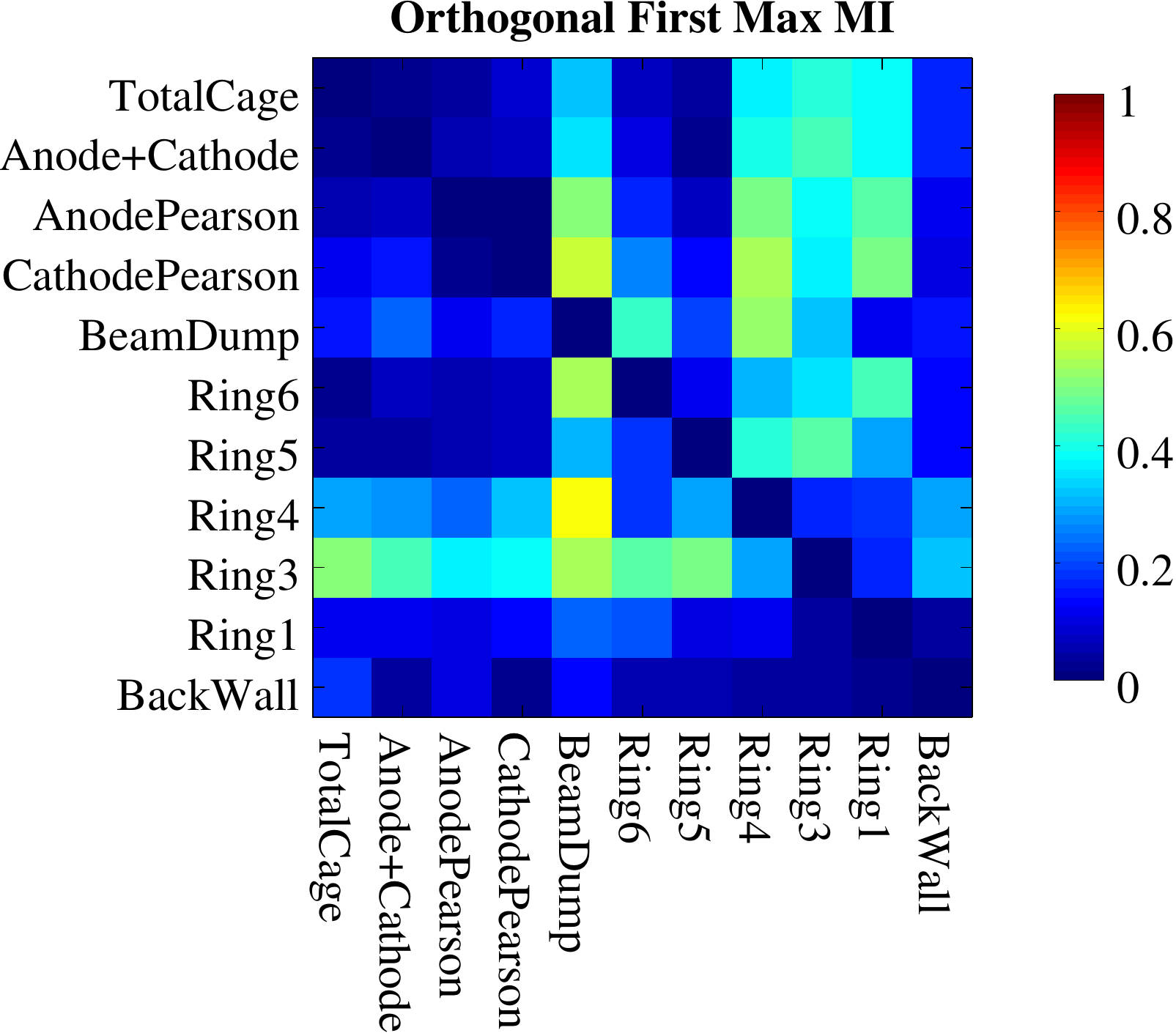}  
  \caption{Comparison of cross-mapping correlation coefficients resulting from different embedding lag choices for 11 variable Hall thruster data set. Top figure represent the lag correlations optimized for maximum average cross correlation coefficient, $\rho_{\tilde{i}|j}$, of the row variable, $\tilde{i}$, reconstructed from the column variable, $\tilde{j}$. Middle figures show these correlation coefficients for the other lag selection criteria. The bottom row figures show the corresponding clamped deficit of correlation coefficient relative to the maximum from the swept lag, $\Delta \rho_{\tilde{i}|j}$.}
  \label{Fig:EXP_CCM}
\end{figure}

Figure \ref{Fig:EXP_CCM} compares the correlation coefficients extracted from the experimental data set.  The baseline case, referred to as ``maximum swept lag'', results from sweeping the lag from 40ns-40$\mu$s in steps of 1$\mu$s and picking the lag that maximizes the sum  $\rho_{\tilde{X}|Y}+\rho_{\tilde{Y}|X}$ and represents the optimal effective correlation coefficient over the rage of lags tested.  The other cases shown include a single step of 40ns lag, the lesser lag of the two first minima of mutual information for the $i$ and $j$ data sets from all possible lags less than 40$\mu$s, the lesser lag of the two first local maxima in the orthogonalized coordinates from the $i$ and $j$ sets, and the lesser lag of the two global maxima of mutual information from the $i$ and $j$ sets in orthogonalized coordinates for all possible lags less than 40$\mu$s of the data sets.

Note that the swept maxima only approximately maximizes the correlation as the sweep skips 24 possible lag lengths per step due to the computational cost of computing the 80 reconstructions per pair rather than 2000. This coarseness is sufficient for qualitatively demonstrating the impact of the different lag choices but does allow for correlation coefficients slightly larger than the peak determined from the sweep.

The figure includes both the bi-directional correlation coefficients as well as the deficit of these correlation coefficient for the indicated lag relative the swept lag, $\Delta \rho_{\tilde{i}|j}$.  This deficit is the clamped difference between the swept coefficient and the coefficient resulting from the indicated lag clamped to zero as $\Delta \rho_{\tilde{i}|j}=max(0,\rho^{Swept}_{\tilde{i}|j}-\rho_{\tilde{i}|j})$ so that the deficit can be plotted with the same $[0,1]$ colormap. This is required because the coarseness of the sweep potentially results in some averages that exceed the sweep maximum identified as well as the fact that the sweep maximizes on the bi-directional average of coefficients which means potentially one of the two coefficients may exceed the swept value while the average is still below the swept average.  The visualized $\Delta \rho_{\tilde{i}|j}$ is 

Figure \ref{Fig:EXP_CCM} shows that selecting the lesser of global maximum orthogonal mutual information lags provides a correlation array that is considerably more similar to the result of sweeping the lag over a wide range than does the naive single step lag or either local extrema of mutual information based lags. The failure of the local criteria results from sensitivity of the numerical derivative of the mutual information with respect to lag length in determining the first local extrema from finite noisy data. Examining plots of mutual information versus lag length for some of the noisier signals highlights this sensitivity where the first extrema criteria is triggered relatively early at lags where the orthogonalized coordinates still indicate a significant deficit of mutual information compared to the global maxima due to the high frequency noise similar to the noise added to the chaotic model equations of the previous sections. Unlike with the noisy Lorenz example of Subsection \ref{SubSec:LorenzImpact}, the noise and limited data length of the experimental data seems to have impacted the first local maximum criteria as well as the first local minimum criteria. This will be explored more fully with a specific example in the following Section \ref{SubSec:DemoExp}.

\subsection{Demonstrative Example of Experimental Lag Choice Impact}
\label{SubSec:DemoExp}
To demonstrate the impact of nontrivial lag choice on CCM correlation, consider the impact on the ``AnodePearson'' and ``Ring1'' signal combinations. 
The correlation coefficient for reconstructions of ``AnodePearson'' current from the ``Ring1'' current triples between the lag identified by the sweep of lags and a simple lag of one timestep.  The pair is also notable because of the distinct asymmetry depending on the reconstruction direction.

Physically, it is reasonable to expect that information flows from the tightly coupled Hall thruster circuit of which the ``AnodePearson'' signal is a major component downstream to the noisy ``Ring1'' signals measured on the wall slightly behind the thruster face.  Consistent with the expectations of CCM, the non-obvious implications of this is that the neighbors of the ``Ring1'' signal perform better at selecting points for reconstructing ``AnodePearson'' than the converse. Because ``AnodePearson'' causes an larger effect on ``Ring1'', the reconstruction of ``AnodePearson'' from ``Ring1'' neighbors is more highly correlated to the true ``AnodePearson'' signal than ``Ring1'' is to the reconstruction of ``Ring1''.   Further exploration of the physical implications of the identified CCM correlations from Figure \ref{Fig:EXP_CCM} are left to Reference \cite{Eptempest2018} as the purpose of this work is simply to explore the impact of lag choice on the ability of CCM to identify strong correlations in finite noisy data.

\begin{figure}
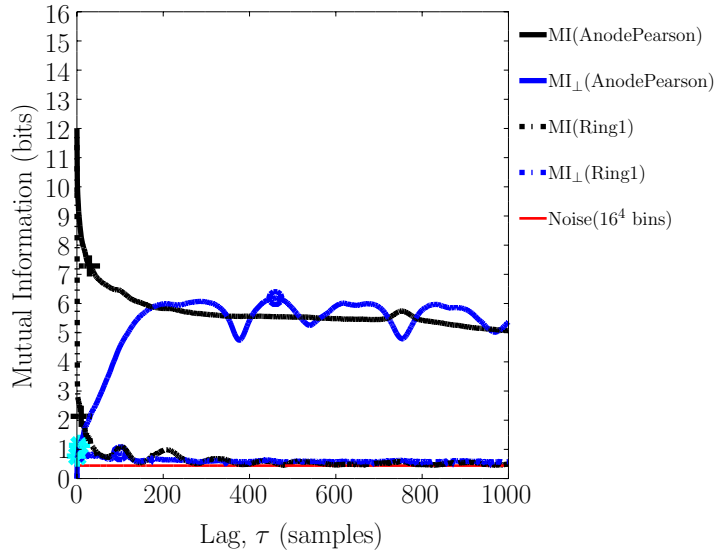

\scalebox{0.5}{\include{EXP_CCM_MIVsLag_PlusFirstMax_AnodePearson-Ring1}}
  \caption{Mutual information as a function of time lag for ``AnodePearson'' and ``Ring1'' signals extracted from experimental Hall thruster data set in both lag and orthogonalized coordinates. first local minimum (black), global orthogonalized maxima (blue), and first local orthogonalized maxima (cyan) points are marked with symbols. The noise floor is also included.}
  \label{Fig:EXP_CCM_MIVsLag_AnodePearson-Ring1}
\end{figure}

Table \ref{Tab:CompareImpact} provides a comparison of the correlation coefficients achieved for the reconstructions by choice of time lag. The relationship between these choices of lag and the measures of mutual information within possible four dimensional lags of the two signals across lags ranging from 1 to 1000 samples can be seen in Figure \ref{Fig:EXP_CCM_MIVsLag_AnodePearson-Ring1}.

\begin{table}
  \scalebox{0.7}{\begin{tabular}{r||c||c|c|c|c|c|c|c}
      CCM $\rho_{\tilde{i}|j}$ & Max Swept & Min$_1$ \MI(X) & Min$_1$ \MI(Y) & Max \MI$_\perp$(X) & Max \MI$_\perp$(Y) & Max$_1$ \MI$_\perp$(X) & Max$_1$ \MI$_\perp$(Y)\\ \hline
      $\tilde{i}=$ AnodePearson, $j=$ Ring1 & 0.67 & 0.47 & 0.22 & 0.33 & 0.62 & 0.21 & 0.21 \\ 
      $\tilde{i}=$ Ring1, $j=$ AnodePearson & 0.32 & 0.28 & 0.26 & 0.33 & 0.27 & 0.25 & 0.25 \\ \hline
      Lag Used (Samples)  & 76   & 29   & 10   & 460  & 100  & 6    & 1\\
  \end{tabular}}
  \caption{Comparison of cross mapping correlation coefficient as a function of lag choice for ``AnodePearson''-''Ring1'' pair.}
  \label{Tab:CompareImpact}
\end{table}

In Figure \ref{Fig:EXP_CCM_MIVsLag_AnodePearson-Ring1}, it is first clear that the ``Ring1'' data is considerably noisier and less informative about the dynamics of the system with only approximately 1-bit of mutual information in the orthogonalized coordinate maximum where short time redundancy has been removed.  That only exceeds the finite data noise floor of 1/2-bit for the 125k points on $16^4$ bins. The ability for the cross mapping to reconstruct the ``AnodePearson'' data with any degree of confidence is remarkable considering how nearly indistinguishable the ``Ring1'' signal appears to be relative to noise. 

\begin{figure}
\includegraphics[width=\textwidth]{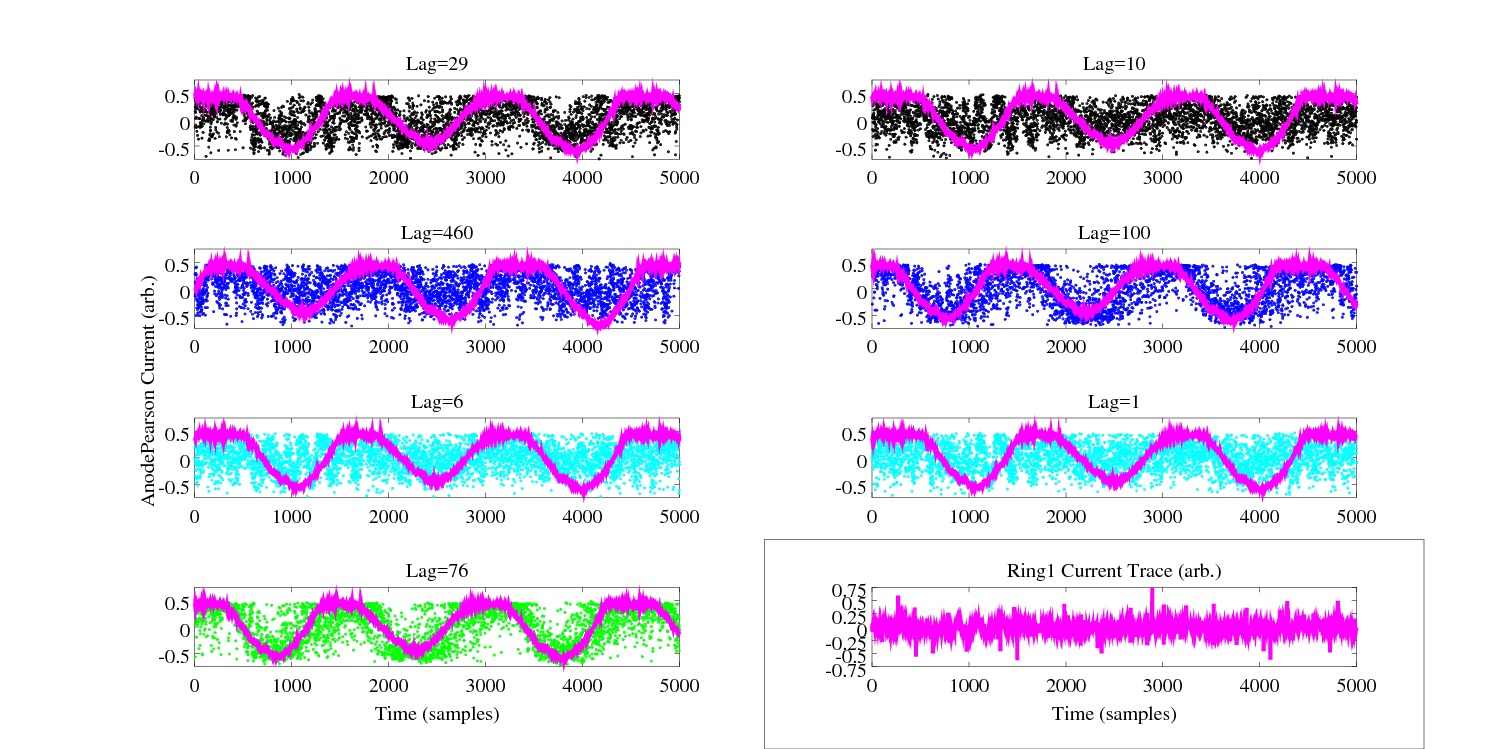}
  \caption{Comparison of original and reconstructed ``AnodePearson'' current using various cross mapped lags of ``Ring1'' current depicted in lower right inset. The original signals are denoted by magenta curves.  The reconstructions points correspond to the lags indicated by criteria on mutual information from Figure \ref{Fig:EXP_CCM_MIVsLag_AnodePearson-Ring1}. The top row (black) are lags based off the original first local minima in \MI in time lag coordinates.  The second row (blue) are lags based off the global maxima in orthogonal coordinates.  The third row (cyan) are based off of the first local maxima in orthogonal coordinates.  In each case the left side are lags identified from the ``AnodePearson'' data while the right side is based off ``Ring1'' data. The reconstruction in the fourth row (green) is the reference reconstruction for the lag identified from the maximum average correlation in the sweep of lags. }  
  \label{Fig:EXP_CCM_Trace-AnodePearson-Ring1}
\end{figure}

To provide a better understanding of the what the difference between a 0.22 and 0.67 correlation coefficient looks like in terms of reconstructions, Figure \ref{Fig:EXP_CCM_Trace-AnodePearson-Ring1} shows ``AnodePearson(Ring1)''-reconstructions using the various lags identified as candidate lags for reconstruction including the sweep over all possible lags between $1-1000$ samples.

\begin{figure}
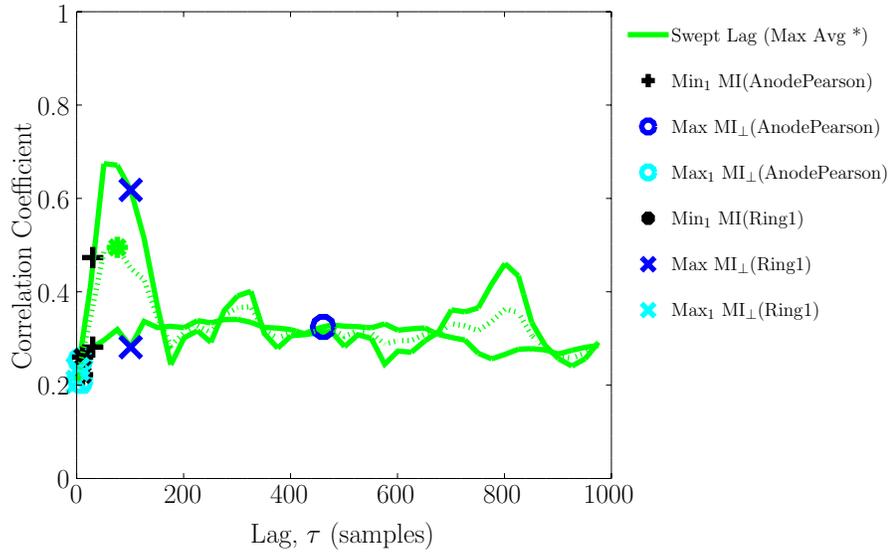

\scalebox{0.5}{\include{EXP_CCM_CorrelationVsLag_PlusFirstMax_AnodePearson-Ring1}}
  \caption{Correlation coefficient pairs and average versus lag length for ``AnodePearson''-''Ring1'' experimental data pair with points marked corresponding to potential lag choice criteria from Table \ref{Tab:CompareImpact}.}
  \label{Fig:EXP_CCM_CorrelationVsLag_AnodePearson-Ring1}
\end{figure}

The lag based off maximizing the mean correlation coefficient resulted from sweeping lags between 1 and 1000 samples in steps of 25.  The relationship of the correlation coefficients as a function of lag for the ``AnodePearson'' ``Ring1'' pair can be seen in Figure \ref{Fig:EXP_CCM_CorrelationVsLag_AnodePearson-Ring1}. 

\begin{figure}
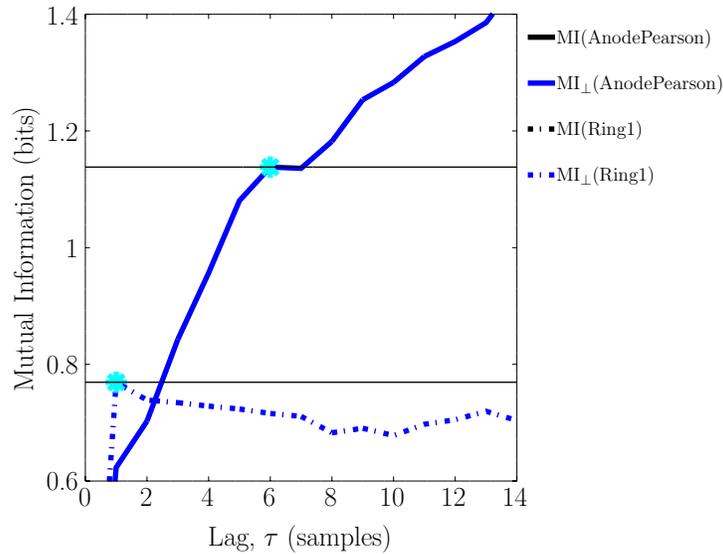

\scalebox{0.5}{\include{EXP_CCMFirstMax_MIVsLag_PlusFirstMax_ZoomIn_AnodePearson-Ring1}}
  \caption{Zoom of initial mutual information as a function of time lag for ``AnodePearson'' and ``Ring1'' signals extracted from experimental Hall thruster data set in both lag and orthogonalized coordinates. First local orthogonalized maxima (cyan) points are marked with symbols with added horizontal lines to visually emphasize that the points are in fact early local maxima.}
  \label{Fig:EXP_CCM_MIVsLag_AnodePearson-Ring1_ZoomIn}
\end{figure}

The reason the correlation results for the first local maximum in orthogonalized coordinates are considerably worse than those with the global maximum condition is that the local maximum criteria is triggered early as a result of fluctuations in the mutual information curves.  Figure \ref{Fig:EXP_CCM_MIVsLag_AnodePearson-Ring1_ZoomIn} shows a zoomed version of Figure \ref{Fig:EXP_CCM_MIVsLag_AnodePearson-Ring1} in the vicinity of the lags identified by the first local maxima. The local maxima identified are triggered by fluctuations in the mutual information curves.  On the anode trace, the identified maximum is a minor single lag deviation.  Schemes could be devised to handle these sorts of special cases, but it demonstrates the general sensitivity of any 'local' criterion when compared to a global criterion. 

\section{Conclusion}

From several potential alternative criteria considered, the shorter of the two global maxima of mutual information in the stretched orthogonal coordinates is proposed as a robust lag criterion for convergence cross mapping. Like Fraser and Swinney's first local minimum criterion, this criterion identifies a timescale neither too short for which the embedding has significant redundance nor too long resulting in irrelevance.  However, the global maximum \MI criterion is significantly more robust to noise and algorithmic error due to finite data and discretization sizes than other methods based on a local criterion.  Though alternative prescriptions can sometimes identify stronger correlations, as was the case with the shorter of first local maxima applied to the noisy Lorenz model equation, the sensitivity of local criteria demonstrated throughout this work makes automating their use challenging, particularly when applied to noisy, finite datasets where pairwise relationships need to be investigated.  The sensitivity in identification of local extrema suggests that the benefits of improved robustness are likely to warrant selection of the global instead of local criteria.  

Regardless of the lag choice prescription, the act of transforming into the discrete Legendre coordinate system for directly removing the short lag redundant portion of observed mutual information for continuous nonlinear systems certainly provides a means of measuring mutual information that is more relevant to the identification of coupled system dynamics. Artificially high mutual information values at short lags were shown to be a numerical artifact of the measurement algorithm.
With this understanding, potential criteria resulting from both local and global maxima of mutual information in these coordinates were developed and shown to identify lags which produced higher average correlations for CCM. Furthermore, lags resulting from maximizing mutual information are conceptually more satisfying than the local minimum criterion because they represent lags where nonrandom structure in the data is most evident which in turn is potentially meaningful with respect to low dimensional projections of causal system dynamics if it exists.

It is recognized that the examples of the work, both model and experimental, appear particularly well suited only to very low dimensional embedding.  Nevertheless, the evidence for the use of the global maximum criterion, which remains only a heuristic, implies it is an attractive option in this context, particularly considering the stronger conceptual motivation for the choice in terms of maximally emphasizing nonrandom structure in the data.  More work is required to assess the performance of these techniques across a wider array of systems and for higher dimensional dynamics.  Of particular interest for future inquiry is the impact of these techniques on systems with multiple time scales and whether these tools can aid the decomposition of such dynamics into their constituent scales.  It is also expected that any approach to tackle higher dimensional systems with finite data will likely require concepts from the multiview embedding technique to overcome the curse of dimensionality.

\begin{acknowledgments}
The authors like to acknowledge the support provided by the Air Force Office of Scientific Research (AFOSR) grants No. FA9550-18RQCOR107 (PO: Dr. Fahroo), No. FA9550-17RQCOR497 (PO:Pokines) and the National Research Council Grant No. FA9550-17-D0001.
\end{acknowledgments}





\bibliography{database_all}

\end{document}